\documentclass[11pt]{article}
\usepackage{jheppub}
\usepackage{mathtools}

\usepackage{hyperref}
\hypersetup{
	unicode,	
	colorlinks,
	citecolor=[rgb]{0,.3,.5}, 
	linkcolor=[rgb]{0,.3,.5},
	urlcolor=[rgb]{0,.3,.5}, 
	bookmarksopen=true,
	bookmarksopenlevel=\maxdimen,
}

\usepackage{amsmath, amssymb}		
\usepackage{mathrsfs} 			
\usepackage{bm}				
\usepackage{youngtab}  			
\usepackage{multirow}
\usepackage{color}
\usepackage{slashed}

\DeclareMathOperator{\Tr}{\textrm{Tr}}

\numberwithin{equation}{section}

\newcommand{\bbA}{{\mathbb A}}

\newcommand {\cA}{{\cal A}}
\newcommand {\cB}{{\cal B}}

\newcommand {\cD}{{\cal D}}
\newcommand {\cE}{{\cal E}}
\newcommand {\cF}{{\cal F}}

\newcommand {\cH}{{\cal H}}
\newcommand {\cI}{{\cal I}}
\newcommand {\cJ}{{\cal J}}
\newcommand {\cK}{{\cal K}}
\newcommand {\cL}{{\cal L}}
\newcommand {\cM}{{\cal M}}
\newcommand {\cN}{{\cal N}}
\newcommand {\cO}{{\cal O}}

\newcommand {\cV}{{\cal V}}

\def\a{\alpha}
\def\b{\beta}

\def\g{\gamma}

\def\q{\theta}

\def\L{\Lambda}

\newcommand{\rd}{{\mathrm d}}
\newcommand{\ri}{{\mathrm i}}

\newcommand{\ad}{{\dot{\alpha}}}
\newcommand{\bd}{{\dot{\beta}}}
\newcommand{\dalpha}{{\dot{\alpha}}}
\newcommand{\dbeta}{{\dot{\beta}}}
\newcommand{\dgamma}{{\dot{\gamma}}}
\newcommand{\ddelta}{{\dot{\delta}}}

\newcommand{\veps}{\varepsilon}
\newcommand{\eps}{{\epsilon}}
\newcommand{\eol}{\notag \\}

\newcommand{\ul}{\underline}
\newcommand{\pa}{\partial}

\newcommand{\rep}[1]{\mathbf{#1}}
\newcommand{\brep}[1]{\mathbf{\overline{#1}}}

\newcommand{\HC}{\text{h.c.}}

\newcommand{\balpha}{{\boldsymbol \alpha}}
\newcommand{\bbeta}{{\boldsymbol \beta}}
\newcommand{\bgamma}{{\boldsymbol \gamma}}
\newcommand{\bdelta}{{\boldsymbol \delta}}

\newcommand{\SL}{\textrm{SL}}
\newcommand{\SO}{\textrm{SO}}
\newcommand{\SU}{\textrm{SU}}
\newcommand{\gU}{\textrm{U}}

\makeatletter
\g@addto@macro\bfseries{\boldmath}
\makeatother

\title{$\boldsymbol{\cN\!=\!4}$ conformal supergravity: the complete actions}

\author[a]{Daniel Butter,}
\author[b]{Franz Ciceri,}
\author[c]{and Bindusar Sahoo}

\affiliation[a]{
George P. and Cynthia Woods
Mitchell Institute for 
Fundamental Physics and Astronomy,
Texas A\&{}M University,
College Station, TX 77843, USA}

\affiliation[b]{
Max-Planck-Institut f\"ur Gravitationsphysik (Albert-Einstein-Institut) 
M\"uhlenberg 1, 
D-14476 Potsdam,
Germany
}

\affiliation[c]{
Indian Institute of Science Education and Research, 
Vithura, Thiruvananthapuram - 695551, 
India
}

\emailAdd{dbutter@tamu.edu}
\emailAdd{franz.ciceri@aei.mpg.de}
\emailAdd{bsahoo@iisertvm.ac.in}

\abstract{
The most general class of 4D $\cN=4$ conformal supergravity actions depends on a holomorphic function of the scalar fields that parametrize an $\SU(1,1)/\gU(1)$ coset space. The bosonic sector of these actions was presented in a letter \cite{Butter:2016mtk}. Here we provide the complete actions to all orders in the fermion fields. They rely upon a new $\cN=4$ density formula, which permits a direct but involved construction. This density formula also recovers the on-shell action for vector multiplets coupled to conformal supergravity. Applications of these results in the context of Poincar\'e supergravity are briefly discussed.}

\begin{document}
\maketitle

\section{Introduction}

It is well-known that local supersymmetry imposes stringent restrictions on the higher-derivative structure of supergravity theories. Studying the allowed invariants is directly relevant to understand the effective actions arising in supersymmetric compactifications of string theory. The latter play an important role, for instance, in computing corrections to the entropy of certain extremal black holes. Constructing these higher-derivative actions is however notoriously difficult. This is because the addition of higher-derivative terms typically 
triggers the start of an iterative procedure where the transformation rules, and consequently the action, have to be repeatedly modified to ensure closure of the supersymmetry algebra on the fields modulo the equations of motion.  A major exception is when supersymmetry is realized off-shell, since in this case the supersymmetry algebra closes independently of the dynamics considered. The transformation rules are then fixed and supersymmetric higher-derivative actions can be analyzed on their own. 

A standard method for constructing off-shell supergravity actions is to employ a description in which the theory is invariant under local superconformal transformations and coupled to various compensating matter multiplets. These compensating matter fields can optionally be used to fix the extraneous conformal symmetries, and in this way, ensure gauge equivalence with the original Poincar\'e supergravity \cite{Kaku:1978ea, Freedman:2012zz}. In the superconformal formulation, the local Weyl symmetry implies that the conformal mode of the metric, along with its superpartners, must be absent. The resulting so-called Weyl multiplet, turns out (when it exists) to be a unique off-shell multiplet, which contains fewer fields than its Poincar\'e analogue.

In four dimensions, the Weyl multiplet exists up through $\cN=4$, which means it can in fact provide an off-shell formalism even in cases when its Poincar\'e counterpart does not. There is one type of action which involves the Weyl multiplet alone with no other matter multiplets or compensators -- this is the conformal supergravity action which corresponds to the supersymmetrization of the square of the Weyl tensor. This particular action can be considered either as an off-shell action in its own right, or as a higher-derivative correction to Poincar\'e supergravity. It will be the main focus of this paper.

Conformal supergravity actions in four dimensions were first constructed for $\cN\leq 2$ forty years ago \cite{Kaku:1978nz, deWit:1979dzm,Bergshoeff:1980is}. In the $\cN=4$ case, while the full non-linear superconformal transformations of the Weyl multiplet fields were already determined in \cite{Bergshoeff:1980is}, the conformal supergravity action remained largely unstudied beyond the linearized level until this decade.\footnote{Even the 4D $\cN=3$ Weyl multiplet was \textit{terra incognita} until recently \cite{vanMuiden:2017qsh,Hegde:2018mxv}, when its transformation laws were first written down, and the action remains unstudied. In three dimensions, the conformal supergravity actions are Lorentz-Chern-Simons and their complete form for $\cN \leq 6$ were constructed in \cite{Butter:2013rba, Nishimura:2013poa, Kuzenko:2013vha}. The half-maximal $\cN=8$ multiplet possesses at most a pseudo-action \cite{Bergshoeff:2010ui}. In six dimensions, there are three conformal gravity actions, but $(1,0)$ supersymmetry selects out two \cite{Butter:2016qkx,Butter:2017jqu}; the unique $(2,0)$ supersymmetric combination was partly constructed in \cite{Butter:2017jqu} by lifting from $(1,0)$. In five dimensions, there is no pure conformal supergravity action although the Weyl multiplet exists for $\cN=1$.}
Some results were first obtained for the purely bosonic sector of the action by computing the conformal anomaly of  $\cN=4$ vector multiplets in a background of conformal supergravity \cite{Buchbinder:2012uh}. All the terms up to quadratic order in the fermion fields were then derived in \cite{Ciceri:2015qpa} by starting with the linearized action and iteratively adding terms required by supersymmetry. Proceeding to all orders in this manner would however seem practically impossible. This has to do the fact that, unlike for $\cN\leq 2$, the $\cN=4$ Weyl multiplet contains a dimension-1/2 fermion field $\Lambda_i$. This means that the complete action should be fairly involved, with terms up to $\mathcal O(\Lambda^8)$. 

Another singular feature of the $\cN=4$ Weyl multiplet is the presence of dimensionless scalar fields $\phi_\balpha$ parametrizing an $\SU(1,1) / \gU(1)$ coset space. It is therefore possible to consider a whole class of conformal supergravity actions, where the leading Weyl squared term is multiplied by a function of the coset scalars. This modification was already mentioned in \cite{Fradkin:1981jc}, and represents a deviation from the $\cN<4$ cases for which there is a unique action. Such ``non-minimal'' couplings later emerged in the effective action of type IIA strings compactified on $K3\times T^2$ \cite{Harvey:1996ir} in the form of a modular function multiplying the square of the Weyl tensor, as well as in the semiclassical approximation of the microscopic degeneracy formula for certain dyonic $\cN=4$ black holes \cite{LopesCardoso:2004law, Jatkar:2005bh}. In the context of on-shell Poincar\'e supergravity, an analysis of higher-derivative invariants \cite{Bossard:2012xs, Bossard:2013rza} established the existence of similar couplings where the Riemann tensor squared is multiplied by a generic holomorphic function. The corresponding class of conformal supergravity actions however remained entirely unknown, even at the purely bosonic level.

This was resolved in a recent letter \cite{Butter:2016mtk}, in which we reported on the construction of the full $\cN=4$ conformal supergravity
action for the case of a generic holomorphic function. This is the most general case, which encompasses all possible actions, including the ``minimal'' action partially constructed in \cite{Ciceri:2015qpa} which is recovered by setting the function to a constant.
The complete set of bosonic terms was
presented, but the mechanism for its construction was only briefly sketched. The aim of this
paper is to fill in the gaps by presenting the fermionic terms and describing in detail the method that was used. The complete action
turns out to be \emph{significantly} complicated, and the computer algebra program \texttt{Cadabra} \cite{Peeters.Cadabra1,Peeters.Cadabra2} played an essential role in its construction. In the following, we sketch the main lines of our approach.

In general, the problem of constructing off-shell supersymmetric actions can be tackled in various ways. A first option is to build the action by iteratively adding terms to impose supersymmetry order by order in the fields. 
As mentioned above, this method was used for the minimal $\cN=4$ conformal
supergravity action in \cite{Ciceri:2015qpa}.
Another common approach is to employ superspace, which guarantees supersymmetry but requires
integration over Grassmann (fermionic) coordinates. 
The integration must be over
the entire superspace or some invariant subspace (e.g. chiral superspace),
but for sufficiently many supercharges this leads to
rather high dimension expressions corresponding to
terms with a large number of bosonic derivatives.
While there are ways to reduce the number of fermionic dimensions,
for example, by introducing an auxiliary bosonic manifold as with
harmonic/projective superspace or pure spinor superspace, these typically require
infinite numbers of auxiliary fields or other complications.

In this paper, we take a more pragmatic approach which relies on the construction of a generic supersymmetric action principle,
also known as a \emph{density formula},
directly at the component level. Such a density formula is
built upon an abstract multiplet whose component fields appear linearly in the expression, along with some of the supergravity fields such as the vielbein. The multiplet
in question is typically required to obey only very mild supersymmetry constraints
(e.g. chirality), and it is in this sense that the action principle is generic.
If one can build such a multiplet, for example by combining more fundamental constituents,
the action principle can be applied.
This approach is not distinct from superspace: whenever an invariant superspace exists,
the corresponding component action always falls into this type, where the multiplet in question
is identified with the superspace Lagrangian.
The converse does not generally hold, and a density formula may exist in the absence
of a corresponding superspace.

As we will show, there exists such a density formula for
four-dimensional $\cN=4$ theories with local superconformal symmetry.
It is based on an abstract multiplet involving superconformally primary fields $C^{ij}{}_{kl}$,
$\bar C^{ij}{}_{kl} = (C^{kl}{}_{ij})^*$, and $A^{ij}{}_{kl} = (A^{kl}{}_{ij})^*$, 
each in the $\rep{20'}$ of the $\rm SU(4)$ R-symmetry group, 
and whose supersymmetry
transformations into the $\rep{60}$ and $\brep{60}$ are constrained. That is,
\begin{align}
\delta _\epsilon C^{ij}{\!}_{kl} &= \bar\epsilon^m \,\Xi^{ij}{\!}_{kl,m} 
	+ \bar\epsilon_m\,\Xi^{ij,m}{\!}_{kl} \,, \nonumber \\
\delta_\epsilon A^{ij}{\!}_{kl} &= \bar\epsilon^m \,\Omega^{ij}{\!}_{kl,m}
	+ \bar\epsilon_m\,\Omega^{ij,m}{\!}_{kl} \,,
\label{eq:deltaCA}
\end{align}
for fermions $\Xi$ and $\Omega$ whose traceless parts are fixed as
\begin{align}
[\Xi^{i j}{\!}_{kl,m}]_{\brep {60}} = [2\Lambda_m A^{i j}{}_{kl}]_{\brep{60}}~, \qquad 
[\Xi^{ij,m}{\!}_{kl}]_{\rep{60}} = 0~, \qquad
[\Omega^{i j}{}_{kl,m}]_{\brep{60}} &= [\Lambda_m \bar C^{ij}{}_{kl}]_{\brep{60}}
\label{eq:Bigconstraint}
\end{align}
in terms of the dimension-1/2 fermion $\Lambda_i$ of the Weyl multiplet.
This implies that the fields $C^{ij}{}_{kl}$ and $A^{ij}{}_{kl}$ must
be intricately related. The density formula is
\begin{align}\label{E:DensityFormula}
e^{-1} \cL &= F
	+ 2 \,\bar \psi_{\mu i} (\Omega^{\mu i} + \gamma^\mu \Omega^i)
	+ 2 \,\bar \psi_{\mu}{}^i (\Omega^{\mu}{}_i + \gamma^\mu \Omega_i)
	\eol & \quad
	+ \frac{1}{8} \bar \psi_{[\mu i} \gamma^\mu \psi_{\nu]}{}^j \, \cE^\nu{}^i{}_j
	- \frac{1}{8} \Big(i \,\bar \psi_{\mu i} \gamma^{\mu\nu} \psi_{\nu j} \cE^{i j}
		+ \frac{i}{2} \veps^{\mu\nu \rho\sigma} \bar\psi_{\mu i} \psi_{\nu j} \cE_{\rho\sigma}{}^{i j}
		+ \HC
	\Big)
	\eol & \quad
	+ \frac{i}{8} \veps^{\mu\nu\rho\sigma} \Big(
		\bar\psi_{\mu i} \psi_{\nu j}
			\,\bar \psi_{\rho k} \gamma_\sigma \rho_{r s}{}^k \,\veps^{i j r s}
		+ 2\,\bar\psi_{\mu i} \psi_{\nu j}\,
			\bar \psi_{\rho}{}^k \gamma_\sigma \kappa{}^{i j}{}_{k}
		+ \HC
	\Big)
	\eol & \quad
	- \frac{i}{4} \veps^{\mu\nu\rho\sigma} \Big(
		\bar\psi_{\mu i} \psi_{\nu j} \bar\psi_{\rho k} \psi_{\sigma l}
		\,\veps^{k l r s} \,C^{i j}{}_{r s}
		+ 2 \,\bar\psi_{\mu}{}_i \psi_{\nu}{}_j \bar\psi_{\rho}{}^k \psi_{\sigma}{}^l
		\,A^{i j}{}_{k l}
		+ \HC \Big)\,.
\end{align}
The fields $C^{ij}{}_{kl}$ and $A^{ij}{}_{kl}$ of lowest Weyl weight multiply
four gravitini $\psi$, while higher weight fields of the abstract multiplet
appear with fewer gravitini: these include
fermions $\rho_{ij}{}^k$, $\kappa^{ij}{}_k$, $\Omega_{a i}$ and $\Omega_i$,
as well as bosons $\cE_a{}^{i}{}_j$, $\cE_{a b}{}^{ij}$, $\cE^{i j}$, and $F$. They
descend from the fields $C^{ij}{}_{kl}$ and $A^{ij}{}_{kl}$ via
supersymmetry in a manner that will be described in due course. Their superconformal
transformations, which leave \eqref{E:DensityFormula} invariant, turn out to be determined
entirely by the basic supersymmetry constraints \eqref{eq:Bigconstraint}. Provided
such fields can be constructed out of more fundamental constituents, invariant actions follow. The composite field $F$ then contains all the bosonic terms  of these actions.

This is by no means the only possible density formula for $\cN=4$,
but it turns out to be sufficient for our needs. Once one specifies the form of the basic fields $C^{ij}{}_{kl}$ and $A^{ij}{}_{kl}$ in terms of the $\cN=4$ Weyl multiplet fields, it allows the direct construction
of the class of superconformal Weyl squared actions which depend on a generic holomorphic function.
As a bonus, we will show that the locally superconformal super Yang-Mills (SYM) action
\cite{deRoo:1984zyh, deRoo:1985jh}
can also be described by the same density formula, albeit at the on-shell level only.

The paper is arranged as follows. In section \ref{S:Density},
we describe the superform action principle that leads to the density formula
\eqref{E:DensityFormula}. In section \ref{S:AllActions}, we explain how to apply
the density formula to build the conformal supergravity actions, by explicitly solving
the supersymmetry constraints on the basic fields $C^{ij}{}_{kl}$ and $A^{ij}{}_{kl}$.
We also give a proof that this must lead to the most general class of
actions. In section \ref{S:Results} we review the bosonic action presented
in \cite{Butter:2016mtk} and give the covariant (non-gravitino) two-fermion contributions. These
terms already are quite substantial in number. Because of the sheer complexity of
the full action, we cannot present it explicitly here. Instead, we include an addendum file
in the \texttt{arXiv} submission that contains the full set of terms.
In section \ref{S:Origins}, we speculate on the origin of the density formula and show
how it can recover the on-shell SYM action, suggesting that it may have broader
applications than conformal supergravity alone.
Two appendices are included. The first summarizes the conventions we use
and gives technical details on the $\cN=4$ superspace we employ. The second elaborates
on certain details of the analysis of superspace Bianchi identities.

\section{Superform action principle}\label{S:Density}

The superform action principle is a common method to construct supersymmetric
invariants. In the superspace literature, it was proposed by Gates et al.
under the name ``ectoplasm'' \cite{Gates:1997kr,Gates:1997ag}, but is equivalent to how actions
are built in the rheonomic or group manifold approach to supersymmetric theories,
see \cite{DAuria:1982mkx} and the review in \cite{RheonomicBook}.
The approach itself has nothing to do with supersymmetry \emph{per se}, but
can be applied to any local symmetry that can be interpreted as a diffeomorphism
in some higher dimensional spacetime. The basic approach is as follows. Suppose
one has some $D$-dimensional manifold $\cM$ and some $d$-dimensional
submanifold $M$ over which one wants to integrate a $d$-form $J$.
The action integral
\begin{align}
S = \int_{M} J
\end{align}
transforms under arbitrary diffeomorphisms on $\cM$ as
\begin{align}
\delta_\xi J = \imath_\xi \rd J + \rd \imath_\xi J \quad\implies\quad
\delta S = \int_{M} \imath_\xi \rd J + \int_{\pa M} \imath_\xi J~.
\end{align}
The second term vanishes if $M$ is closed, or if we put fall-off conditions at infinity.
In either case, the condition for invariance is that the $d$-form $J$ should
be closed, $\rd J = 0$.
Schematically, we may decompose a general diffeomorphism into pieces normal and tangent
to $M$. Invariance under tangent diffeomorphisms follows if $M$ is closed,
while invariance under normal diffeomorphisms means that the precise embedding of
$M$ into $\cM$ is irrelevant: this latter condition is what requires
$J$ to be a closed form on $\cM$.

The key idea is to think of the submanifold $M$
as the full spacetime and $\cM$ as a larger manifold where we have geometrized
some gauge symmetry (such as supersymmetry). Finding a gauge invariant Lagrangian
is then translated into finding a closed $d$-form $J$.
This is distinct from the procedure of superspace integration, where one
integrates over the anti-commuting $\theta$ coordinates of chiral or full superspace.
Such actions typically exist only for simple superspaces; but where they do, they can
always be identified with some appropriate closed $d$-form $J$.

Let us now consider the supersymmetric case in some detail. Superspace comes equipped
with a supervielbein $E^A = (E^a, E^{\ul\alpha})$, which generically has coordinate legs 
in all of superspace,
\begin{align}
E^A = \rd x^\mu \, E_\mu{}^A + \rd \q^{\mathfrak{m}} \,E_{\mathfrak{m}}{}^A~,
\end{align}
where $\q^{\mathfrak{m}}$ are Grassmann coordinates. For now, we will use $\ul\alpha$ to denote a general
tangent space spinor index, which will later decompose into an irreducible spinor
and an $R$-symmetry index. 
When restricted to $\theta=\rd\theta=0$, $E^a$ is identified with the vielbein
1-form $e^a$, while $E^{\ul\alpha}$ is identified with $\frac{1}{2} \psi^{\ul\alpha}$
where $\psi^{\ul\alpha}$ is the gravitino 1-form in spacetime.
For the rest of this section, we will abuse notation somewhat and identify 
$e^a \equiv E^a$ and $\frac{1}{2} \psi^{\ul\alpha} \equiv E^{\ul\alpha}$ without taking
$\theta=\rd\theta=0$ so that these 1-forms live on the entire superspace.

We are interested in supersymmetric actions in $d=4$ spacetime corresponding to 4-forms $J$
that possess a covariant expansion in terms of $E^A$,\footnote{This precludes Chern-Simons type actions, which would require other connections
to appear explicitly, but these can straightforwardly be included. For the
actions we will be discussing, there will be no need to consider such cases.}
\begin{align}
J = \frac{1}{4!} E^A E^B E^C E^D J_{DCBA}~,
\end{align}
where $J_{DCBA}$ are covariant superfields.
The action is constructed by integrating the four-form $J$ over the bosonic spacetime 
manifold at $\theta=0$,
\begin{align}
S &= \frac{1}{4!} \int \rd x^\mu \rd x^\nu \rd x^\rho \rd x^\sigma\, E_\mu{}^A E_\nu{}^B E_\rho{}^C E_\sigma{}^D J_{DCBA} \Big\vert_{\theta=0} \eol
	&= -\frac{i}{4!} \int \rd^4x \, \veps^{\mu \nu \rho \sigma} E_\mu{}^A E_\nu{}^B E_\rho{}^C E_\sigma{}^D J_{DCBA} \Big\vert_{\theta=0} ~.\label{eq:ap}
\end{align}
We use conventions where $\veps_{abcd}$ and $\veps_{\mu\nu\rho\sigma}$ are imaginary, 
see appendix \ref{app:N=4CSG}.
The 4-form $J$ is assumed to be invariant under the other gauge symmetries. In the case of
$\cN=4$ conformal supergravity, this means $J$ should be a Lorentz scalar, inert
under Weyl transformations, a singlet under the $\rm SU(4) \times U(1)$ R-symmetry,
and a conformal primary -- that is, annihilated by $S$-supersymmetry and conformal boosts ($K$). All but the last condition are easy to realize, by
taking the terms $J_{DCBA}$ to transform in the obvious way under Lorentz
transformations and homogeneously under Weyl and $R$-symmetries with certain weights.
Because the one-forms $e^a$ and $\psi^{\ul\alpha}$ are $K$-inert, the
components $J_{DCBA}$ will also be $K$-inert; however, because $\psi^{\ul\alpha}$
transforms into $e^a$ under $S$-supersymmetry, some of the components $J_{DCBA}$
will transform into each other in a complementary way to leave $J$ unchanged.
We will elaborate on this in due course.

In the superform approach, invariance of the action follows if $J$ is a closed
form in superspace. Because $J$ is assumed to be gauge-invariant,
closure is equivalent to covariant closure, $\nabla J = 0$, where $\nabla$ carries
all of the gauge connections, except for the gravitino which is now interpreted
as part of the supervielbein.  For $\cN=4$ conformal supergravity, these connections
are associated with the local symmetries mentioned above.
For further details on the Weyl multiplet fields and their superconformal transformation rules, we refer to appendix \ref{app:N=4CSG}.

Defining the torsion 2-form as the covariant
exterior derivative of the supervielbein,
\begin{align}
T^A  \equiv \nabla E^A = \frac{1}{2} E^B E^C T_{CB}{}^A~,
\end{align}
the condition of covariant closure of $J$ amounts to
\begin{align}\label{eq:JClosure}
\nabla J = \frac{1}{4!} E^A E^B E^C E^D E^F \Big(\nabla_F J_{DCBA}
	+  2 \,T_{FD}{}^G J_{GCBA} \Big) = 0~.
\end{align}
The construction of a supersymmetric action principle then amounts to finding 
covariant tensors $J_{DCBA}$ that satisfy this equation.

Let us focus our attention on the lowest Weyl weight terms,
corresponding to $J_{\ul{\delta\gamma\beta\alpha}}$, and make the
extremely restrictive ansatz that only Lorentz scalars appear at this level.
This implies the following three structures in terms of the gravitino one-forms
$\psi_{\alpha}{}^i = \rd x^\mu \,\psi_{\mu\, \alpha}{}^i$,
with $\SU(4)$ index $i=1,\cdots,4$ and chiral spinor index $\alpha=1,2$:
\begin{subequations}
\begin{alignat}{2}
J_{\psi_R^4} &= \frac{1}{4} \bar \psi_i \psi_j \,\bar\psi_k \psi_l\,
	\veps^{klrs} \,C^{ij}{}_{rs}
		&\, &= - \frac{i}{4} \, \rd^4x\, \veps^{\mu\nu\rho\sigma} \,
		\bar\psi_{\mu i} \psi_{\nu j} \bar\psi_{\rho k} \psi_{\sigma l}
		\,\veps^{k l r s} \,C^{i j}{}_{r s}		
		~, \\
J_{\psi_L^4} &= \frac{1}{4} \bar \psi^i \psi^j \, \bar\psi^k \psi^l\,
	\,\veps_{klrs} \,\bar C^{rs}{}_{ij}
	&\, &= - \frac{i}{4} \, \rd^4x\, \veps^{\mu\nu\rho\sigma} \,
		\bar\psi_{\mu}{}^i \psi_{\nu}{}^j \bar\psi_{\rho}{}^k \psi_{\sigma}{}^l
		\,\veps_{k l r s} \,\bar C^{r s}{}_{i j}		
	~, \\
J_{\psi_L^2 \psi_R^2} &= \bar \psi_i \psi_j\, \bar\psi^k \psi^l \, A^{ij}{}_{kl}
	&\, &= - i\, \rd^4x\, \veps^{\mu\nu\rho\sigma} \,
	\bar\psi_{\mu}{}_i \psi_{\nu}{}_j \bar\psi_{\rho}{}^k \psi_{\sigma}{}^l
		\,A^{i j}{}_{k l}
	~.
\end{alignat}
\end{subequations}
We will frequently as above employ Dirac notation to suppress spinor indices.
The fields $C^{ij}{}_{kl}$ and $A^{ij}{}_{kl}$ are assumed to be supercovariant and
$S$-invariant. The four gravitini of like chirality multiplying $C^{ij}{}_{kl}$
imply that it transforms only in the $\rep{20'}$ representation. We shall further assume
the same is true for $A^{i j}{}_{k l}$.\footnote{A contribution to $A^{ij}{}_{kl}$
in the $\rep{15}$ or $\rep{1}$ could be removed by adding an appropriate total derivative.
} 
While we have made a number of strong assumptions
about the form of the action principle \eqref{eq:ap}, we will find that it nevertheless leads to the most general
conformal supergravity action (see section \ref{sec:unique}) .

We now derive the supersymmetry transformation properties of these fields in order to construct 
an invariant action principle. In principle, this can be done directly using a tangent space
decomposition as in \eqref{eq:JClosure}, but this can become unwieldy with
the (anti)symmetrizations on the tangent space indices. An abstract form-based approach
is more efficient. The idea is to decompose the covariant exterior derivative into
formal pieces corresponding to the various torsion tensors and covariant derivatives
that appear in \eqref{eq:JClosure}.
For example, if $\Phi$ is a covariant superfield (such as $J_{DCBA}$), we may decompose
\begin{align}
\nabla \Phi = e^a \nabla_a \Phi 
	+ \frac{1}{2} \psi^{\alpha i} \nabla_{\alpha i} \Phi
	+ \frac{1}{2} \psi_{\dalpha i} \nabla^{\dalpha i} \Phi
	\equiv \nabla_1 \Phi + \nabla^L_{1/2} \Phi + \nabla^R_{1/2} \Phi,
\end{align}
with the numerical subscripts on $\nabla$ denoting the Weyl weights of the operators
in tangent space. We use $L$ and $R$ superscripts to denote the left and right-handed
spinor derivative operations. For the one-forms, $e^a$, $\psi_\alpha{}^i$, and $\psi^\dalpha{}_i$,
exterior covariant differentiation generates the torsion tensors,
\begin{align}
\nabla e^a &=
	-\frac{1}{2} \psi^{\alpha k} (\gamma^a)_{\alpha\dalpha} \psi^{\dalpha}_k
	\equiv t_0 e^a~, \\
\nabla \psi_{\alpha}{}^i &=
	\frac{1}{2} e^a e^b R(Q)_{b a}{}_{\alpha}{}^i
	- \frac{1}{4} T_{a b}{}^{i j} e^c (\gamma^{ab} \gamma_c)_{\alpha \dalpha} \psi^{\dalpha}{}_j
	+ \frac{1}{4} \veps^{i j k l} \psi_{\dalpha j} \psi^{\dalpha}{}_k \Lambda_{\alpha l} \eol
	&\equiv t^L_{3/2} \psi_\alpha{}^i
		+ t^L_1 \psi_\alpha{}^i
		+ t^L_{1/2} \psi_\alpha{}^i~,
\end{align}
and similarly for $\nabla \psi^{\dalpha}{}_i$.
In both expressions above, the first equality gives the expansion of the
superspace torsion tensors as defined in Appendix \ref{app:N=4CSG}. The second equality 
introduces shorthand $t_n$ for the various torsion components, with subscripts denoting their
Weyl weights. Note the appearance of the covariant fields $\Lambda_{\alpha i}$, $T_{ab}{}^{ij}$,
and the gravitino curvature $R(Q)_{ab}{}_\alpha{}^i$ of the Weyl multiplet.
Their properties as well as those of the other supercovariant fields
of the Weyl multiplet are also summarized in Appendix \ref{app:N=4CSG}.

On a given four-form $J$ decomposed in tangent space,
the closure condition \eqref{eq:JClosure} can now be written
\begin{align}\label{E:N4ActionMaster}
\Big(\nabla_1  + \nabla^L_{1/2}  + \nabla^R_{1/2} 
	+ t_0 
	+ t^L_{1/2} 
	+ t^R_{1/2} 
	+ t^L_{1} 
	+ t^R_{1} 
	+ t^L_{3/2} 
	+ t^R_{3/2} \Big) J = 0~.
\end{align}
The advantage of this formalism is that it lets one more easily decompose the five-form $\nabla J$
into terms with five gravitini, four gravitini, etc., and check that each batch of terms
separately vanishes.

\subsection{Solving the $\psi^5$ Bianchi identities}
We begin by solving the part of \eqref{E:N4ActionMaster} involving five dotted spinor indices.
In form notation, this corresponds to the part of $\nabla J$ involving five right-handed
gravitini, which is simply
\begin{align}
0 = \nabla^R_{1/2} J_{\psi_R^4}~.
\end{align}
It is easy to check that the five gravitino term appearing in this expression
is in the $\rep{60}$, so the content of this identity is
\begin{align}\label{eq:QbC60}
[\nabla^{\dalpha m} C^{ij}{}_{kl}]_{\rep{60}} = 0~,
\end{align}
which implies
$\nabla^{\dalpha m} C^{ij}{}_{kl}
	= \delta^m_{[k} \,\Upsilon^{\dalpha\, ij}{}_{l]} 
	+ \delta^{[i}_{[k} \,\Upsilon^{\dalpha j] m}{}_{l]}$
for some spinor $\Upsilon^{\dalpha}{}^{ij}{}_k$ in the $\rep{20}$. This field
will not appear in the action principle, although it is part of the multiplet that is used
to define it.

The $\psi_L \psi_R^4$ Bianchi identity reads
\begin{align}\label{eq:psi5.b}
0 = \nabla^L_{1/2} J_{\psi_R^4} + t^L_{1/2} J_{\psi_L^2 \psi_R^2} + t_0 J_{e \psi_R^3}~.
\end{align}
Projecting onto the $\brep{60}$, only the first two terms contribute, leading to
\begin{align}
[\nabla_{\alpha m} C^{ij}{}_{rs}]_{\brep{60}} = [2 \Lambda_{\alpha m} A^{ij}{}_{rs}]_{\brep{60}}
\end{align}
We solve this constraint by introducing a fermion $\rho_{\alpha\, ij}{}^k$ in the $\brep{20}$,
\begin{align}\label{eq:QC60}
\nabla_{\alpha m} C^{ij}{}_{rs} &=
	\delta^{[i}_{m} \rho_{\alpha\, rs}{}^{j]} + \delta^{[i}_{[r} \rho_{\alpha s] m}{}^{j]}
	+ 2 \,\Lambda_{\alpha m} A^{ij}{}_{rs}
	+ 4 \,\delta^{[i}_m \Lambda_{\alpha p} A^{j] p}{}_{rs}
	+ 4 \,\Lambda_{\alpha p} \,\delta^{[i}_{[r} A^{j] p}{}_{s] m}~.
\end{align}
We have included a certain factor of $\Lambda_{\alpha m} A^{ij}{}_{rs}$ in the $\brep{20}$
on the right-hand side so that the remaining part of the Bianchi identity \eqref{eq:psi5.b}
implies that
\begin{align}
J_{e \psi_R^3}
	&= -\frac{1}{8} \bar \psi_i \psi_j \,\psi_{\dalpha k}\, e^a \, (\gamma_a)^{\dalpha \alpha}
	\veps^{ij rs} \rho_{\alpha rs}{}^k~.
\end{align}

There is a subtlety at this stage associated with the cohomology of the operators appearing in
\eqref{E:N4ActionMaster}. In particular, $t_0$ satisfies $(t_0)^2 = 0$, so $J_{e \psi_R^3}$ 
can be shifted by terms that are $t_0$-exact.\footnote{Actually, the terms need only be
$t_0$-closed, but one can show for $\cN=4$ supersymmetry that all $t_0$-closed terms are
$t_0$-exact \cite{Brandt:2010tz}.} Such terms generally correspond to parts of
total derivatives, so they can be discarded. This is also incidentally the reason to take
$A^{ij}{}_{kl}$ in the $\rep{20'}$ as smaller representations lead to $t_0$-exact contributions.

The final Bianchi identity at this dimension involves $\psi_L^2 \psi_R^3$:
\begin{align}
0 = t^R_{1/2} J_{\psi_R^4} + \nabla^R_{1/2} J_{\psi_L^2 \psi_R^2} + t_0 J_{e \psi_L \psi_R^2}~.
\end{align}
This implies that
\begin{align}\label{eq:QbA60}
[\nabla^{\dalpha m} A^{ij}{}_{kl}]_{\rep{60}} &=
	[\Lambda^{\dalpha m} C^{ij}{}_{kl}]_{\rep{60}}
\end{align}
which we solve as
\begin{align}
\nabla^{\dalpha m} A^{ij}{}_{kl} &=
	\Lambda^{\dalpha m} C^{ij}{}_{kl}
	+ \delta^{[i}_{[k} \Big(
	\kappa^{\dalpha j] m}{}_{l]}
	-4 \Lambda^{\dalpha p} C^{j] m}{}_{p l]}
	\Big)
	+ \delta^m_{[k} \Big(
	\kappa^{\dalpha i j}{}_{l]}
	-4 \Lambda^{\dalpha p} C^{i j}{}_{p l]}
	\Big)
\end{align}
for some fermion $\kappa^{\dalpha\, ij}{}_k$ in the $\rep{20}$, which
appears at the next level in the action principle
\begin{align}
J_{e \psi_L \psi_R^2}
	&= -\frac{1}{4} \bar \psi_i \psi_j\, \psi^{\alpha k} e^a\, (\gamma_a)_{\alpha \dalpha}\,
		\kappa^{\dalpha \,i j}{}_k~.
\end{align}
The other Bianchi identities follow by complex conjugation.

At this point, we emphasize that the supersymmetry constraints
\eqref{eq:QbC60}, \eqref{eq:QC60}, and \eqref{eq:QbA60} amount to the basic
constraints \eqref{eq:Bigconstraint} mentioned in the introduction.
It will turn out that all the other constraints we encounter are consequences
of these.

\subsection{Solving the $e \psi^4$ Bianchi identities}\label{sec:2.2}
The $e \psi_R^4$ Bianchi identity reads
\begin{align}\label{E:BI.epb4}
0 = \nabla_1 J_{\psi_R^4}
	+ \nabla^R_{1/2} J_{e \psi_R^3}
	+ t^L_{1/2} J_{e \psi \psi_R^2}~.
\end{align}
Because the expression involves only quantities we have already defined,
it leads only to new constraints on the fields:
\begin{align}
\Big[\tfrac{1}{4} (\gamma_a)^{\dalpha \alpha} \nabla_\dalpha{}^{[k} \rho_{\alpha\, rs}{}^{l]}
	+ 2 \nabla_a C^{kl}{}_{rs}
	- \tfrac{1}{2} (\gamma_a)_{\alpha\dalpha} \Lambda^{\alpha}_{[r} \kappa{}^{\dalpha kl}{}_{s]}
\Big]_{\rep{20'}} = 0~,\quad
[\nabla^{\dbeta (m} \rho_{\alpha \, i j}{}^{k)}]_{\rep{45}} = 0~.
\end{align}
It turns out that these conditions (as well as other ones we will find shortly)
are not independent of the basic constraints
\eqref{eq:Bigconstraint}. They actually follow using closure of the supersymmetry
algebra, which is a welcome result. Establishing this result is rather technical,
so we delay an explicit discussion to appendix \ref{app:BianchiAnalysis}.

The $e \psi_L \psi_R^3$ Bianchi identity is a bit more involved:
\begin{align}
0 = \nabla^L_{1/2} J_{e \psi_R^3}
	+ \nabla^R_{1/2} J_{e \psi_L \psi_R^2}
	+ t_0 J_{e^2 \psi_R^2}
	+ t^L_{1/2} J_{e \psi_L^2 \psi_R}
	+ t^R_{1} J_{\psi_R^4}
	+ t^L_1 J_{\psi_L^2 \psi_R^2}~.
\end{align}
It possesses a $t_0$ term, which can be used to determine $J_{e^2 \psi_R^2}$
up to a $t_0$-exact piece.
This term generically involves the $\rep{6}$ and the $\rep{10}$, so constraints
can be found by first requiring the $\rep{64}$ to vanish. No other SU(4) representations
are present. Because the Lorentz structure of $e \psi_L \psi_R^3$ is already quite complicated,
a number of constraints emerge when projected upon irreducible representations:
\begin{align}
\Big[\veps^{ijrs} \nabla_{(\alpha l} \rho_{\beta) rs}{}^k
	- \veps^{ijrs} \Lambda_{(\alpha l} \kappa{}_{\beta) rs}{}^k
	+ 8 (\gamma^{bc})_{\a\b} T_{bc}{}^{m k} A^{i j}{}_{l m}\Big]_{\rep{64}} = 0~, \\
\Big[\nabla^{(\dalpha k} \kappa{}^{\dbeta) i j}{}_l
	+ 2 (\gamma^{bc})^{\dalpha \dbeta} T_{b c}{}_{lm} \veps^{klrs} C^{i j}{}_{rs}\Big]_{\rep{64}} = 0~,\\
\Big[\veps^{ijrs} \nabla^\alpha_l \rho_{\alpha rs}{}^k
	- \veps^{ijrs} \Lambda^\alpha_l \kappa_{\alpha r s}{}^k
	+ 2 \nabla_\dalpha^k \kappa^{\dalpha i j}{}_l \Big]_{\rep{64}} = 0~.
\end{align}
For the $\rep{6}$ and the $\rep{10}$, there are two Lorentz representations each,
one leading to a constraint and one that allows the determination of $J_{e^2\psi_R^2}$.
The constraints are
\begin{align}
3 \,\veps^{ijrs} \nabla^{\gamma}_k \rho_{\gamma rs}{}^k
	+ 4 \,\nabla_{\dgamma}^{k} \kappa^{\dgamma\, i j}{}_k
	+ 2 \,\veps^{ijrs} \Lambda^\gamma_k \kappa_{\gamma rs}{}^k
	&= 0~, \\
\Big(
	\nabla_{(\alpha l} \rho_{\beta) r s}{}^{(k}
	+ 2\Lambda_{(\alpha l} \kappa_{\beta) rs}{}^{(k}
	\Big) \veps^{j)lrs} &= 0~.
\end{align}
The representations appearing in $J_{e^2 \psi_R^2}$ are
\begin{align}
J_{e^2 \psi_R^2}
	= \frac{1}{16} \bar \psi_i \gamma_{a b} \psi_j\, e^a e^b\, \cE^{i j}
	+ \frac{1}{16} \bar \psi_i \psi_j \,e^a e^b\, \cE_{a b}{}^{i j}
\end{align}
where the fields $\cE^{ij} = \cE^{(ij)}$ and $\cE_{ab}{}^{ij} = \cE_{[ab]}{}^{[ij]}$
are determined by supersymmetry as
\begin{align}
\cE_{a b}{}^{i j} &= - \frac{1}{20} (\gamma_{a b})^{\alpha \beta}
	\Big(
		3 \,\veps^{i j r s} \nabla_{\alpha k} \rho_{\beta r s}{}^k
		+ 2 \,\veps^{ijrs}  \Lambda_{\alpha k} \kappa{}_{\beta r s}{}^k
		- 16\, T_{cd}{}^{rs} (\gamma^{cd})_{\a\b}\, A^{i j}{}_{rs}
	\Big)
	\eol & \qquad
	-\frac{1}{5} (\gamma_{a b})_{\dalpha \dbeta} \Big(
	\nabla^{\dalpha l} \kappa{}^{\dbeta i j}{}_l
	-3 \,T_{cd}{}_{kl} (\gamma^{cd})^{\ad\bd}\, \veps^{klrs} C^{i j}{}_{rs} \Big)~,\\
\cE^{i j} &= \frac{1}{6} \Big(\nabla^{\alpha}_l \rho_{\alpha r s}{}^{(i}
	+ 2\Lambda^\alpha_l \kappa_{\alpha rs}{}^{(i}\,\Big)\veps^{j)lrs} ~.
\end{align}
Note that $\cE_{ab}{}^{ij}$ is a complex tensor with both self-dual and anti-self-dual parts.

Finally, we have the $e \psi_L^2 \psi_R^2$ Bianchi identity:
\begin{align}
0 = \nabla^L_{1/2} J_{e \psi_L \psi_R^2}
	+ \nabla^R_{1/2} J_{e \psi_L^2 \psi_R}
	+ \nabla_1 J_{\psi_L^2 \psi_R^2}
	+ t_0 J_{e^2 \psi_L \psi_R}
	+ t^L_{1/2} J_{e \psi_L^3}
	+ t^R_{1/2} J_{e \psi_R^3 }~.
\end{align}
The terms in the $\rep{45}$, $\brep{45}$, and the $\rep{20'}$ cannot be
cancelled by $t_0 J_{e^2 \psi_L \psi_R}$. These lead to the constraints
\begin{align}
\Big[\nabla_{\beta (k} \kappa^{\dalpha \,i j}{}_{l)}
	- \Lambda_{\beta (k} \rho^{\dalpha\, i j}{}_{l)} \Big]_{\brep{45}} = 0~, \qquad
	\Big[\nabla^{\dbeta (k} \kappa{}_{\alpha \,i j}{}^{l)}
	- \Lambda^{\dbeta (k} \rho{}_{\alpha \,i j}{}^{l)}\Big]_{\rep{45}} &= 0~, \\
\frac{1}{16} (\gamma_a)^{\dalpha \alpha}
	\Big[\nabla_{\alpha i} \kappa_{\dalpha}{}^{kl}{}_j
	+ \nabla_{\dalpha}^k \kappa_{\alpha}{}_{ij}{}^l
	- 3 \Lambda_{\alpha i} \rho_{\dalpha}{}^{kl}{}_j
	- 3 \Lambda_{\dalpha}^k \rho_{\alpha}{}_{ij}{}^l
	\Big]_{\rep{20'}}
	+ \nabla_a A^{kl}{}_{ij} &= 0~.
\end{align}
The terms in the $\rep{15}$ that cannot be cancelled by $t_0 J_{e^2 \psi_L \psi_R}$ require
\begin{align}
\nabla_\dalpha^k \kappa_{\alpha}{}_{k j}{}^i
+ 3 \Lambda_\dalpha^k \rho_{\alpha}{}_{k j}{}^i
+ \nabla_{\alpha k} \kappa_{\dalpha}{}^{k i}{}_j
+ 3 \Lambda_{\alpha k}  \rho_{\dalpha}{}^{k i}{}_j
	= 0~.
\end{align}
The remaining terms are cancelled if we choose 
\begin{align}
J_{e^2 \psi_L \psi_R}
	&= \frac{i}{32} e^a e^b \,\bar \psi_i \gamma^c \psi^j\, \veps_{abcd} \,\cE^d{\,}^i{}_j
\end{align}
for a pseudoreal field $\cE_a{}^i{}_j$ in the $\rep{15}$ given by
\begin{align}
i (\gamma^a)_{\alpha \dalpha} \cE_a{}^i{}_j
	&=  \nabla_\dalpha^k \kappa_{\alpha}{}_{k j}{}^i
	+ 3 \Lambda_\dalpha^k \rho_{\alpha}{}_{k j}{}^i
	- \nabla_{\alpha k} \kappa_{\dalpha}{}^{k i}{}_j
	- 3 \Lambda_{\alpha k}  \rho_{\dalpha}{}^{k i}{}_j~.
\end{align}

\subsection{Solving the remaining Bianchi identities}
At this point, we trust that the procedure is fairly clear, if tedious. As we have already
mentioned, new constraints that emerge at each level turn out to be consequences of
the closure of the superconformal algebra and the basic constraints \eqref{eq:Bigconstraint}.
The only independent information we determine at each level is the specific form
of the new terms that we need to add and how they are related by supersymmetry to
terms that have already appeared.

The $e^2 \psi_R^{3}$ Bianchi identity only leads to constraints, so we omit its
discussion. The $e^2 \psi_L \psi_R^2$ Bianchi requires the introduction of new terms
\begin{align}
J_{e^3 \psi_R}=\frac{i}{3} e^{a} e^{b} e^{c} \, \bar{\psi}_{k}{\Omega}^{d\,k} \,\veps_{abcd}
	+ \frac{i}{3} e^{a} e^{b} e^{c} \,\bar{\psi}_{k} \gamma^{d} {\Omega}^{k} \, \veps_{abcd}~,
\end{align}
where $\Omega_a{}^{\dalpha k}$ is gamma-traceless. These fermions are determined to be
\begin{align}
\Omega_a{}^{\dalpha k} &= 
	- \frac{i}{256} (\gamma^{bc} \gamma_{a})^{\dalpha \beta} \nabla_{\beta l} \cE_{bc}{}^{l k}
	+ \frac{i}{256} \varepsilon^{ijkm} \cE_{bc ij} (\gamma^{bc} \gamma_{a})^{\dalpha \beta} \Lambda_{\beta m} 
~,\\
\Omega_\alpha{}^{k} &= 
	- \frac{i}{40}\nabla_{\alpha  l} \cE^{ l k}
	- \frac{1}{640} (\gamma^{a})_{\alpha \dbeta} \nabla^{\dbeta l} \cE_a{}^{k}{}_{ l}
	- \frac{i}{40} T_{ab}{}^{ij} (\gamma^{ab})_\alpha{}^\beta \kappa_{\beta}{}_{ij}{}^{k}~.
\end{align}

The last piece of the action principle is a Lorentz scalar,
\begin{align}
J_{e^4}=\frac{i}{4!} e^a e^b e^c e^d \,F \,\varepsilon_{abcd}~,
\end{align}
and it is determined from the $e^3 \psi_L \psi_R$ Bianchi identities as
\begin{align}
F = -\frac{1}{4}\Big(\nabla^\alpha{}_{k} \Omega_\alpha{}^{k} 
		+ {\nabla}_{\dalpha}{}^{k} \Omega^{\dalpha}{}_{k}\Big)
	-\frac{i}{32}\Big(T_{ab\,ij} \cE^{ab\, ij} - T_{ab}{}^{ij} \cE^{ab}{}_{ij}\Big)~.
\end{align}

\subsection{Summary}
The full 4-form $J$, rewritten as a Lagrangian, gives the density formula
\begin{align}
e^{-1} \cL &= F
	+ 2 \,\bar \psi_{\mu i} (\Omega^{\mu i} + \gamma^\mu \Omega^i)
	+ 2 \,\bar \psi_{\mu}{}^i (\Omega^{\mu}{}_i + \gamma^\mu \Omega_i)
	\eol & \quad
	+ \frac{1}{8} \bar \psi_{[\mu i} \gamma^\mu \psi_{\nu]}{}^j \, \cE^\nu{}^i{}_j
	- \frac{1}{8} \Big(i \,\bar \psi_{\mu i} \gamma^{\mu\nu} \psi_{\nu j} \cE^{i j}
		+ \frac{i}{2} \veps^{\mu\nu \rho\sigma} \bar\psi_{\mu i} \psi_{\nu j} \cE_{\rho\sigma}{}^{i j}
		+ \HC
	\Big)
	\eol & \quad
	+ \frac{i}{8} \veps^{\mu\nu\rho\sigma} \Big(
		\bar\psi_{\mu i} \psi_{\nu j}
			\,\bar \psi_{\rho k} \gamma_\sigma \rho_{r s}{}^k \,\veps^{i j r s}
		+ 2\,\bar\psi_{\mu i} \psi_{\nu j}\,
			\bar \psi_{\rho}{}^k \gamma_\sigma \kappa{}^{i j}{}_{k}
		+ \HC
	\Big)
	\eol & \quad
	- \frac{i}{4} \veps^{\mu\nu\rho\sigma} \Big(
		\bar\psi_{\mu i} \psi_{\nu j} \bar\psi_{\rho k} \psi_{\sigma l}
		\,\veps^{k l r s} \,C^{i j}{}_{r s}
		+ 2 \,\bar\psi_{\mu}{}_i \psi_{\nu}{}_j \bar\psi_{\rho}{}^k \psi_{\sigma}{}^l
		\,A^{i j}{}_{k l}
		+ \HC \Big)~.\label{eq:densityf}
\end{align}
The basic constraints are imposed on the constituents $C$ and $A$, which must obey
\begin{subequations}\label{eq:cons}
\begin{align}
0 &= [ \nabla^{\dalpha m} C^{ij}{}_{kl}]_{\rep{60}} =
[ \nabla^{\dalpha m} A^{ij}{}_{kl} - \Lambda^{\dalpha m} C^{ij}{}_{kl}]_{\rep{60}} =
[\nabla^{\dalpha m} \bar C^{ij}{}_{rs} - 2 \Lambda^{\dalpha m} A^{ij}{}_{rs}]_{\rep{60}}\label{eq:cons2}
~, \\
0 &= [\nabla_{\alpha m} \bar C^{ij}{}_{kl}]_{\brep{60}} =
[\nabla_{\alpha m} A^{ij}{}_{kl} - \Lambda_{\alpha m} \bar C^{ij}{}_{kl}]_{\brep{60}} =
[\nabla_{\alpha m} C^{ij}{}_{rs} - 2 \Lambda_{\alpha m} A^{ij}{}_{rs}]_{\brep{60}}\label{eq:cons1}
\end{align}
\end{subequations}
The other fields appearing in the density formula are defined by
\begin{subequations}
\begin{align}
\rho_{\alpha\, rs}{}^i &=  \nabla_{\alpha k} C^{k i}{}_{rs} + 2 \,\Lambda_{\alpha k} A^{k i}{}_{rs}~, \\
\kappa_{\alpha rs}{}^i &= \nabla_{\alpha k} A^{k i}{}_{rs} + 3 \,\Lambda_{\alpha k} \bar C^{k i}{}_{rs}~, \\
\cE^{i j} &= \frac{1}{6} \Big(\nabla^{\alpha}_l \rho_{\alpha r s}{}^{(i}
	+ 2\Lambda^\alpha_l \kappa_{\alpha rs}{}^{(i}\,\Big)\veps^{j)lrs} ~,\\
\cE_{a b}{}^{i j} &= - \frac{1}{20} (\gamma_{a b})^{\alpha \beta}
	\Big(
		3 \,\veps^{i j r s} \nabla_{\alpha k} \rho_{\beta r s}{}^k
		+ 2 \,\veps^{ijrs}  \Lambda_{\alpha k} \kappa{}_{\beta r s}{}^k
		- 16\, T_{cd}{}^{rs} (\gamma^{cd})_{\a\b}\, A^{i j}{}_{rs}
	\Big)
	\eol & \qquad
	-\frac{1}{5} (\gamma_{a b})_{\dalpha \dbeta} \Big(
	\nabla^{\dalpha l} \kappa{}^{\dbeta i j}{}_l
	-3 \,T_{cd}{}_{kl} (\gamma^{cd})^{\ad\bd}\, \veps^{klrs} C^{i j}{}_{rs} \Big)~,\\
\cE_a{}^i{}_j &=
	- \frac{i}{2} (\gamma_{a})^{\dalpha \alpha}  \nabla_\dalpha^k \kappa_{\alpha}{}_{k j}{}^i
	- \frac{3i}{2} (\gamma_{a})^{\dalpha \alpha}  \Lambda_\dalpha^k \,\rho_{\alpha}{}_{k j}{}^i
	- \HC
	~, \\
\Omega_a{}^{\dalpha k} &= 
	- \frac{i}{256} (\gamma^{bc} \gamma_{a})^{\dalpha \beta} \nabla_{\beta  l} \cE_{bc}{}^{ l k}
	+ \frac{i}{256} \varepsilon^{ijkm} \cE_{bc ij} (\gamma^{bc} \gamma_{a})^{\dalpha \beta} \Lambda_{\beta m} 
~,\\
\Omega_\alpha{}^{k} &= 
	- \frac{i}{40}\nabla_{\alpha  l} \cE^{ l k}
	- \frac{1}{640} (\gamma^{a})_{\alpha \dbeta} \nabla^{\dbeta l} \cE_a{}^{k}{}_{ l}
	- \frac{i}{40} (\gamma^{ab})_\alpha{}^\beta T_{ab}{}^{ij} \kappa_{\beta}{}_{ij}{}^{k}~, \\
F &= -\frac{1}{4}\Big(\nabla^\alpha{}_{k} \Omega_\alpha{}^{k} 
		+ {\nabla}_{\dalpha}{}^{k} \Omega^{\dalpha}{}_{k}\Big)
	- \frac{i}{32}\Big(T_{ab\,ij} \cE^{ab\, ij} - T_{ab}{}^{ij} \cE^{ab}{}_{ij}\Big)~.\label{eq:tr2}
\end{align}\label{eq:tr1}\end{subequations}
Their Weyl ($w$) and chiral ($c$) weight as well as their SU(4) representation and algebraic properties are summarised in Table \ref{table:composites}.

\begin{table}[t]
  \begin{center}
 \begin{tabular}{ c | c l c c c }
 \hline 
 \hline
   Type & Fields & Properties & $\mathrm{SU}(4)$ & $w$ & $c$ \\
 \hline 
 \multirow{6}{*}{Bosonic} 
 & $A^{ij}{}_{kl}$ &  $ A^{ij}{}_{kl}= (A^{kl}{}_{ij})^*$ & $\mathbf{20}^\prime$ & 2 & 0 \\
  & $C^{ij}{}_{kl}$  &  & $\mathbf{20}^\prime$ & 2 & $-2$ \\
 &$\mathcal{E}^{ij}$  & 
  & $\mathbf{10}$ & 1 & $-1$\\
 & $\mathcal{E}_{ab}{}^{ij}$ &  $\mathcal{E}_{(ab)}{}^{ij}=\,0$
  & $\mathbf{6}$ & 1 & $-1$ \\
 & $\mathcal{E}_a{}^{i}{}_j$  && $\mathbf{15}$ & 1 & $0$ \\
  & $F$ &  $F= F^*$ & $\mathbf{1}$ & 0 & 0\\
 \hline
\multirow{4}{*}{Fermionic} 
& $\rho_{ij}{}^k$  & $\gamma_5 \,\rho_{ij}{}^k=\rho_{ij}{}^k$ & $\overline{\mathbf{20}}$ & $\tfrac{3}{2}$ & $-\tfrac32$ \\
& $\kappa_{ij}{}^k$  & $\gamma_5\,\kappa_{ij}{}^k=\kappa_{ij}{}^k$  & $\overline{\mathbf{20}}$ & $\tfrac32$ & $\tfrac12$ \\
& $\Omega^i$  & $\gamma_5 \,\Omega^i=\Omega^i$ & $\mathbf{4}$ & $\tfrac12$ & $-\tfrac12$ \\
& $\Omega_a{}^i$  & $\gamma_5\, \Omega_a{}^i=-\Omega_a{}^i\,,\;\gamma^a\,\Omega_a{}^i=0$ & $\mathbf{4}$ & $\tfrac12$ & $-\tfrac12$ \\[1mm]
 \hline
 \hline
\end{tabular}
\end{center}
 \caption{Fields appearing in the density formula.}\label{table:composites}
\end{table}

There is one final important check that is necessary. In the construction of $J$, we
assumed that it was gauge invariant in order to exchange closure for covariant closure.
While it is manifestly invariant under Lorentz, Weyl, and R-symmetry transformations,
invariance under $S$-supersymmetry and conformal boosts must be verified. These are
consequences of closure of the superconformal algebra.  Provided $C^{ij}{}_{kl}$
and $A^{ij}{}_{kl}$ are $S$ and $K$-invariant, one can show that each of their
descendants appearing in the Lagrangian are also $K$-invariant. Their $S$-supersymmetry
transformations are given by
\begin{gather}
\delta_S \rho_{ij}{}^k = 8 \,\eta_l \,C^{l k}{}_{i j}~, \qquad
\delta_S \kappa^{ij}{}_k = 8 \,\eta^l \, A^{i j}{}_{l k}~, \eol
\delta_S \cE_{ab}{}^{ij} = 3 \,\veps^{ijrs} \, \bar\eta_k \gamma_{ab} \rho_{rs}{}^k
	+ 4 \, \bar\eta^k \gamma_{ab} \kappa^{ij}{}_k~, \qquad
\delta_S \cE^{ij} = 2 \, \bar\eta_k \rho_{rs}{}^{(i} \veps^{j)krs}~, \eol
\delta_S \cE_a{}^i{}_j = 8i\, \bar\eta_k \gamma_a \kappa^{k i}{}_j
	- 8 i \, \bar\eta^k \gamma_a \kappa_{k j}{}^i~, \eol
\delta_S \Omega^i = - \frac{3i}{8} \cE^{ij}\, \eta_j 
	+ \frac{i}{32} \cE_{ab}{}^{ij}\,\gamma^{ab} \eta_j~, \eol
\delta_S \Omega_a{}^i = 
	- \frac{i}{16} \cE_{ab}{}^{ij} \gamma^b \eta_j
	- \frac{i}{32} \veps_{abcd}\, \cE^{bc}{}^{ij} \gamma^d \eta_j
	- \frac{3}{32} \cE_a{}^i{}_j\, \eta^j 
	+ \frac{1}{32} \cE^b{}^i{}_j\, \gamma_{ab} \eta^j ~, \eol
\delta_S F = -8 \bar \eta_i \Omega^i -8 \bar \eta^i \Omega_i~.
\end{gather}

\section{Building all $\cN=4$ conformal supergravity actions}\label{S:AllActions}

In this section, we provide the foundation for the construction of all
$\cN=4$ conformal supergravity actions. Making use of the density formula
built in the previous section, we only need to specify the lowest Weight covariant
fields $C^{ij}{}_{kl}$ and $A^{ij}{}_{kl}$. Our goal will be to build
candidates for these composites, using only the fields of the Weyl multiplet (summarized in Appendix \ref{app:N=4CSG}) as
our constituents. Once such composite fields are specified, the supersymmetry
transformations \eqref{eq:tr1} can be used to build all of the
other composite fields appearing in the density formula. Invariance under local
superconformal transformations is then guaranteed. 

This approach is however not \textit{a priori} guaranteed to lead to all possible conformal supergravity actions.
To establish that the class we construct is actually exhaustive, we show in section \ref{sec:unique} that its supercurrent (the multiplet containing the energy-momentum tensor) corresponds to the most general supercurrent of conformal supergravity.

\subsection{Ansatz for $C^{ij}{}_{kl}$ and $A^{ij}{}_{kl}$}\label{sec:3.1}
Let us attempt to construct $C^{ij}{}_{kl}$ and $A^{ij}{}_{kl}$
out of the constituent fields of the $\cN=4$ Weyl multiplet. We will need
basic ``building blocks'' $X^{ij}{}_{kl}$ corresponding to $S$-invariant
combinations of Weyl weight two in the $\rep{20'}$.
It turns out that there are essentially four possible combinations
which we denote $X_{n}$, with $n=1,\cdots,4$ indicating the degree of
homogeneity in the covariant Weyl multiplet fields appearing 
within:\footnote{The $X_n$ combinations \eqref{eq:X}, which obey \eqref{E:DX.Constraints}
below, should coincide with the similarly-named functions in eq. (3.59) of
\cite{Bossard:2013rza}, in the context of on-shell $\cN=4$ supergravity.}
\begin{align}
X_1{}^{ij}{}_{kl} &= D^{ij}{}_{kl}~,\nonumber \\
X_2{}^{ij}{}_{kl} &= 
2 \bar \L_{[k} \chi^{ij}{}_{l]} + 2 \delta^{[i}_{[k} \bar \L_m \chi^{j]m}{}_{l]}
	- \frac{1}{4} \veps^{ijmn} E_{mk} E_{nl}
	\eol & \quad
	+ \frac{1}{2} T^{ij} \cdot T^{mn} \veps_{klmn} - \frac{1}{12} \delta_k^{[i} \delta_{l}^{j]} T^{pq}\cdot T^{rs} \veps_{pqrs}~,\nonumber \\
X_3{}^{ij}{}_{kl} &= 
	- \frac{1}{4} \bar \L_k \g_{cd} \L_l T_{cd}{}^{ij}
	+ \frac{1}{2} \delta^{[i}_{[k} \bar \L_{l]} \g_{cd} \L_m \, T_{cd}{}^{j]m}
	- \frac{1}{12} \delta^{[i}_k \delta^{j]}_l \bar\L_m \g_{cd} \L_n T_{cd}{}^{mn}
	\eol & \quad
	+ \frac{1}{4} \veps^{ijmn} E_{m [k} \,\bar\L_{l]} \L_n~,\nonumber \\
X_4{}^{ij}{}_{kl} &=
	-\frac{1}{24} \veps^{ijmn} \,\bar\Lambda_k \Lambda_m \, \bar\Lambda_n \Lambda_l~.\ \label{eq:X}
\end{align}
$X_1$ is real, and we simply denote it as $D^{ij}{}_{kl}$ from now on.
$X_2$, $X_3$, and $X_4$ are complex.
We then choose the following ansatz for $C^{ij}{}_{kl}$ and $A^{ij}{}_{kl}$:
\begin{equation} \label{eq:hb}
\begin{aligned}
C^{ij}{}_{kl}
	&= C_1^{(-2)} D^{ij}{}_{kl}
	\\ & \quad
	+ C_2^{(0)} X_2{}^{ij}{}_{kl}
	+ C_2^{(-4)} \bar X_2{}^{ij}{}_{kl}
	\\ & \quad
	+ C_3^{(+2)} X_3{}^{ij}{}_{kl}
	+ C_3^{(-6)} \bar X_3{}^{ij}{}_{kl}
	\\ & \quad
	+ C_4^{(+4)} X_4{}^{ij}{}_{kl}
	+ C_4^{(-8)} \bar X_4{}^{ij}{}_{kl}~, 
\end{aligned}
\begin{aligned}
A^{ij}{}_{kl}
	&=
	A_1^{(0)}\, D^{ij}{}_{kl}
	\\ & \quad
	+ A_2^{(+2)} X_2{}^{ij}{}_{kl}
	+ \bar A_2^{(-2)} \bar X_2{}^{ij}{}_{kl}
	\\ & \quad
	+ A_3^{(+4)} X_3{}^{ij}{}_{kl}
	+ \bar A_3^{(-4)} \bar X_3{}^{ij}{}_{kl}
	\\ & \quad
	+ A_4^{(+6)} X_4{}^{ij}{}_{kl}
	+ \bar A_4^{(-6)} \bar X_4{}^{ij}{}_{kl}~,
\end{aligned}
\end{equation}
with $\bar C^{ij}{}_{kl}$ given by complex conjugation.
The factors $C_1^{(-2)}$, $A_1^{(0)}$, $A_2^{(+2)}$, etc., are functions of the coset scalars $\phi_\balpha\,,\phi^\balpha$, and their superscript correspond to their $\gU(1)$ charge. Their complex conjugates are denoted by $\bar C_1^{(+2)}$, $\bar A_1^{(0)}$, $\bar A_2^{(-2)}$, etc.

The supersymmetry constraints \eqref{eq:Bigconstraint} on $C^{ij}{}_{kl}$ and $A^{ij}{}_{kl}$ should then become
constraints on these functions of the coset scalars. This is indeed the case since
the four combinations $X_n$ turn out to transform into each other under supersymmetry
when we restrict to the largest $\SU(4)$ representations:
\begin{equation}\label{E:DX.Constraints}
\begin{alignedat}{2}
[\nabla_{\alpha m} D^{ij}{}_{kl} ]_{\rep{60}} &= 0~, \\
[\nabla_{\alpha m} X_2{}^{ij}{}_{kl} + \Lambda_{\alpha m} D^{ij}{}_{kl}]_{\rep{60}} &= 0~,
&\quad
[\bar\nabla^{\dalpha m} X_2{}^{ij}{}_{kl} - 2 \bar\Lambda^{\dalpha m} X_3{}^{ij}{}_{kl}]_{\brep{60}} &= 0~,\\
[\nabla_{\alpha m} X_3{}^{ij}{}_{kl} + \Lambda_{\alpha m} X_2{}^{ij}{}_{kl}]_{\rep{60}} &= 0~,
&\quad
[\bar\nabla^{\dalpha m} X_3{}^{ij}{}_{kl} - 6 \bar\Lambda^{\dalpha m} X_4{}^{ij}{}_{kl}]_{\brep{60}} &= 0~,\\
[\nabla_{\alpha m} X_4{}^{ij}{}_{kl} + \Lambda_{\alpha m} X_3{}^{ij}{}_{kl}]_{\rep{60}} &= 0~,
&\quad
[\bar\nabla^{\dalpha m} X_4{}^{ij}{}_{kl} ]_{\brep{60}} &= 0~, \\
[\Lambda_{\alpha m} X_4{}^{ij}{}_{kl}]_{\rep{60}} &= 0~.
\end{alignedat}
\end{equation}
The final condition arises because five $\Lambda_i$'s cannot be placed into the $\rep{60}$.
From these results, one can derive a set of differential equations on the coset functions and
search for a solution. To explain these conditions, we first make a brief detour to discuss the structure
of the coset space geometry.

\subsection{$\SU(1,1)/\gU(1)$ coset fields and derivatives}\label{sec:3.2}

The coset scalar fields $\phi_\balpha$ and $\phi^\balpha$ present in the $\cN=4$ Weyl multiplet
may be understood as constrained coordinates describing an $\SU(1,1)$ group manifold.
In addition to the constraint,
$\phi^\balpha \phi_\balpha = 1$, they are related by complex conjugation
$\phi^\balpha = \eta^{\balpha \bbeta} (\phi_\bbeta)^*$ with
$\eta^{\balpha\bbeta} = \textrm{diag}(1,-1)$, which ensures
that the coordinates describe a three-dimensional real manifold. Because they may be identified
up to a local $\gU(1)$ transformation, $\phi^\balpha \sim e^{i \lambda} \phi^\balpha$,
they actually describe an $\SU(1,1) / \gU(1)$ coset space.

Let us introduce the three derivative operators associated with the group manifold.
It is convenient to define these as
\begin{align}
\cD^{++} &\equiv -\phi^\balpha \veps_{\balpha\bbeta} \frac{\pa}{\pa \phi_\bbeta}~, \quad
\cD^{--} \equiv \phi_\balpha \veps^{\balpha\bbeta} \frac{\pa}{\pa \phi^\bbeta}~, \quad
\cD^0 \equiv \phi^\balpha \frac{\pa}{\pa \phi^\balpha} - \phi_\balpha \frac{\pa}{\pa \phi_\balpha}~.\label{eq:cosetder}
\end{align}
The first two derivatives were denoted
$\cD \equiv \cD^{++}$ and $\cD^\dag \equiv \cD^{--}$
in \cite{Butter:2016mtk}. 
We will return to this notation later on, but in the next few
subsections, it will be useful to keep the $\gU(1)$ charge explicit.
$\cD^0$ measures the $\gU(1)$ charge of the fields, while $\cD^{++}$ and $\cD^{--}$ convert
$\phi_{\balpha}$ to $\phi^\balpha$ and vice-versa.
These satisfy the SU(1,1) algebra
\begin{align}
[\cD^0, \cD^{++}] = 2 \cD^{++}~, \qquad
[\cD^0, \cD^{--}] = -2 \cD^{--}~, \qquad
[\cD^{++}, \cD^{--}] = \cD^0~.
\end{align}

It is sometimes convenient to work with complex coordinates on the coset
space $\SU(1,1) / \gU(1)$ directly. We define
$S = \phi_2 / \phi_1$ and $\bar S = -\phi^2 / \phi^1$
and introduce the phase $e^{2i \psi} = \phi^1 / \phi_1$ to describe the $\gU(1)$.
The coset space, parametrized by $S$ with $0 \leq |S| < 1$, is the Poincar\'e disk.
In these coordinates,
\begin{align}
\cD^{++} &= -e^{2 i \psi} \Big(
	(1 - S \bar S) \pa_S + \frac{i}{2} \bar S \pa_\psi
\Big)~, \quad
\cD^{--} = (\cD^{++})^*~, \quad
\cD^0 = -i \pa_\psi~.
\end{align}
$\cD^{++}$ and $\cD^{--}$ may be thought of as modified
holomorphic and anti-holomorphic derivatives. For example, a function that
has vanishing $\gU(1)$ charge that is also annihilated by $\cD^{--}$ is necessarily
holomorphic in $S$. We will abuse terminology somewhat and refer to a function of
\emph{any} $\gU(1)$ charge that is annihilated by $\cD^{--}$ as a holomorphic function.

\subsection{Solving for $C^{ij}{}_{kl}$ and $A^{ij}{}_{kl}$}\label{sec:solAC}
Let us now analyze the conditions for supersymmetry \eqref{eq:cons2}, \eqref{eq:cons1} on $C^{ij}{}_{kl}$ and $A^{ij}{}_{kl}$.
The condition $[\bar \nabla^{\dalpha m} C^{ij}{}_{kl}]_{\brep{60}} = 0$ implies that
\begin{equation}
\begin{aligned}
\cD^{--} C_2^{(0)} &= 0~,\\
\cD^{--} C_3^{(+2)} &= 2\, C_2^{(0)}~, \\
\cD^{--} C_4^{(+4)} &= 6 \,C_3^{(+2)}~,
\end{aligned}
\qquad
\begin{aligned}
\cD^{--} C_1^{(-2)} &= -C_2^{(-4)}~, \\
\cD^{--} C_2^{(-4)} &= -C_3^{(-6)}~, \\
\cD^{--} C_3^{(-6)} &= -C_4^{(-8)}~.
\end{aligned}
\end{equation}
The first equation implies that $C_2^{(0)}$ must be holomorphic in $S$. Let us denote
\begin{align}
C_2^{(0)} \equiv -\frac{i}{2} \cH(S)~, \qquad \bar C_2^{(0)} \equiv \frac{i}{2} \bar \cH(\bar S)~.
\end{align}
The holomorphic function $\cH(S)$ will turn out to govern the conformal supergravity
action. The numerical factor has been chosen to give the same normalization as
\cite{Butter:2016mtk}.

The second column of equations defines $C_2^{(-4)}$, $C_3^{(-6)}$ and $C_4^{(-8)}$
as successive derivatives of $C_1^{(-2)}$.
The remaining equations for $C_3^{(+2)}$ and $C_4^{(+4)}$ imply that
\begin{align} \label{E:HfromW4}
-\frac{i}{2} \cH(S) \equiv C_2^{(0)} = \frac{1}{12} (\cD^{--})^2 C_4^{(+4)}
\end{align}
and so $C_4^{(+4)}$ can be interpreted as a potential for $\cH(S)$. 

The other constraints are a bit more involved. The constraint
$[\nabla_{\alpha m} C^{ij}{}_{rs}]_{\rep{60}} = [2 \Lambda_{\alpha m} A^{ij}{}_{rs}]_{\rep{60}}$
and its complex conjugate imply the following set of coupled equations
\begin{subequations}\label{E:BlockEqs1}
\begin{alignat}{3}
-\cD^{++} C_1^{(-2)} - C_2^{(0)} &=& 2 \,A_1^{(0)}\,\,\, &= -\cD^{--}  \bar C_1^{(+2)} - \bar C_2^{(0)} \label{E:BlockEqs1a}\\
-\cD^{++} C_2^{(0)} - C_3^{(+2)} &=&\, 2 \,A_2^{(+2)} &= -\cD^{--} \bar C_2^{(+4)} \label{E:BlockEqs1b}\\
-\cD^{++} C_3^{(+2)} - C_4^{(+4)} &=&\, 2 \,A_3^{(+4)} &= -\cD^{--} \bar C_3^{(+6)} + 2\, \bar C_2^{(+4)} \label{E:BlockEqs1c}\\
            &\phantom{=}&\, 2 \,A_4^{(+6)} &= -\cD^{--} \bar C_4^{+8} + 6\, \bar C_3^{(+6)}~.\label{E:BlockEqs1d}
\end{alignat}
\end{subequations}
The constraint $[\nabla^{\dalpha m} A^{ij}{}_{kl}]_{\rep{60}} = [\bar\Lambda^{\dalpha m} C^{ij}{}_{kl}]_{\rep{60}}$
leads to
\begin{subequations}
\begin{alignat}{2}
C_2^{(0)} &= -\cD^{--} A_2^{(+2)}~, &\qquad
C_1^{(-2)} &= -\cD^{--} A_1^{(0)} - \bar A_2^{(-2)}~, \label{E:BlockEqs2a} \\
C_3^{(+2)} &= -\cD^{--} A_3^{(+4)} + 2 \,A_2^{(+2)}~, &\qquad
C_2^{(-4)} &= -\cD^{--} \bar A_2^{(-2)} - \bar A_3^{(-4)}~, \\
C_4^{(+4)} &= -\cD^{--} A_4^{(+6)} + 6 \,A_3^{(+4)}~, &\qquad
C_3^{(-6)} &= -\cD^{--} \bar A_3^{(-4)} - \bar A_4^{(-6)}~.
\end{alignat}
\end{subequations}

To disentangle these equations, it is helpful to start with the second
equation of \eqref{E:BlockEqs2a} as the definition of $C_1^{(-2)}$ and insert it into 
\eqref{E:BlockEqs1a}. The result reads
\begin{align}
\cD^{++} \cD^{--} A_1^{(0)} + \frac{i}{2} (\cH - \bar \cH) = 2 \,A_1^{(0)}~.
\end{align}
This suggests to write $A_1^{(0)} = \cB + \frac{i}{4} (\cH - \bar \cH)$
where $\cB$ is a real function obeying
\begin{align}\label{E:DDB}
\cD^{++} \cD^{--} \cB = 2 \,\cB~.
\end{align}
We will discuss this equation shortly, but one immediate implication is
$\cD^{--} (\cD^{++})^2 \cB = 0$.

Next, we need to equate both solutions for $A_2^{(+2)}$ in \eqref{E:BlockEqs1b}.
This implies that
\begin{align}\label{E:C44}
C_4^{(+4)} - 2 (\cD^{++})^2 \cB = \frac{3i}{2} (\cD^{++})^2 \cH - \frac{1}{2} \cD^{++} \cD^{--} C_4^{(+4)}~.
\end{align}
To disentangle $\cB$ from this equation, we write
\begin{align}
C_4^{(+4)} = -6i \, \cI^{(+4)} -\frac{3i}{2} (\cD^{++})^2 \cH + 2 (\cD^{++})^2 \cB
\end{align}
in terms of a new quantity $\cI^{(+4)}$. Then \eqref{E:C44} and \eqref{E:HfromW4} are
equivalent to the two equations
\begin{align}\label{E:DDGamma}
\cI^{(+4)} &= - \frac{1}{2} \cD^{++} \cD^{--} \cI^{(+4)} 
    - \frac{1}{4} (\cD^{++})^2 (\cD^{--})^2 \cI^{(+4)}~, \\
\label{E:HfromGamma}
\cH &= (\cD^{--})^2 \cI^{(+4)}~.
\end{align}
Furthermore, one can show that provided $\cI^{(+4)}$ obeys \eqref{E:DDGamma},
the quantity \eqref{E:HfromGamma} is automatically holomorphic.

One finds using the remaining equations that each of the coset scalar functions is
determined and all of the required equations are satisfied. It is actually remarkable
because the system is overdetermined, with numerous overlapping differential
equations. The coset functions appearing in $C^{ij}{}_{kl}$ are
\begin{subequations}
\begin{alignat}{2}
C_1^{(-2)} &= -\cD^{--} \cB 
    + \frac{i}{4} \cD^{--} \bar \cH 
    + \frac{i}{2} \cD^{++} \bar \cI^{(-4)}~, & \quad & \quad \\
C_2^{(0)} &= -\frac{i}{2} \cH~, & \quad
C_2^{(-4)} &= -\cD^{--} C_1^{(-2)}~, \\
C_3^{(+2)} &= -i \cD^{--} \cI^{+4} + \frac{i}{2} \cD^{++} \cH~, & \quad
C_3^{(-6)} &= (\cD^{--})^2 C_1^{(-2)}~, \\
C_4^{(+4)} &= -6i \,\cI^{(+4)} - \frac{3i}{2} (\cD^{++})^2 \cH + 2 (\cD^{++})^2 \cB~, & \quad
C_4^{(-8)} &= -(\cD^{--})^3 C_1^{(-2)}~.
\end{alignat}
\end{subequations}
while those in $A^{ij}{}_{kl}$ become
\begin{subequations}
\begin{align}
A_1^{(0)} &= \cB + \frac{i}{4} (\cH - \bar \cH)~,\\
A_2^{(+2)} &= \frac{i}{2} \cD^{--} \cI^{(+4)}~, \\
A_3^{(+4)}
	&= -(\cD^{++})^2  \cB 
        -i \cD^{++} \cD^{--} \cI^{(+4)} 
        - \frac{i}{4} (\cD^{++})^2 \cH ~, \\
A_4^{(+6)} &=
	2 (\cD^{++})^3 \cB
	+ \frac{i}{2} (\cD^{++})^3 \cH
	+ \frac{3i}{2} (\cD^{++})^2 \cD^{--} \cI^{(+4)}~.
\end{align}
\end{subequations}

There are two linearly independent solutions described above. One involves a function
$\cB$ obeying \eqref{E:DDB}. This equation is just the massive Laplace equation 
on $\SU(1,1) / \gU(1)$. Remarkably, the action generated from $\cB$ turns out to be a
total derivative, which is not at all obvious at this stage. We will justify this claim
via a formal argument in section \ref{sec:unique} below. We have also checked this by
analyzing the bosonic part of the action it generates.

The second solution involves a function
$\cI^{(+4)}$ obeying \eqref{E:DDGamma}, which is used to construct
a holomorphic function $\cH(S)$ from \eqref{E:HfromGamma}. At first glance,
this does not quite match our expectations as it is not at all obvious that a suitable
$\cI^{(+4)}$ can be constructed for any given $\cH(S)$, nor is it clear that the action will depend on
$\cH(S)$ alone. The latter claim will be justified in the next subsection.
For the former claim, observe that if we can find some function
$\cK$ obeying $\cD^{++} \cD^{--} \cK = \cH$ for any $\cH(S)$, then $\cI^{(+4)}$ can be chosen as
$\cI^{(+4)} = -\frac{1}{2} (\cD^{++})^2 \cK$.  The required equation for $\cK$ is
simply the inhomogeneous Poisson equation on the coset space $\SU(1,1)/\gU(1)$: in
Poincar\'e disk coordinates, it reads $(1- S \bar S)^2 \pa_S \pa_{\bar S} \cK = \cH(S)$.
We have denoted the function $\cK$ because, although complex, it is defined
up to K\"ahler transformations where it is shifted by purely holomorphic or
anti-holomorphic functions.

\subsection{Uniqueness of the conformal supergravity action}\label{sec:unique}

Before moving on to construct the explicit action based on the Weyl multiplet fields, let us first justify an important claim that was made previously: the most general action of
$\cN=4$ conformal supergravity is parametrized by a single holomorphic function of
the coset scalars. Our path to this result involves establishing the
uniqueness of the corresponding supercurrent. A similar line of argument
was used in \cite{Butter:2016qkx} to establish the uniqueness of the 6D $\cN=(1,0)$ conformal
supergravity actions.

Any classically superconformal theory possesses a supercurrent, which is the
multiplet containing the stress-energy tensor. The bottom component of this
multiplet is a S-invariant pseudoreal field $J^{ij}{}_{kl}$ transforming in the $\rep{20'}$
representation with Weyl weight $w=2$. The defining constraint
is that  $[ \nabla^{\dalpha m} J^{ij}{}_{kl}]_{\rep{60}} = 0$, and similarly
for its complex conjugate \cite{Bergshoeff:1980is}.

When the theory is coupled to conformal supergravity -- or in this case 
is itself conformal supergravity -- then $J^{ij}{}_{kl}$ may be identified
with the functional variation of the action with respect to $D^{kl}{}_{ij}$.
Its weight and the restriction on its supersymmetry variation follow rather
easily.

Suppose the theory in question involves the Weyl multiplet alone. Then
$J^{ij}{}_{kl}$ can only have an expansion of the form
\begin{align}\label{E:Jcurrent}
J^{ij}{}_{kl} &= J_1^{(0)} D^{ij}{}_{kl}
	\eol & \quad
	+ J_2^{(+2)} X_2{}^{ij}{}_{kl} + \bar J_2^{(-2)} \bar X_2{}^{ij}{}_{kl}
	\eol & \quad
	+ J_3^{(+4)} X_3{}^{ij}{}_{kl} + \bar J_3^{(-4)} \bar X_3{}^{ij}{}_{kl}
	\eol & \quad
	+ J_4^{(+6)} X_4{}^{ij}{}_{kl} + \bar J_4^{(-6)} \bar X_4{}^{ij}{}_{kl}
\end{align}
involving the same tensors $X_{n}$ appearing in \eqref{eq:X}.
These constitute the $S$-invariant combinations that are of Weyl weight 2.
If we require that the supersymmetry transformation of $J^{ij}{}_{kl}$ does
not involve either the $\rep{60}$ or  the $\brep{60}$, we find a sequence of equations
\begin{alignat}{2}
J_2^{(+2)} &= -\cD^{++} J_1^{(0)}~, & \qquad
\cD^{--} J_2^{(+2)} &= 0~, \eol
J_3^{(+4)} &= (\cD^{++})^2 J_1^{(0)}~, & \qquad
\cD^{--} J_3^{(+4)} &= 2\, J_2^{(+2)}~, \eol
J_4^{(+2)} &= -(\cD^{++})^3 J_1^{(0)}~, & \qquad
\cD^{--} J_4^{(+6)} &= 6 \,J_3^{(+4)}~.
\end{alignat}
The two equations involving $J_2^{(+2)}$ imply that $\cD^{--} \cD^{++} J_1^{(0)} = 0$, which
means $J_1^{(0)}$ is the sum of a holomorphic and anti-holomorphic piece. Calling
these pieces $\cJ(S)$ and $\bar \cJ(\bar S)$, we have
\begin{gather}
J_1^{(0)} = \cJ + \bar \cJ~,  \quad
J_2^{(+2)} = -\cD^{++} \cJ~,  \quad
J_3^{(+4)} = (\cD^{++})^2 \cJ~, \quad
J_3^{(+6)} = -(\cD^{++})^3 \cJ~.
\end{gather}
The supercurrent is parametrized by a single holomorphic function.
This implies that the action whose variation yields this supercurrent
must also be parametrized by a single holomorphic function.
Since $J^{ij}{}_{kl}$ can be identified as the functional variation of 
the action with respect to $D^{ij}{}_{kl}$, the action
we are discussing must  possess the term
$(\cJ + \bar \cJ)\, D^{ij}{}_{kl} D^{kl}{}_{ij}$.

It is then enough to show that for the class of actions we have constructed,
the coefficient of $D^{ij}{}_{kl} D^{kl}{}_{ij}$ involves only $\cH(S)$ and not
$\cI^{(+4)}$ or $\cB$.  The apparent dependence on $\cI^{(+4)}$ must just be
an artefact of how we built the conformal supergravity action and can be removed by extracting a total derivative.
Similarly, the independent action that can be constructed out of $\cB$
must be a total derivative, since that action has no supercurrent.

\section{Presentation of results}\label{S:Results}
\allowdisplaybreaks

From the expressions of the lowest dimension composite \eqref{eq:hb} in terms of the Weyl multiplet fields and the supersymmetry transformations rules \eqref{eq:tr1}, we use the computer algebra package \texttt{Cadabra} \cite{Peeters.Cadabra1,Peeters.Cadabra2} to generate the full $\cN=4$ conformal supergravity Lagrangian. The resulting complete expression is attached as a separate file. The purely bosonic part $\mathcal{L}_B$ of the Lagrangian was already presented in \cite{Butter:2016mtk} and can be written as\footnote{The Lagrangian has been rescaled by
a factor of 2 relative to \cite{Butter:2016mtk}.}
\begin{align}
  \label{eq:BosLag}
e^{-1}\mathcal{L}_B &=
         \mathcal{H}\,\Big[\,
	\tfrac{1}{2}\, {R(M)}^{a b c d}\, R(M)^-_{a b c d}
	+ {R(V)}^{a b}{}^{i}{}_{j} \,R(V)^-_{a b}{}^{j}{}_{i}
	+ \tfrac{1}{8}\, {D}^{i j}{}_{k l} \,{D}^{k l}{}_{i j}
	+ \tfrac{1}{4}\, {E}_{i j} \,D^2 E^{i j}\,
	\nonumber\\[1mm]
	& \qquad\;\;\;
	- 4\, {T}_{a b}{\!}^{i j} {D}^{a}{D}_{c}\,{T}^{cb}{\!}_{i j}- {\bar P}^{a} {D}_{a}{D}_{b}{P}^{b}
        + P^2 \bar P^2 + \tfrac{1}{3} (P^a \bar P_a)^2
        - \tfrac{1}{6}\,{P}^{a} {\bar P}_{a} \,{E}_{i j}\,{E}^{i j} 
       \nonumber\\[2mm]
        & \qquad\;\;\;
          - 8\, {P}_{a} \,{\bar P}^{c}\, {T}^{a b}{\!}_{i j} \,{T}_{b c}{\!}^{i j} 
        - \tfrac{1}{16}\, {E}_{i j} \,{E}^{j k} \,{E}_{k l} \,{E}^{li} 
	+ \tfrac{1}{48}\, ({E}_{i j}\, {E}^{i j})^2
	+ {T}^{ab}{\!}_{i j}\, {T}_{a b\, k l}\, {T}^{c d\,i j} \,{T}_{c d}{}^{k l}
	\nonumber\\[2mm]
        & \qquad\;\;\;
        	- {T}^{ab}{\!}_{i j}\,{T}_{c d}{\!}^{jk}\, {T}_{a b\, k l}\,{T}^{c d\,l i}
        - \tfrac{1}{2}\, {E}^{i j} \,{T}^{a b\,k l}\, {R(V)}_{a b}{\!}^{m}{}_{i}\, {\varepsilon}_{j k l m}
        + \tfrac{1}{2}\, {E}_{i j}\, {T}^{ab}{}_{k l}\,
                {R(V)}_{a b}{\!}^{i}{}_{m} \,{\varepsilon}^{j k l m}  
	\nonumber\\[2mm]
        & \qquad\;\;\;
	- \tfrac{1}{16}\, {E}_{i j} {E}_{k l} \,{T}^{a b}{\!}_{m n}\,
        {T}_{a b \,p q}\,
        {\varepsilon}^{i k m n}\, {\varepsilon}^{j l p q} 
	- \tfrac{1}{16}\, {E}^{i j} {E}^{k l} \,{T}^{a b\,m n} \,{T}_{a b}{}^{p q} \,{\varepsilon}_{i k m n} \,{\varepsilon}_{j l p q} 
	\nonumber\\[2mm]
	& \qquad\;\;\;
	- 2 \,{T}^{a b\,i j}\,\big( {P}_{[a} {D}_{c]}{{T}_{b}{}^{c\,k l}}
        +\tfrac16\,{P}^{c} {D}_{c}{{T}_{a b}{}^{k l}}+\tfrac13\,{T}_{a
          b}{}^{k l} {D}_{c}{{P}^{c}}\big)\,  {\varepsilon}_{i j k l} 
          \nonumber\\[2mm]	
	& \qquad\;\;\;
	- 2 \,{T}^{a b}{\!}_{i j} \,\big({\bar P}_{[a} {D}_{c]}{T}_{b}{}^{c}{}_{ k l}
        -\tfrac12\, {\bar P}^{c} {D}_{c}{{T}_{a b \,k l}}\big)\,
        {\varepsilon}^{i j k l}   
        \,	\Big] \nonumber\\[1mm]
        & \;\;\;\,
	+ \mathcal{D}{\mathcal{H}} \,  \Big[\,
	 \tfrac{1}{4}\, {T}_{a b}{}^{i j}\, {T}_{c d}{}^{k l} {R(M)}^{a b c d} \,{\varepsilon}_{i j k l}
	+{E}_{i j} \,{T}^{a b\,i k} \,{R(V)}_{a b}{\!}^{j}{}_{k}
	+{T}^{a b\,i j} \,{T}_{a}{}^{c\,k l} \,{R(V)}_{b c}{\!}^{m}{}_{k} \,{\varepsilon}_{i j l m}
	\nonumber\\[1mm]
        & \qquad\;\;\;\;\;\;\;\;\;
        - \tfrac{1}{24}\, {E}_{i j} \,{E}^{i j} \,{T}^{a b\,k l}\, 
        {T}_{a b}{}^{m n}\, {\varepsilon}_{k l m n}
        - \tfrac{1}{6}\, {E}^{i j} \,{T}_{a b}{}^{k l}\, {T}^{a c\,m n} \,{T}^b{}_{c}{}^{p q} \,{\varepsilon}_{i k l m} \,{\varepsilon}_{j p q n}\,\,
        \nonumber\\[1mm]
        & \qquad\;\;\;\;\;\;\;\;\;
        -\tfrac{1}{8}\, {D}^{i j}{}_{k l} \,\big(\,{T}^{a b\,m n} 
        \,{T}_{a b}{}^{k l}\, {\varepsilon}_{i j m n}
        - \tfrac{1}{2}\, {E}_{i m}\, {E}_{j n}\, {\varepsilon}^{k l m n}\,\big)
        \Big]	\nonumber\\[1mm] 
        & \;\;\;\; 
        + {\mathcal{D}}^2\mathcal{H}\, \Big[ \tfrac{1}{32}\,{T}^{a b \, i
          j}\,{T}^{c d \, p q}\, {T}_{a b}{}^{m n} \,{T}_{c d}{}^{k
          l}\, {\varepsilon}_{i j k l} \,{\varepsilon}_{m n p q}
        -\tfrac{1}{64}\,{T}^{a b \, i j}\,{T}^{c d \, p q}\, {T}_{a
          b}{}^{k l} \,{T}_{c d}{}^{m n} \, {\varepsilon}_{i j k l}
        \,{\varepsilon}_{m n p q}
	\nonumber\\[1mm]
        & \qquad\;\;\;\;\;\;\;\;\;\;\, 
         +\tfrac{1}{6}\,
        {E}_{i j} \,{T}_{a b}{}^{i k} \,{T}^{a c \, jl}\,
        {T}^{b}{}_{c}{}^{m n} \,{\varepsilon}_{k l m n} 
        + \tfrac{1}{384}\, {E}_{i j} \,{E}_{k l}
        \,{E}_{m n} \,{E}_{p q} \,{\varepsilon}^{i k m p}\,
        {\varepsilon}^{j l n q}
        \nonumber\\[1mm]
        & \qquad\;\;\;\;\;\;\;\;\;\;\,
        -\tfrac{1}{8}\, {E}_{i j} \,{E}_{k l} \,{T}_{a b}{}^{i k}
        \,{T}^{a b \, j l}  \, \Big]
	\nonumber\\
        &  \;\;\;\;	
	+ 2 \,\mathcal{H} \,e_a{}^\mu f_\mu{}^c\,\eta_{cb} \, \Big[\,
	 {P}^{a} \,{\bar P}^{b}  - P^d\, \bar P_d \, \eta^{a b}\,
	\Big]+\text{h.c.}\
\end{align}
The coset derivatives $\cD \equiv \cD^{++}$ and $\cD^\dagger \equiv \cD^{--}$ are
defined in \eqref{eq:cosetder}; here we use the notations of \cite{Butter:2016mtk}
where the U(1) charges are suppressed. $D_a$ is the fully supercovariant derivative (including
the gravitino connection) and it coincides with the projection to components of the superspace
derivative $\nabla_a$ defined in \eqref{eq:covdd}. All covariant fields of the Weyl
multiplet play a role in the action, including $E^{ij}$ and $D^{ij}{}_{kl}$, as well as
the $\SU(4)$ curvature $R(V)_{ab}{}^i{}_j$ and the Lorentz curvature $R(M)_{abcd}$.
$P_a$ is the supercovariant vielbein on the coset space and is defined in \eqref{eq:def.Pa}.

From the point of view of the density formula \eqref{eq:densityf}, this Lagrangian corresponds to the bosonic part of the composite field $F$. As explained in section \ref{sec:solAC}, the conformal supergravity Lagrangian obtained from our action principle a priori depends on the real and complex functions of the coset scalars $\cB$ and $\cI$, respectively.  It was further argued in section \ref{sec:unique} that the dependence on these functions can be removed by extracting a total derivative. The elimination of $\cI$ in this way however typically generates terms that depend on $\mathcal{H}$ through \eqref{E:HfromGamma}. These terms can modify the structure of the density formula \eqref{eq:densityf}. In particular, at the purely bosonic level this total derivative introduces a dependence on the bare K-gauge field. This explains the last term in \eqref{eq:BosLag} whose presence also ensures the invariance of the kinetic term for the coset scalars under conformal boosts. The expression \eqref{eq:BosLag} is then fully invariant all the bosonic symmetries.

A very stringent check of our result can be performed by setting the function $\mathcal{H}$ to a constant. The Lagrangian is then invariant under rigid SU(1,1) transformations and the bosonic terms \eqref{eq:BosLag} reduce precisely to the result of \cite{Ciceri:2015qpa}. In this case, the bare K-gauge field can also be eliminated by extracting a total derivative and writing a kinetic term for the coset scalars which is invariant under conformal boosts up to fermionic terms. For any other holomorphic function, the rigid SU(1,1) invariance is broken.

Let us now present all the supercovariant terms which are quadratic in the fermion fields. They are still all contained in the field $F$ and for legibility we will decompose them according to the number of coset derivatives acting on the holomorphic function $\cH$. Once again, all the terms depending on the function $\cI$ can be eliminated by splitting off a total derivative. The terms which do not depend on derivatives of $\cH$ read
\begin{align}
\cH&\Big[\bar R(Q)_{ab}^i\,R(S)^{ab}_i-\tfrac34\bar\chi^{ij}{}_k\slashed{D} \chi_{ij}{}^k-\tfrac14\bar\chi_{ij}{}^k\slashed{D}\chi^{ij}{}_k-\tfrac{3}{8}\bar{\Lambda}^{i}\left(\slashed{D}D^2+D^{2}\slashed{D}-\slashed{D}^{3}\right)\Lambda_{i}\nonumber\\[1mm]
&-\tfrac{1}{8}\bar{\Lambda}_{i}\left(\slashed{D}D^2+D^{2}\slashed{D}-\slashed{D}^{3}\right)\Lambda^{i}-\tfrac{1}{8}\bar{\chi}^{lm}{}_{k}\gamma\cdot \slashed{D}T^{ij}\Lambda^{k}\varepsilon_{ijlm} -\tfrac{1}{8}\bar{\chi}^{k}{}_{lm}\gamma\cdot \slashed{D}T_{ij}\Lambda_{k}\,\varepsilon^{ijlm}\nonumber\\[1mm]
& +\tfrac{1}{8}\bar{\chi}^{lm}{}_{k}\gamma\cdot T^{ij}\slashed{D}\Lambda^{k}\varepsilon_{ijlm}+\tfrac{1}{8}\bar{\chi}^{k}{}_{lm}\gamma\cdot T_{ij}\slashed{D}\Lambda_{k}\varepsilon^{ijlm}-\tfrac14 E^{ij}\bar\chi^{kl}{}_i\chi^{mn}{}_j\varepsilon_{klmn}\nonumber\\[1mm]
&-\tfrac14 E_{ij}\,\bar\chi^i{}_{kl}\chi^j{}_{mn}\,\varepsilon^{klmn}-\tfrac12 \bar R(Q)^{ab\,i}\slashed{D} T_{ab}{}^{kl}\Lambda^j \varepsilon_{ijkl}-\tfrac12\bar R(Q)^{ab}_{i}\slashed{D} T_{ab\,kl}\,\Lambda_j\, \varepsilon^{ijkl}\nonumber\\[1mm]
&-D_c\bar R(Q)^{ab\,i}\gamma_c\Lambda^j T_{ab}{}^{kl}\varepsilon_{ijkl}-\tfrac12 D_c \bar R(Q)^{ab}_{i}\gamma_c\Lambda_j T_{ab \,kl}\,\varepsilon^{ijkl}+2\,D_{a}\bar{\Lambda}_{i}R(Q)^{ab\,i}P_{b}\nonumber\\[1mm]
& +\tfrac{1}{2}T_{ij}\cdot T_{kl}\,\bar{\Lambda}_{m}\chi^{kl}{}_{n}\,\varepsilon^{ijmn}+\tfrac{1}{2}T^{ij}\cdot T^{kl}\bar{\Lambda}^{m}\chi_{kl}{}^{n}\varepsilon_{ijmn}+\tfrac{1}{2}P^{a}\bar{\Lambda}_{k}\gamma\cdot T^{ij}\gamma_{a}\chi^{k}{}_{ij}\nonumber\\[1mm]
&+\tfrac{1}{2}\bar{P}^{a}\bar{\Lambda}^{k}\gamma\cdot T_{ij}\gamma_{a}\chi^{ij}{}_{k}-\tfrac{1}{12} E^{ij}E^{kl}\bar\Lambda_i\chi^{mn}{}_k\,\varepsilon_{jlmn}-\tfrac{1}{12} E_{ij}E_{kl}\bar\Lambda^i\chi^k{}_{mn}\,\varepsilon^{jlmn}\nonumber\\[1mm]
&-\tfrac16 E^{ij}\bar P^a\bar\Lambda^k\gamma_a\chi^{lm}{}_{i}\,\varepsilon_{jklm}-\tfrac16 E_{ij}P^a\bar\Lambda_m\gamma_a\chi^i{}_{kl}\,\varepsilon^{jklm}-\tfrac16 E^{ij} D_a E_{jk}\,\bar\Lambda_i\gamma^a\Lambda^k\nonumber\\[1mm]
&-\tfrac{1}{48}E^{ij}D_a E_{ij}\,\bar\Lambda_k\gamma^a\Lambda^k-\tfrac{1}{48}E_{ij}D_a E^{ij}\bar\Lambda_k\gamma^a\Lambda^k-\tfrac{1}{16} E^{ij}E_{ij}\bar\Lambda_k\gamma^a D_a\Lambda^k\nonumber\\[1mm]
&+\tfrac{1}{48} E^{ij}E_{ij}\,D_a\bar\Lambda_k\gamma^a\Lambda^k+\tfrac{1}{12} E^{ik} E_{ij}\,\bar\Lambda_k\gamma^a D_a\Lambda^j-\tfrac14 E_{ik} E^{ij}D_a\bar\Lambda_j\gamma^a\Lambda^k\nonumber\\[1mm]
&+\tfrac{5}{12}D_a E^{ij} P^a \bar\Lambda_i\Lambda_j +\tfrac{5}{12}D_a E_{ij} \bar{P}^a \bar\Lambda^i\Lambda^j -\tfrac{1}{3}E^{ij}P^a\bar\Lambda_i\gamma_{ab}D_b\Lambda_j +\tfrac{1}{3}E^{ij}D_{a}P^{a}\bar{\Lambda}_{i}\Lambda_{j} \nonumber\\[1mm]
&+\tfrac{1}{3}E_{ij}\,D_{a}\bar{P}^{a}\bar{\Lambda}^{i}\Lambda^{j}-\tfrac{1}{4}\bar{\Lambda}^{i}\gamma^{a}\Lambda_{i}D_{b}\bar{P}_{a}P^{b}-\tfrac{5}{12}\bar{\Lambda}^{i}\gamma^{a}\Lambda_{i}D_{b}P_{a}\bar{P}^{b}+\tfrac{2}{3}\bar{\Lambda}^{i}\gamma^{a}D_{b}\Lambda_{i}\bar{P}_{a}P^{b}\nonumber\\[1mm]
 & -\tfrac{2}{3}D_{a}\bar{\Lambda}^{i}\gamma^{b}\Lambda_{i}\bar{P}^{a}P^{b}+\tfrac{1}{12}\bar{\Lambda}^{i}\gamma^{b}\Lambda_{i}D_{a}\bar{P}^{a}P^{b}-\tfrac{1}{12}\bar{\Lambda}^{i}\gamma^{b}\Lambda_{i}D_{a}{P}^{a}\bar{P}^{b}-\tfrac{2}{3}D_{a}\bar{\Lambda}^{i}\gamma^{a}\Lambda_{i}\bar{P}^{b}P_{b}\nonumber\\[1mm]
& +\tfrac{4}{3}\bar{\Lambda}_{i}\gamma_{a}D_{b}\Lambda^{i}\bar{P}_{c}P_{d}\varepsilon^{abcd}-{2}T_{ab \,ij} T^{ac\,ik}\bar\Lambda_k\gamma^bD_c\Lambda^j-2T_{ab \,ij} D_{c}T^{ac\,ik}\bar\Lambda_k\gamma^b\Lambda^j\nonumber\\[1mm]
&-2 \,T_{ab}{}_{ij} T^{ac}{}^{ij}D_{c}\bar\Lambda_k\gamma^b\Lambda^k-{2}\, D_{c}T_{ab}{}_{ij} T^{ac}{}^{ij}\bar\Lambda_k\gamma^b\Lambda^k +\tfrac13 P_a T^{ab}{}_{ij}\bar\Lambda_kD_b\Lambda_l\varepsilon^{ijkl}\nonumber\\[1mm]
&+\tfrac13\bar P_a T^{ab\,ij}\bar\Lambda^k D_b\Lambda^l\varepsilon_{ijkl}-\tfrac12 P_cD^b T_{ab \,ij}\bar\Lambda_k\gamma^{ac}\Lambda_l\varepsilon^{ijkl}-\tfrac12\bar P^c D_b T^{ab\,ij}\bar\Lambda^k\gamma_{ac}\Lambda^l\varepsilon_{ijkl}\nonumber\\[1mm]
&-\tfrac13 D^c P_b T^{ab}{}_{ij}\bar\Lambda_k\gamma_{ac}\Lambda_l\varepsilon^{ijkl}-\tfrac13 D^c \bar P_b T^{ab\,ij}\bar\Lambda^k\gamma_{ac}\Lambda^l\varepsilon_{ijkl}+\tfrac16 E^{ij} D_b T^{ab\,lm}\bar\Lambda_i\gamma_a\Lambda^k\varepsilon_{jklm}\nonumber\\[1mm]
&-\tfrac16 E_{ij} D_b T^{ab}{}_{lm}\bar\Lambda_k\gamma_a\Lambda^i\varepsilon^{jklm}-\tfrac13 E_{ij} T^{ab}{}_{kl} D_b\bar\Lambda_m\gamma_a\Lambda^i\varepsilon^{jklm}+\tfrac13 E^{ij} T^{ab\,kl}\bar\Lambda_i\gamma_aD_b \Lambda^m\varepsilon_{jklm}\nonumber\\[1mm]
&-\tfrac12 D_b E^{km} T^{ab\,ij}\bar\Lambda_k\gamma_a\Lambda^l\varepsilon_{ijlm}+D_b E_{km} T^{ab}{}_{ij}\bar\Lambda_l\gamma_a\Lambda^k\varepsilon^{ijlm}-\tfrac{1}{24}E^{ij}T_{ab\, kl} T^{ab}{}_{mn}\bar\Lambda_i\Lambda_j\varepsilon^{klmn}\nonumber\\[1mm]
&-\tfrac{1}{24}E_{ij} T_{ab}{}^{kl} T^{ab\,mn}\bar\Lambda^i\Lambda^j\varepsilon_{klmn}-\tfrac13 E_{ij} P^a T_{ab}{}^{ik}\bar\Lambda_k\gamma^b\Lambda^j +\tfrac13 E^{ij} P^a T_{ab\, ik}\bar\Lambda_j \gamma^b\Lambda^k\nonumber\\[1mm]
& -\tfrac16 P^c P_c T_{ab}{}^{ij}\bar\Lambda_i\gamma^{ab}\Lambda_j-\tfrac16 \bar P^c\bar P_c T_{ab ij}\bar\Lambda^i\gamma^{ab}\Lambda^j -\tfrac{1}{24} T_{ab}{}^{ij} T^{ab\,kl}\bar\Lambda_m\gamma_c\Lambda^m P^c\varepsilon_{ijkl}\nonumber\\[1mm]
&+{\tfrac{1}{24}} T_{ab}{}_{ij} T^{ab}{}_{kl}\bar\Lambda_m\gamma_c\Lambda^m \bar{P}^c\varepsilon^{ijkl}-\tfrac{1}{2}\bar{\Lambda}_{i}\gamma^{a}\Lambda^{j}D^{b}R(V)_{ab}{}^{i}{}_{j}\Big]+\text{h.c.}\,.\label{eq:res2f1}
\end{align}
This result can once more be checked by setting the function $\cH$ to a constant. In this case, the above expression indeed reduces again to the result of \cite{Ciceri:2015qpa}. The remaining terms which are quadratic in fermions depend on derivatives of $\cH$. Those with a single derivative can be written as
\begin{align}
\cD \cH&\Big[\tfrac14 \bar \chi^{ij}{}_k\gamma_a\chi^k{}_{ij}\bar P^a-\tfrac12 \bar\chi^{ik}{}_l\chi^{jl}{}_k E_{ij}-\tfrac14\bar \chi^{kl}{}_m\gamma^{ab}\chi^{mn}{}_k T_{ab}{}^{ij}\varepsilon_{ijln}-\tfrac12 E_{ij} \bar R(Q)_{ab}^i R(Q)^{ab\,j}\nonumber\\[1mm]
&\;\;-\tfrac34 T_{ab}{}^{ij} \bar R(Q)^{ab\,k}\chi^{lm}{}_k\varepsilon_{ijlm}-\tfrac12 D^{ij}{}_{kl}\bar\Lambda_i\chi^{kl}{}_j-\tfrac12 R(V)_{ab}{}^j{}_k\bar\Lambda_i\gamma^{ab}\chi^{ik}{}_j-\tfrac14 E_{ij} E^{ik}\bar\Lambda_l\chi^{jl}{}_k\nonumber\\[1mm]
&\;\;-\tfrac14 E_{ij}\bar\Lambda_k\slashed{D}\chi^i{}_{lm}\varepsilon^{jklm}+\tfrac{1}{24} E^{ij} T_{ab}{}^{kl}\bar\Lambda_i\gamma^{ab}\chi^{mn}{}_j\varepsilon_{klmn}
+\tfrac18 E^{ij} T_{ab}{}^{kl}\bar\Lambda_m\gamma^{ab}\chi^{mn}{}_i\varepsilon_{jkln}\nonumber\\[1mm]
&\;\; +\tfrac{1}{4}\bar{\Lambda}_{i}\gamma\cdot T^{jk}\slashed{D}\chi^{i}{}_{jk}-\tfrac{1}{12}\bar{\chi}^{lm}{}_{k}\gamma\cdot T^{ij}\bar{\slashed{P}}\Lambda^{k}\varepsilon_{ijlm}+\tfrac{1}{8}\bar{\chi}^{k}{}_{lm}\gamma\cdot T_{ij}\bar{\slashed{P}}\Lambda_{k}\varepsilon^{ijlm}-\tfrac{3}{8}\bar{\Lambda}^{i}\gamma_{a}D^2 \Lambda_{i}\bar{P}^{a}\nonumber\\[1mm]
&\;\;+\tfrac13 E^{ij} T_{ab}{}^{kl} \bar\Lambda_i R(Q)^{ab\,m}\varepsilon_{jklm}+\tfrac16 T^{ab}{}_{ij} \bar P^c\bar\Lambda_k\gamma_c R(Q)_{ab l}\varepsilon^{ijkl}-\tfrac34 T_{ab}{}^{ij}\bar P^c \bar R(Q)^{ab\,k}\gamma_c\Lambda^l\varepsilon_{ijkl}\nonumber\\[1mm]
&\;\;+\tfrac{1}{3}\bar{\Lambda}_{i}R(Q)_{ab}^{i}\bar{P}^{a}P^{b}-\tfrac{7}{6}\bar{\Lambda}_{i}R(Q)^{ab\,j}R(V)_{ab}{}^{i}{}_{j}-\tfrac{3}{2}\bar{\Lambda}_{i}T^{ij}\cdot R(S)_{j}+\tfrac{1}{4}\bar{\Lambda}_{i}\gamma^{cd}R(Q)^{ab\,i}R(M)_{abcd}\nonumber\\
&\;\; +\tfrac{1}{8}\bar{\Lambda}^{i}\gamma_{d}\Lambda_{j}R(V)_{ab}{}^{j}{}_{i}\bar{P}_{c}\varepsilon^{abcd}+\tfrac{1}{6}\bar{\Lambda}_{k}\Lambda_{l}T^{ab}{}_{ij}R(V)_{ab}{}^{k}{}_{m}\varepsilon^{ijlm}+\tfrac{8}{3}\bar{\Lambda}^{i}\gamma^b \Lambda_{i}P^a \bar{P}_a \bar{P}_{b}+\tfrac{7}{24}D_{a}\bar{\Lambda}^{i}\gamma_b D_c \Lambda_{i}\bar{P}_{d}\varepsilon^{abcd}\nonumber\\[1mm]
&\;\; -\tfrac{7}{12}\bar{\Lambda}^{i}\gamma^{b}D_a D_b \Lambda_{i}\bar{P}^{a}-\tfrac{1}{4}\bar{\Lambda}^{i}\slashed{D} \Lambda_{i}D_a \bar{P}^{a}-\tfrac{1}{2}\bar{\Lambda}^{i}\gamma^{a}D_b \Lambda_{i}D_a \bar{P}^{b}+\tfrac{1}{24}\bar{\Lambda}_{i}\gamma^{a}D^2 \Lambda^{i}\bar{P}_{a}\nonumber\\[1mm]
& \;\;+\tfrac{1}{24}D^{b}\bar{\Lambda}_{i}\gamma^{a}D_b \Lambda^{i}\bar{P}_{a}-\tfrac{1}{12}\bar{\Lambda}_{i}\gamma^{b}D_a D_b \Lambda^{i}\bar{P}^{a}-\tfrac{1}{12}\bar{\Lambda}_{i}\slashed{D}\Lambda^{i}D_a \bar{P}^{a}-\tfrac{1}{6}\bar{\Lambda}_{i}\gamma^{a}D_b \Lambda^{i}D_a \bar{P}^{b}\nonumber\\[1mm]
& \;\;-\tfrac{1}{8}D_a \bar{\Lambda}_{i}\slashed{D} \Lambda^{i}\bar{P}^{a}+\tfrac{1}{8}D_a \bar{\Lambda}^{i}\slashed{D} \Lambda_{i}\bar{P}^{a}-\tfrac{1}{6}\bar{\Lambda}^{i}\gamma^a \Lambda_{i}D_a D_b \bar{P}^{b}-\tfrac{17}{12}\bar{\Lambda}^{i}\gamma^a \Lambda_{i}P_a \bar{P}^b \bar{P}_{b}\nonumber\\[1mm]
&\;\;+\tfrac56 E_{ij} T_{ab\,kl}\bar\Lambda_m\gamma^b\Lambda^i\bar P^a\varepsilon^{jklm}+\tfrac{1}{18} E^{ij} T_{ab}{}^{kl}\bar\Lambda_i\gamma^b\Lambda^m\bar P^a\varepsilon_{jklm}
+\tfrac16 E_{ij}T_{ab}{}^{ik}D^b\bar\Lambda_k\gamma^a\Lambda^j\nonumber\\[1mm]
&\;\;+\tfrac16D^b E_{jk} T_{ab}{}^{ij}\bar\Lambda_i\gamma^a\Lambda^k+\tfrac13 E_{ij}T_{ab}{}^{ik}\bar\Lambda_k\gamma^a D^b\Lambda^j-\tfrac13 E_{ij}D^b T_{ab}{}^{ik}\bar\Lambda_k\gamma^a\Lambda^j\nonumber \\[1mm]
&\;\;-T_{ab ij}\bar\Lambda_k\gamma^{ac}D^b D_c \Lambda_l\varepsilon^{ijkl}+\tfrac16 D_a D^c T_{bc \,kl}\bar\Lambda_i\gamma^{ab}\Lambda_j\varepsilon^{ijkl}-\tfrac16 T^{ab}{}_{ij}D_a\bar\Lambda_kD_b \Lambda_l\varepsilon^{ijkl}\nonumber\\[1mm]
&\;\;-\tfrac16 D_a T^{ab}{}_{ij}\bar\Lambda_kD_b\Lambda_l\varepsilon^{ijkl}+\tfrac{1}{12} D_a T_{bc\, ij}\bar\Lambda_k\gamma^{ab}D^c\Lambda_l\varepsilon^{ijkl}-\tfrac{1}{12}D^c T_{ac\, ij}\bar\Lambda_k\gamma^{ab}D_b\Lambda_l\varepsilon^{ijkl}\nonumber\\[1mm]
&\;\;-\tfrac13 T_{ab\, ij}D^b\bar\Lambda_k\gamma^{ac}D_c\Lambda_l\varepsilon^{ijkl}+\tfrac23 P_a T^{ab\,ij}\bar\Lambda_iD_b\Lambda_j-\tfrac12 D^c T_{bc}{}^{ij} P_a\bar\Lambda_i\gamma^{ab}\Lambda_j\nonumber\\[1mm]
&\;\;-\tfrac{1}{24}P^a D_a T_{bc}{}^{ij}\bar\Lambda_i\gamma^{bc}\Lambda_j+\tfrac13 P_a T_{bc}{}^{ij}\bar\Lambda_i\gamma^{ab}D^c\Lambda_j+\tfrac{5}{12} P^a T_{bc}{}^{ij}\bar\Lambda_i\gamma^{bc}D_a\Lambda_j\nonumber\\[1mm]
&\;\;-\tfrac{5}{9}T_{ac}{}^{ij} \bar P^a\bar P_b\bar\Lambda^k\gamma^{bc}\Lambda^l\varepsilon_{ijkl}+\tfrac{7}{12}T_{ab ij}\bar P^b P_c \bar\Lambda_k\gamma^{ac}\Lambda_l\varepsilon^{ijkl}-\tfrac13 E_{ij}E^{ik}\bar\Lambda_k\gamma_a\Lambda^j\bar P^a\nonumber\\[1mm]
&\;\;+\tfrac{1}{48}E_{ij} E^{ij}\bar\Lambda_k\gamma^a\Lambda^k\bar P_a -{\tfrac{2}{3}}T_{ac}{}^{kl}T^{ab}{}^{ij}\bar{\Lambda}_{k}\gamma_{b}D^{c}\Lambda^{m}\varepsilon_{ijlm}+\tfrac{1}{8}T^{kl}\cdot T^{ij}\bar{\Lambda}_{m}\slashed{D}\Lambda^{m}\varepsilon_{ijkl}\nonumber \\[1mm]
&\;\; +\tfrac{1}{3}T_{ac}{}^{kl}T^{ab}{}^{ij}D^{c}\bar{\Lambda}_{k}\gamma_{b}\Lambda^{m}\varepsilon_{ijlm}-\tfrac{1}{24}T^{kl}\cdot T^{ij}\bar{\Lambda}^{m}\slashed{D}\Lambda_{m}\varepsilon_{ijkl}-\tfrac{1}{3}T_{ac}{}^{ij}D_{b}T^{ab}{}^{kl}\bar{\Lambda}_{k}\gamma^{c}\Lambda^{m}\varepsilon_{ijlm}\nonumber \\[1mm]
&\;\; +\tfrac{1}{6}T_{ac}{}^{ij}D_{b}T^{ab}{}^{kl}\bar{\Lambda}_{m}\gamma^{c}\Lambda^{m}\varepsilon_{ijkl}+\tfrac{1}{3}T_{ab}{}^{ij}D_{c}T^{ab}{}^{kl}\bar{\Lambda}_{k}\gamma^{c}\Lambda^{m}\varepsilon_{ijlm}+{\tfrac{1}{3}}T_{ab}{}^{ij}D_{c}T^{ab}{}^{kl}\bar{\Lambda}_{m}\gamma^{c}\Lambda^{m}\varepsilon_{ijkl}\nonumber \\[1mm]
&\;\;+\tfrac{1}{8}D^2 E^{ij}\bar{\Lambda}_{i}\Lambda_{j}+\tfrac{1}{2}\bar{P}^{a}\bar{P}_{a} E_{ij}\bar{\Lambda}^{i}\Lambda^{j}-\tfrac{1}{6}\bar{P}^{a}{P}_{a} E^{ij}\bar{\Lambda}_{i}\Lambda_{j}-\tfrac{1}{24}\bar{\Lambda}_{i}\gamma\cdot R(V)^{k}{}_{j}\Lambda_{k}E^{ij}\nonumber\\[1mm]
&\;\; +\tfrac{1}{24} E_{ij}E^{ij}E^{kl}\bar\Lambda_k\Lambda_l -\tfrac{1}{16}E_{ij} E^{ik}E^{jl}\bar\Lambda_k\Lambda_l-\tfrac{7}{288}E^{ij} E^{kl}T_{ab}{}^{mn}\bar\Lambda_i\gamma^{ab}\Lambda_k\varepsilon_{jlmn}\nonumber\\[1mm]
&\;\;+\tfrac{1}{16}T_{ab\, ij} T^{ab}{}_{kl}T_{cd}{}^{ij}\bar\Lambda_m\gamma^{cd}\Lambda_n\varepsilon^{klmn}+\tfrac{1}{16}T_{ab\, ij}T^{ab}{}_{kl} T_{cd}{}^{kl}\bar\Lambda_m\gamma^{cd}\Lambda_n\varepsilon^{ijmn}\nonumber\\[1mm]
&\;\;-\tfrac18E_{ij}T^{ab}{}_{kl}T_{ab\,mn}\bar\Lambda_p\Lambda_q\varepsilon^{iklp}\varepsilon^{jmnq}\Big]+\text{h.c.}\,,\label{eq:res2f2}
\end{align}
while those with two and three derivatives of the function read
\begin{align}
\mathcal{D}^2 \cH&\Big[\tfrac14 E_{ij} T_{ab}{}^{ik}\bar\Lambda_l\gamma^{ab}\chi^{jl}{}_k-\tfrac58 E_{ij}T_{ab}{}^{ik}\bar\Lambda_k R(Q)^{ab\,j}-\tfrac18 E_{ij}E_{kl}\bar\Lambda_m\chi^{ik}{}_n\varepsilon^{jlmn}\nonumber\\[1mm]
&\;\;+\tfrac{1}{4}\bar{\Lambda}_{k}\chi^{mn}{}_{l}T^{ij}\cdot T^{kl}\varepsilon_{ijmn}+\tfrac{1}{4}\bar{\Lambda}_{m}\gamma^{b}{}_{c}\chi^{mn}{}_{i}T_{ab}{}^{ij}T^{ac\,kl}\varepsilon_{jkln}+\tfrac{5}{64}\bar{\Lambda}_{m}\gamma^{cd}R(Q)_{ab}^{m}\,T^{ab\,ij}T_{cd}{}^{kl}\varepsilon_{ijkl}\nonumber\\[1mm]
&\;\; +\tfrac{5}{8}\bar{\Lambda}_{i}R(Q)_{bc}^{m}\,T^{ab\,ij}T_{a}{}^{c\,kl}\varepsilon_{jklm}-\tfrac{3}{4}\bar{\Lambda}^{i}\gamma^a D_b \Lambda_{i}\bar{P}_a \bar{P}^{b}-\tfrac{5}{24}\bar{\Lambda}^{i}\slashed{D}\Lambda_{i}\bar{P}^a \bar{P}_{a}-\tfrac{5}{12}\bar{\Lambda}^{i}\gamma_a \Lambda_{i}\bar{P}^a D_b \bar{P}^{b}\nonumber\\[1mm]
& \;\;-\tfrac{1}{2}\bar{\Lambda}^{i}\gamma^b \Lambda_{i}\bar{P}^a D_a \bar{P}_{b}+\tfrac{1}{12}\bar{\Lambda}_{i}\gamma^a D_b \Lambda^{i}\bar{P}_a \bar{P}^{b}-\tfrac{1}{8}\bar{\Lambda}_{i}\slashed{D} \Lambda^{i}\bar{P}^a \bar{P}_{a}-\tfrac{1}{6} E_{ij}\bar P^b\bar\Lambda_k\gamma^a\Lambda^iT_{ab}{}^{jk}\nonumber\\[1mm]
&\;\;-\tfrac16 T_{ab\,ij}\bar P^a\bar\Lambda_kD^b\Lambda_l\varepsilon^{ijkl}+\tfrac{1}{24}\bar{P}_{a}D^c T_{bc ij}\bar\Lambda_k\gamma^{ab}\Lambda_l\varepsilon^{ijkl}-\tfrac{7}{12} T_{ab \,ij}\bar P_c\bar\Lambda_k\gamma^{ac}D^b\Lambda_l\varepsilon^{ijkl}\nonumber\\[1mm]
&\;\;-\tfrac18 T_{ab\, ij}D_c\bar P^b\bar\Lambda_k\gamma^{ac}\Lambda_l\varepsilon^{ijkl}+\tfrac16 T_{bc}{}^{ij}P^a\bar P_a\bar\Lambda_i\gamma^{bc}\Lambda_j +\tfrac{1}{48} T_{ab}{}^{ij} T^{ab}{}^{kl}\bar P_c\bar\Lambda_m\gamma^c\Lambda^m\varepsilon_{ijkl}\nonumber\\[1mm]
&\;\;-\tfrac{1}{2}\bar{\Lambda}_{i}\gamma_{b}{}^{c}\Lambda_{k}R(V)_{ac}{}^{k}{}_{j}T^{ab\,ij}+\tfrac{1}{2}\bar{\Lambda}_{i}\Lambda_{k}R(V)_{ab}{}^{k}{}_{j}T^{ab\,ij}+\tfrac{1}{16}\bar{\Lambda}_{k}\gamma^{ab}\Lambda_{l}R(V)_{ab}{}^{i}{}_{m}E_{ij}\varepsilon^{jklm}\nonumber\\[1mm]
&\;\;+\tfrac{1}{8}\bar{\Lambda}_{i}\gamma^{cd}\Lambda_{j}R(M)_{abcd}T^{ab\,ij}-\tfrac{1}{16}\bar{\Lambda}_{i}\gamma^{ab}\Lambda_{j}T_{ab}{}^{kl}D^{ij}{}_{kl}+\tfrac{1}{16}\bar{\Lambda}_{j}\Lambda_{n}E_{im}D^{ij}{}_{kl}\varepsilon^{klmn}\nonumber\\[1mm]
&\;\; -\tfrac{1}{48}E_{ij}E^{ij}T_{ab}{}^{kl}\bar\Lambda_k\gamma^{ab}\Lambda_l+\tfrac{1}{24}E_{ij}E^{ik}T_{ab}{}^{jl}\bar\Lambda_k\gamma^{ab}\Lambda_l-\tfrac{1}{48}E^{ij}T^{ab\,kl}T_{ab}{}^{mn}\bar\Lambda_i\Lambda_j\varepsilon_{klmn}\nonumber\\[1mm]
&\;\;-\tfrac{1}{12}E^{ij}T^{ab\,kl}T_{ac}{}^{mn}\bar\Lambda_i\gamma_{b}{}^{c}\Lambda_k\varepsilon_{jlmn}\Big]\nonumber\\[1mm]
+\mathcal{D}^3\cH&\Big[\tfrac18 T_{ab\, ij}\bar P^a \bar P_c \bar\Lambda_k\gamma^{bc}\Lambda_l\varepsilon^{ijkl}-\tfrac{1}{64} E_{ij} E_{kl} T_{ab}{}^{ik}\bar\Lambda_m\gamma^{ab}\Lambda_n\varepsilon^{jlmn}-\tfrac{1}{12}T^{ab\,ij}T_{ac}{}^{kl}T_{b}{}^{c\,mn}\bar\Lambda_i\Lambda_m\varepsilon_{jkln}\nonumber\\[1mm]
&\;\;+\tfrac{1}{64}T^{ab\,ij} T_{ab}{}^{kl}T_{cd}{}^{mn}\bar\Lambda_i\gamma^{cd}\Lambda_j\varepsilon_{klmn}-\tfrac{1}{32}T_{ab}{}^{ij}T^{ac\,kl}T_{cd}{}^{mn}\bar\Lambda_i\gamma^{bd}\Lambda_j\varepsilon_{klmn}\nonumber\\[1mm]
&\;\;+\tfrac{1}{32}T_{ab}{}^{ij}T^{ac\,kl}T_{cd}{}^{mn}\bar\Lambda_k\gamma^{bd}\Lambda_l\varepsilon_{ijmn}+\tfrac{1}{192}E_{ij}E_{kl}E_{mn}\bar\Lambda_p\Lambda_q\varepsilon^{ikmp}\varepsilon^{jlnq}-\tfrac{3}{8}\bar{\Lambda}^{i}\gamma_b \Lambda_{i}\bar{P}_a \bar{P}^{a}\bar{P}^{b}\Big]\nonumber\\[1mm]
&\;\;+\text{h.c.}\label{eq:res2f3}
\end{align}
We emphasize that \eqref{eq:res2f1},\eqref{eq:res2f2} and \eqref{eq:res2f3} correspond to all the terms quadratic in the supercovariant fermion fields. The remaining fermionic terms at this order then necessarily involve bare gravitini and/or S-supersymmetry gauge fields. From the perspective of the density formula \eqref{eq:densityf}, these bare gravitini terms are associated with the composite fields $\Omega^i$, $\Omega_a{}^i$, $\cE^{ij}$, $\cE^{ij}{}_{ab}$ and $\cE_a{}^i{}_j$. However, there will also be contributions coming from the various total derivatives we have extracted in order to write the result as above.

\allowdisplaybreaks[0]

\section{A possible origin of the action principle and on-shell $\cN=4$ SYM}\label{S:Origins}
A curious feature of the action principle we have uncovered is the requirement that
it should be built from three separate superfields $C^{ij}{}_{kl}$, $\bar C^{ij}{}_{kl}$, and $A^{ij}{}_{kl}$,
constrained according to \eqref{eq:Bigconstraint}. In this section, we first show that
they can be derived from a single superfield, provided we are willing
to introduce explicit dependence on the coset scalars. Then we demonstrate that the
on-shell action for $\cN=4$ super Yang-Mills can be recovered using this approach.

\subsection{A more fundamental action principle}
Suppose $\Phi^{i j}{}_{kl}$ is a pseudoreal superfield in the $\rep{20'}$ for which
\begin{align}\label{E:dPhiConstraint}
0 = [\bar\nabla^{\dalpha p} \Phi^{ij}{}_{kl}]_{\brep{60}}
	= [\nabla_{\alpha p} \Phi^{ij}{}_{kl}]_{\rep{60}}~.
\end{align}
Let us take
\begin{align}\label{E:CfromPhi}
C^{ij}{}_{kl} = c^{(-2)} \Phi^{ij}{}_{kl}~, \qquad
A^{ij}{}_{kl} = c^{(0)} \Phi^{ij}{}_{kl}~, \qquad
\bar C^{ij}{}_{kl} = c^{(+2)} \Phi^{ij}{}_{kl}
\end{align}
where the functions $c^{(\pm 2)}$ and $c^{(0)}$ are built from the $\SU(1,1) / \gU(1)$
coset fields $\phi_\balpha$ alone. We have labelled them with their $\gU(1)$ charge.
In order for the constraints \eqref{eq:Bigconstraint} to follow from those on $\Phi^{ij}{}_{kl}$,
the coefficient functions must be chosen to obey
\begin{alignat}{3}
\cD^{--} c^{(-2)} &= 0~, &\qquad
\cD^{--} c^{(0)} &= -c^{(-2)}~, &\qquad
\cD^{--} c^{(+2)} &= -2 \,c^{(0)}~, \eol
\cD^{++} c^{(-2)} &= -2 \,c^{(0)}~, &\qquad
\cD^{++} c^{(0)} &= -c^{(+2)}~, &\qquad
\cD^{++} c^{(+2)} &= 0~.
\end{alignat}
It is not hard to show that the above conditions are uniquely solved by
\begin{align}
c^{(+2)} = c_{\balpha \bbeta} \,\phi^\balpha \phi^\bbeta~, \qquad
c^{(-2)} = c^{\balpha \bbeta} \,\phi_\balpha \phi_\bbeta~, \qquad
c^{(0)}  = -c_{\balpha}{}^{\bbeta} \,\phi^\balpha \phi_\bbeta~,
\end{align}
where $c^{\balpha \bbeta}$ is symmetric, and where we use NW-SE conventions for lowering indices, i.e.
$c_{\balpha \bbeta} \equiv \veps_{\balpha \bgamma} \veps_{\bbeta \bdelta} c^{\bgamma \bdelta}$
and $c_\balpha{}^\bbeta \equiv c^{\bgamma \bbeta} \, \veps_{\bgamma \balpha}$.
In order for $\Phi^{ij}{}_{kl}$ to be pseudo-real, $c^{\balpha\bbeta}$ must be pseudoreal, i.e.
\begin{align}
\eta_{\balpha \bgamma} \eta_{\bbeta \bdelta} (c^{\bgamma \bdelta})^* 
	= c_{\balpha \bbeta} = \veps_{\balpha \bgamma} \veps_{\bbeta \bdelta} c^{\bgamma \bdelta}~.
\end{align}
This implies that $(c^{11})^* = c^{22}$ and $(c^{12})^* = c^{12}$, so that
the $\SU(1,1)$ invariant
\begin{align}
c^{\balpha \bbeta} c_{\balpha \bbeta} = 2 \,c^{11} c^{22} - 2 \,c^{12} c^{12}
\end{align}
can be any real number.
Naturally, the triplet $c^{\balpha\bbeta}$ may be identified
with a vector of $\SO(2,1) \cong \SU(1,1)$, and this vector may be spacelike,
null, or timelike.
This means that any $\Phi^{ij}{}_{kl}$ subject to the condition \eqref{E:dPhiConstraint} along with a
choice of triplet $c^{\balpha\bbeta}$ defines an invariant action.

A natural question to ask is whether the converse relation holds.
Can one build a field $\tilde\Phi^{ij}{}_{kl}$ out of $C^{ij}{}_{kl}$ and $A^{ij}{}_{kl}$, and is it
related to the field $\Phi^{ij}{}_{kl}$ given above? The answer is yes, and it is given by
\begin{align}
\tilde\Phi^{ij}{}_{kl} = a^{(+2)} C^{ij}{}_{kl} - 2 a^{(0)} A^{ij}{}_{kl} + a^{(-2)} \bar C^{ij}{}_{kl}~,
\end{align}
where $a^{(\pm2)}$, $a^{(0)}$ are built out of a pseudoreal $a^{\balpha\bbeta}$ in the
same manner as for the $c$'s. In fact, if $C^{ij}{}_{kl}$ and 
$A^{ij}{}_{kl}$ are given by \eqref{E:CfromPhi}, it is easy to show that
\begin{align}
\tilde \Phi^{ij}{}_{kl} = (a^{(+2)} c^{(-2)} - 2 a^{(0)} c^{(0)} + a^{(-2)} c^{(+2)}) \Phi^{ij}{}_{kl}
	= a^{\balpha \bbeta} c_{\balpha \bbeta} \Phi^{ij}{}_{kl}~,
\end{align}
and so $\tilde \Phi = \Phi$ up to an overall constant. Given any
$c^{\balpha\bbeta}$, one can always find an $a^{\balpha \bbeta}$
(not necessarily unique) so that this overall constant is unity.

We want to apply this observation to two cases. One case is the vector multiplet
action of de Roo \cite{deRoo:1984zyh}, which will be discussed in the next subsection. The other case is the Weyl multiplet action
constructed in the previous sections. Here, the only choice for $\Phi^{ij}{}_{kl}$
is to build it out of the fields $X_n{}^{ij}{}_{kl}$ defined in \eqref{eq:X}. Its Weyl weight and supersymmetry
properties dictate that it be structurally identical to the supercurrent
\eqref{E:Jcurrent} for some holomorphic function $\cJ(S)$. We also must make a
choice for $c^{\balpha\bbeta}$. Once this choice is made, it is straightforward
to use \eqref{E:CfromPhi} to show that
\begin{align}\label{E:HfromJ}
-\frac{i}{2} \cH &= -c^{(-2)} \cD^{++} \cJ~, \eol
\cB &= c^{(0)} \cJ - \frac{1}{2} c^{(-2)} \cD^{++} \cJ + \HC~, \eol
-\frac{i}{2} \cI^{+4}
    &= c^{(+2)} \cD^{++} \cJ - \frac{3}{2} c^{(0)} (\cD^{++})^2 \cJ + \frac{1}{4} c^{(-2)} (\cD^{++})^3 \cJ~.
\end{align}
These in turn obey the required identities \eqref{E:DDB}, \eqref{E:DDGamma} and \eqref{E:HfromGamma}.
Inverting the relationship for $a^{\balpha \bbeta} c_{\balpha \bbeta} = 1$, one can write
\begin{align}
\cJ + \bar \cJ
    &= 
    - \frac{i}{4} a^{(-2)} \cD^{++} \cH
    - \frac{i}{2} a^{(0)} \cH
    - \frac{i}{2} a^{(-2)} \cD^{--} \cI^{+4}  
    \eol & \quad
    - \frac12 (a^{(+2)} \cD^{--} + 2 a^{(0)} + a^{(-2)} \cD^{++}) \cB+\HC
\end{align}
While this approach can be considered more fundamental since it relies on a single superfield,
in the case of conformal supergravity it does not facilitate the construction of the action
because it introduces a spurious dependence on $c^{\balpha \bbeta}$ and obscures the fact 
that the action depends only on $\cH$.

\subsection{Deriving the on-shell SYM action}

Let us now address how a suitable $\Phi^{ij}{}_{kl}$ may be constructed for the on-shell
super-Yang-Mills action.
We first briefly review the structure of on-shell super-Yang-Mills in $\cN=4$ superspace.
This is a relatively straightforward extension of the flat superspace result \cite{Sohnius:1978wk}
and will lead to the component result of de Roo \cite{deRoo:1984zyh}.
Let $\cF$ be a covariantly closed field strength associated with a non-abelian superspace
one-form. We impose the constraints (as in flat superspace)
\begin{align}
\cF_{\alpha i \, \beta j} = 4\, \eps_{\a\b} \bar W_{i j}~, \qquad
\cF^{\dalpha i \, \dbeta j} = 4\, \eps^{\ad\bd} W^{i j}~,\label{eq:cSYM}
\end{align}
where $(W^{ij})^* = \bar W_{ij}$. Requiring that the lowest dimension Bianchi identity
is satisfied leads to the conditions 
\begin{align}\label{eq:SYM.Constrant}
[\nabla_{\alpha k} \bar W_{ij}]_{\rep{20}} = 0~, \qquad
[\bar \nabla^{\dalpha k} \bar W_{ij}]_{\brep{20}} = 
	[\tfrac{1}{2} \Lambda^{\dalpha k} \veps_{ijpq} W^{pq}]_{\brep{20}}~.
\end{align}
In flat superspace, one would take $W^{ij} = -\frac{1}{2} \veps^{ijkl} \bar W_{kl}$ and identify
its lowest component with the six scalars $\phi^{ij}$ of the vector multiplet.
In curved superspace that is not possible because the two superfields carry differing $\gU(1)$ charges,
and the constraints \eqref{eq:SYM.Constrant} are inconsistent with such an assignment.
Instead, one takes\footnote{We follow the same
pseudoreality convention for $\phi^{ij}$ as de Roo \cite{deRoo:1984zyh}, but we denote
his $\Phi$ and $\Phi^*$ as $\Phi^+$ and $\Phi^-$ respectively.}
\begin{align}\label{eq:WfromPhi}
\bar W_{i j} = \Phi^+\, \phi_{ij}~, \qquad
W^{ij} = \Phi^-\, \phi^{ij}~, \qquad
\phi^{i j} = (\phi_{ij})^* =  -\frac{1}{2} \veps^{ijkl} \phi_{kl}
\end{align}
where $\Phi^+ = (\Phi^-)^*$ is a function of the coset scalars with $\gU(1)$ charge $+1$.
The superfield $\phi^{ij}$ here, which is $\gU(1)$ neutral,
will have its lowest component identified with the scalars
of the vector multiplet, so we use the same name for it.
As a consequence of the constraints \eqref{eq:SYM.Constrant}, one can show that
\begin{align}
\cD^{--} \Phi^+ = \Phi^-~, \qquad \cD^{++} \Phi^- = \Phi^+
\end{align}
provided that $\phi^{ij}$ obeys the same constraint as in flat space
\begin{align}\label{eq:SYM.Constraint2}
[\nabla_{\alpha k} \phi^{ij}]_{\rep{20}} = 0~, \qquad
[\bar \nabla^{\dalpha k} \phi^{ij}]_{\brep{20}} = 0~.
\end{align}
This puts the vector multiplet on-shell.
The conditions on $\Phi^+$ and $\Phi^-$ imply that they are
linear in $\phi^\balpha$ and $\phi_\balpha$, respectively,
\begin{align}
\Phi^+ = \bar d_\balpha \phi^\balpha~, \qquad 
\Phi^- = d^\balpha \phi_\balpha~, 
\end{align}
and that the constants $d^\balpha$ and $\bar d_\balpha$ obey
\begin{align}\label{eq:Real.d}
\bar d_\balpha = d^\bbeta \veps_{\bbeta \balpha} = \eta_{\balpha \bbeta} (d^\bbeta)^*~.
\end{align}
The supersymmetry algebra given in \cite{deRoo:1984zyh} corresponds to the choice
$\bar d_\balpha = (1,1)$, and $\Phi^+ = \phi^1 + \phi^2$. It is easy to show that
the most general choice for $d^\balpha$ is
\begin{align}
\bar d_\balpha = (e^{i \delta}, e^{-i \delta})~, \qquad
d^\balpha = (e^{-i \delta}, - e^{i \delta})
\end{align}
as an overall normalization of $d^\balpha$ can always be absorbed into a rescaling
of $\phi^{ij}$.

Now we may attempt to construct an action involving the above on-shell multiplet.
Because of the constraint \eqref{eq:SYM.Constraint2}, it's easy to check that
the superfield
\begin{align}
\Phi^{ij}{}_{kl} \equiv [\Tr \phi^{ij} \phi_{kl}]_{\rep{20'}} 
    = \Tr \phi^{ij} \phi_{kl} - \text{traces}
\end{align}
obeys the constraint \eqref{E:dPhiConstraint}. One can then go about constructing an
action for it as in the previous subsection, by choosing $C^{ij}{}_{kl}$ and
$A^{ij}{}_{kl}$ to obey \eqref{E:CfromPhi} in terms of some $c^{\balpha \bbeta}$
and then applying the density formula.
However, because the multiplet is on-shell, the action is recovered only
modulo equations of motion. Let us focus only on the bosonic part of the action and for
simplicity take the abelian limit. The Lagrangian recovered in this way is
\begin{align}\label{eq:SYM.onshell}
\cL &=
    c^{(-2)} \Big[
    -\frac{i}{(\Phi^-)^2} (F_{ab}^-)^2
    - 2i \frac{\Phi^+}{(\Phi^-)^2} F_{a b} K^{ab+} 
    + \frac{i}{3} D_a (\phi_{ij} \phi^{ij} P^a)
    + i (K_{ab}^+)^2
    - i (K_{ab}^-)^2 \frac{(\Phi^+)^2}{(\Phi^-)^2}
\Big]
    \eol & \quad
    + c^{(+2)} \Big[
    \frac{i}{(\Phi^+)^2} (F_{ab}^+)^2 
    + 2i \frac{\Phi^-}{(\Phi^+)^2} F_{ab} K^{ab+} 
    - \frac{i}{3} D_a (\phi_{ij} \phi^{ij} \bar P^a)
    - i (K_{ab}^-)^2
    + i (K_{ab}^+)^2 \frac{(\Phi^-)^2}{(\Phi^+)^2}
\Big]
    \eol & \quad
    + c^{(0)} \Big[
    \frac{2i}{\Phi^+} F_{ab} K^{ab+} 
    - \frac{2i}{\Phi^-} F_{ab} K^{ab-} 
    + 2i \frac{\Phi^-}{\Phi^+} (K_{ab}^+)^2 
    - 2i \frac{\Phi^+}{\Phi^-} (K_{ab}^-)^2 
    \Big]
\end{align}
where
\begin{align}
K_{ab}^- = T_{ab}{}^{i j} \phi_{ij}~, \qquad
K_{ab}^+ = T_{ab}{}_{i j} \phi^{ij}~.
\end{align}

A conspicuous feature about this on-shell Lagrangian is that no kinetic terms for $\phi^{ij}$
appear. This is a necessary consequence of working with an on-shell multiplet --
it is blind to such terms. However, we have recovered the $A_\mu$ kinetic terms
correctly; this is because there is no way to exploit the equation of motion
for $A_\mu$ without exposing the naked connection and breaking manifest gauge invariance.

Actually, we can reconstruct all of the $A_\mu$-dependent terms. The full Lagrangian 
should formally possess the structure
\begin{align}
\cL = \frac{1}{2} A \cO_1 A + A \cO_2 \phi + \frac{1}{2} \phi \cO_3 \phi~,
\end{align}
for operators $\cO_i$. Putting $\phi$ on-shell leads to
\begin{align}
\cL = \frac{1}{2} A \cO_1 A + \frac{1}{2} A \cO_2 \phi ~.
\end{align}
Therefore, the terms linear in $A_\mu$ in \eqref{eq:SYM.onshell} should be doubled to recover
the corresponding terms in the original Lagrangian. In principle, we could also reconstruct
$\cO_3$ using the equation of motion that follows from supersymmetry.\footnote{This argument
implies that the $\cO(\phi^2)$ terms in \eqref{eq:SYM.onshell} must vanish.
This is indeed the case (up to a total derivative) provided the additional constraint
\eqref{eq:Cab} holds.}

This observation raises an important point. The conditions
\eqref{eq:WfromPhi} together with \eqref{eq:SYM.Constraint2} imply that the vector
multiplet obeys on-shell constraints following from the closure of the supersymmetry algebra.
These have the structure of equations of motion, but they are not necessarily the field
equations that follow from the Lagrangian \eqref{eq:SYM.onshell}. In particular,
the on-shell constraints depend only on $d^\balpha$ while the field equations depend on $c^{\balpha \bbeta}$,
too.  Focusing on the $A_\mu$ on-shell constraint (and for simplicity dropping
terms involving the vector multiplet scalars), we find
\begin{align}
0 = D^b F_{a b} 
    + 2 F_{a b}^- \bar P^b \frac{\Phi^+}{\Phi^-}
    + 2 F_{a b}^+ P^b \frac{\Phi^-}{\Phi^+}
    + \cO(\phi^{ij})~.
\end{align}
The field equation that derives from \eqref{eq:SYM.onshell} is on the other hand proportional to
\begin{align}
0   &= 
    \Big(
        \frac{1}{2} \frac{c^{(+2)}}{(\Phi^+)^2}
        + \frac{1}{2} \frac{c^{(-2)}}{(\Phi^-)^2}
    \Big) D^b F_{a b} 
    - F_{ab}^- \bar P^b  \cD^{++} \Big(\frac{c^{(-2)}}{(\Phi^-)^2}\Big)
    + F_{ab}^+ P^b  \cD^{--} \Big(\frac{c^{(+2)}}{(\Phi^+)^2}\Big)
    + \cO(\phi^{ij})~.
\end{align}
It is not hard to show that these two equations are equivalent only if
\begin{align}\label{eq:Cab}
c^{\balpha \bbeta} \bar d_\balpha \bar d_\bbeta = 0 \quad \implies \quad
c^{\balpha \bbeta} = d^{(\balpha} b^{\bbeta)}
\end{align}
where $b^{\balpha}$ obeys the same pseudoreality condition \eqref{eq:Real.d} as $d^{\balpha}$.
We can parametrize it as
\begin{align}
\bar b_\balpha = (\lambda \,e^{i \beta}, \lambda \,e^{-i \beta})~, \qquad
b^\balpha = (\lambda \, e^{-i \beta}, - \lambda \,e^{i \beta})~.
\end{align}
The action given by de Roo \cite{deRoo:1984zyh} corresponds to the choice
$b^{\balpha} = -\frac{i}{4} (1,1)$ and $d^{\balpha} = (1,-1)$. This can be
seen both by looking at the leading bosonic terms \eqref{eq:SYM.onshell} or by examining
the four gravitino terms.\footnote{The Lagrangian given in eq. (3.16) of \cite{deRoo:1984zyh}
appears to have a typo. We believe the last four gravitino term should be removed
to match the four gravitino terms we have found. This can also independently be checked by
reconstructing the equation of motion of $\phi_{ij}$.} Note that if $b^\balpha \propto d^\balpha$, then the
Lagrangian becomes a total derivative. We should add that the physical significance
of $b^\balpha$ is that it parametrizes the on-shell dual field strength $G_{ab}$ constructed
by taking $\veps_{abcd} \,\pa \cL / \pa F_{cd}$, just as $d^\balpha$ parametrizes $F_{ab}$.

While this appears to be a generalization of de Roo's action,
this is not actually the case. What is happening is that we have parametrized
the action and the supersymmetry transformations in terms of the
three real parameters $\beta$, $\lambda$ and $\delta$, which are precisely the same degrees of freedom associated
with an $\SL(2,\mathbb R)$ duality transformation. In other words, we have
constructed the duality orbit of de Roo's action.

\section{Conclusions and outlook}

In this paper, we have explicitly constructed an $\cN=4$ density formula using the superform method. Invariance under the local $\cN=4$ superconformal symmetries is ensured provided the lowest Weyl weight fields satisfy the set of constraints \eqref{eq:cons}, and that the remaining fields are defined via the supersymmetry transformation rules \eqref{eq:tr1}. We then showed that, by expressing these fields in terms of those of the $\cN=4$ Weyl multiplet such that the constraints are satisfied, the density formula leads to a class of $\cN=4$ conformal supergravity actions parametrized by a holomorphic function of the coset scalars. Based on the uniqueness of the $\cN=4$ supercurrent, we further argued that this must correspond to the most general class of maximal conformal supergravity actions. We presented its expression up to terms which are quadratic in the covariant fermion fields. A stringent check of this result is that when the function is set to a constant, it recovers \cite{Ciceri:2015qpa}. For ergonomic considerations, the complete action is given explicitly in an addendum file. As a second application of the density formula, we also re-derived an on-shell sector of the action constructed in \cite{deRoo:1984zyh} for a vector multiplet in a background of conformal supergravity.

An intriguing feature of the density formula we employed is that it seems it could be derived from a single superfield $\Phi^{ij}{}_{kl}$ and a constant $c^{\balpha\bbeta}$. 
The properties of this superfield resemble those of a $G$-analytic superfield in $(4,2,2)$ superspace \cite{HartwellHowe:Npq,HoweHartwell:Survey}, but it cannot be a Lagrangian in that superspace because it has the wrong dimension. Perhaps it can be used to build an action principle in $(4,2,2)$ superspace along the lines of \cite{Berkovits:2006ik}.

The construction of the full class of $\cN=4$ conformal supergravity actions opens up various perspectives on the higher-derivative structure of the Poincar\'e theory. As was shown already long ago in \cite{deRoo:1984zyh}, $\cN=4$ Poincar\'e supergravity at the two-derivative level can be described as a system of six vector multiplets coupled to conformal supergravity. The standard Poincar\'e action is recovered after gauge fixing the conformal symmetries and integrating out the various auxiliary fields of the Weyl multiplet. It is now possible to consider the class of actions constructed in this paper as a deformation of the two-derivative conformal setup. In this case the transition to the Poincar\'e theory is non-trivial as the field equations of the auxiliary fields have now become non-linear. This requires to integrate out the fields through an iterative procedure, which will result in an infinite power series of the spin-1 field strengths and their derivatives. We will show in an upcoming paper that this procedure can be carried out consistently and leads to a class of supersymmetric higher-derivative Poincar\'e invariants which depends on the holomorphic function of the coset scalars. The procedure can also be applied to describe Poincar\'e supergravity coupled to vector multiplets.

These higher-derivative Poincar\'e couplings are relevant from several point views. When considered on-shell, they could be directly compared with the results obtained in \cite{Bossard:2013rza}. It would also be interesting to see if they could be embedded in the formalism of \cite{Bossard:2011ij} where higher-derivative corrections are described as deformations of the twisted self-duality constraint relating the spin-1 field strengths to their magnetic duals. Another application concerns the matching of subleading corrections to the microscopic entropy of $\cN=4$ black holes obtained via state counting. From the supergravity point of view, some of these corrections are known to originate from the class of couplings considered in this paper, and could be calculated by considering the induced modifications to the area law as in \cite{LopesCardoso:1998tkj, LopesCardoso:2000qm} or \cite{Sen:2005wa, Sen:2007qy}, or perhaps by using more recent localization techniques along the lines of \cite{Dabholkar:2010uh, Murthy:2015yfa}. These approaches have so far relied on a truncated $\cN=2$ setting and it should be interesting to reconsider these results in a fully $\cN=4$ supersymmetric formalism.

Finally, these invariants might clarify the ultraviolet properties of the Poincar\'e theory. Explicit loop computations have revealed a divergence at four loops \cite{Bern:2013uka} which is believed to be connected to the presence of a potential anomaly in the duality symmetry of the theory. It was however shown recently that there exists a finite counterterm, whose leading term includes the square of the Riemann tensor multiplied by a holomorphic function of the coset scalars, and which cancels the anomalous contribution of the graphs up to two loops \cite{Carrasco:2013ypa, Bern:2017rjw, Bern:2019isl}. The consequences of this counterterm for the finiteness of the Poincar\'e theory at four loops however remain to be explored. While these amplitude computations rely on a description of the counterterm via the double copy construction, its explicit supersymmetric expression should follow from the class of invariants constructed in this paper, provided the correct holomorphic function is chosen.

\acknowledgments

We thank Bernard de Wit for helpful discussions and for comments on the manuscript. We also thank Guillaume Bossard for helpful discussions, and Kasper Peeters for correspondence and
for his work on \texttt{Cadabra}.
The work of DB is supported in part by NSF grants PHY-1521099 and the Mitchell Institute
for Fundamental Physics and Astronomy. The work of BS is supported by SERB grant CRG/2018/002373, Government of India. BS and FC would like to thank the Max Planck Institute for gravitational physics (Albert Einstein Institute) and IISER Trivandrum, respectively, for hospitality during the course of this work.

\appendix

\section{The Weyl multiplet in $\cN=4$ superspace}
\label{app:N=4CSG}

The Weyl multiplet of $\cN=4$ conformal supergravity was introduced in \cite{Bergshoeff:1980is}. The gauge field content involves a vielbein $e_\mu{}^a$, gravitino $\psi_\mu{}^i$, spin connection $\omega_\mu{}^{ab}$, dilatation connection $b_\mu$, $\SU(4) \times \gU(1)$ R-symmetry gauge fields $V_\mu{}^i{}_j$ and $a_\mu$, S-supersymmetry connection $\phi_\mu{}_{i}$, and a special conformal (K) connection $f_\mu{}^a$. Together these fields gauge the superconformal algebra $\mathfrak{su}(2,2|4)$. Constraints are imposed on the various curvatures such that the spin connection, $\gU(1)$ gauge field, S-supersymmetry connection, and special conformal connection are algebraically determined in terms of the other fields. This leaves the vielbein, gravitino, and $\SU(4)$ gauge field as the independent connections; the dilatation connection, while also independent, is pure gauge and can be eliminated by a special conformal transformation (\textit{i.e.} a conformal boost).

Additional supercovariant fields are required to complete the multiplet. The scalar fields $\phi_\balpha$ parametrize the coset $\SU(1,1) / \gU(1)$, obeying $\phi^\balpha = \eta^{\balpha \bbeta} (\phi_{\bbeta})^*$ for $\eta^{\balpha \bbeta} = \textrm{diag}(1,-1)$
and $\phi^{\balpha} \phi_\balpha = 1$ with $\balpha=1,2$. Under supersymmetry, they transform into a chiral fermion $\Lambda_{i}$. This in turn transforms into two bosonic fields: a scalar field $E_{ij}$ and an anti-self-dual field $T_{ab}{}^{ij}$ in the $\brep{10}$ and the $\rep{6}$ of SU(4), respectively. At a higher Weyl weight, one finds a chiral fermion $\chi{}^{ij}{}_k$ in the $\rep{20}$. At the top of the multiplet lies a scalar field $D^{ij}{}_{kl}$ transforming in the $\rep{20'}$. The field content consists of 128+128 degrees of freedom and is summarized in Table \ref{table:WeylMultiplet}, where we also give the Weyl weight $w$ and the $\gU(1)$ chiral weight $c$.  
Note that only the positive chirality fermions are presented in the table; the negative chirality fermions transform in conjugate $\SU(4)$ representations and with opposite chiral $\gU(1)$ weights.
The superconformal transformation rules of the various fields can be found in \cite{Bergshoeff:1980is}.

\begin{table}[t!]
\begin{center}
\begin{tabular}{ c | c l c c c }
\hline 
\hline
 & Field & & $\mathrm{SU}(4)$ & $w$ & $c$ \\
\hline
                & $e_\mu{}^a$ & vierbein & $\rep{1}$ & $-1$ & 0 \\
Gauge fields    & $\psi_{\mu}{}^i$ & gravitino & $\rep{4}$ & $-\frac12$ & $-\frac12$ \\
(Independent)   & $b_{\mu}$ &  dilatation gauge field& $\rep{1}$ & 0 & 0 \\
                & $V_{\mu}{}^i{}_j$ & $\mathrm{SU}(4)$ gauge field & $\rep{15}$ & 0 & 0 \\
\hline
                & $\omega_{\mu}{}^{ab}$  & spin connection & $\rep{1}$ & 0 & 0 \\
Gauge fields    & $a_\mu$ & $\mathrm{U}(1)$ gauge field & $\rep{1}$ & 0 & 0 \\
(Composite)     & $\phi_\mu{}_i$ & S-gauge field & $\brep{4}$ & $\tfrac12$ & $\frac{1}{2}$ \\
                & $f_{\mu}{}^a$ & K-gauge field & $\rep{1}$ & 1 & 0 \\
\hline
                & $\phi_{\balpha}$ & & $\rep{1}$ & 0 & $-1$ \\
                & $\Lambda_{i}$  & & $\brep{4}$ & $\tfrac{1}{2}$ & $-\tfrac32$ \\
Covariant fields  & $E_{ij}$ & & $\brep{10}$ & 1 & $-1$ \\
                & $T_{ab}{}^{ij}$  & & $\rep{6}$ & 1 & $-1$\\
                & $\chi{}^{ij}{}_k$ & & $\rep{20}$ & $\tfrac32$ & $-\tfrac12$ \\
                & $D^{ij}{}_{kl}$  & & $\rep{20'}$ & 2 & $0$ \\
\hline
\hline
\end{tabular}
\end{center}
\caption{Independent and composite fields of the $\cN=4$ Weyl multiplet.}
\label{table:WeylMultiplet}
\end{table}

Below we will discuss a formulation of $\cN=4$ superspace that corresponds precisely
to the superspace version of the $\cN=4$ Weyl multiplet. The result,
which we call $\cN=4$ conformal superspace, is constructed in direct analogy with the $\cN=1,2$ cases \cite{Butter:2009cp, Butter:2011sr}, to which we refer for further details such as the construction of superspace torsion tensors and curvatures.
This superspace can be shown to be equivalent, via a degauging procedure and a
redefinition of the $\gU(1)$ connection, to the $\cN=4$ superspace
introduced by Howe \cite{Howe:1981gz}, up to differences in conventions
and some field redefinitions.
Our conventions are similar to the original reference \cite{Bergshoeff:1980is}. We take $\eta_{a b} = \textrm{diag}(-1,1,1,1)$ and $\veps_{abcd}$ imaginary with $\gamma_{abcd} = \veps_{abcd} \,\gamma_5$. We also employ two-component notation. The dictionary for gamma matrices is
\begin{align}
\gamma_5 =
\begin{pmatrix}
\delta_\alpha{}^\beta & 0 \\
0 & - \delta^\dalpha{}_\dbeta           
\end{pmatrix}~, \qquad
\gamma_a =
\begin{pmatrix}
0 & (\gamma_a)_{\alpha \dbeta} \\
(\gamma_a)^{\dalpha \beta} & 0
\end{pmatrix}
\end{align}
where $\alpha=1,2$ and $\dalpha=1,2$ denote left-handed and right-handed spinor indices. The matrices $(\gamma_a)_{\alpha \dbeta}$ are equivalent to $i (\sigma_a)_{\alpha \dbeta}$ where $\sigma_a$ obey the same relations as in Wess and Bagger \cite{Wess:1992cp}.

A 4D $\cN=4$ superspace is a supermanifold parametrized by local coordinates
$z^M = (x^\mu, \theta^{\mathfrak{m}}, \bar\theta_{\dot{\mathfrak{m}}})$.
Along with superdiffeomorphisms (which include spacetime diffeomorphisms and
supersymmetry at the component level), the superspace admits additional
structure group symmetries -- Lorentz transformations ($M_{ab}$), Weyl dilatations
($\mathbb D$), chiral $\gU(1)$ rotations ($\mathbb A$),
$\SU(4)$ transformations ($I^i{}_j$), special conformal transformations
($K_a$), and S-supersymmetry ($S_\alpha{}^i$ and $\bar
S^\dalpha{}_i$). Connection one-forms are associated with each of these
generators. These include a supervielbein $E_M{}^A$ associated
with covariant superdiffeomorphisms. The full supercovariant derivative in
superspace is given by
\begin{align}
\nabla_A &\equiv
	E_A{}^M \Big(
	\pa_M - \frac{1}{2} \Omega_M{}^{a b} M_{a b}
	- B_M \mathbb D - A_M \mathbb A - \cV_M{}^i{}_j \,I^j{}_i
	\eol & \qquad \qquad
	- \tfrac{1}{2} \Phi_M{}^{\alpha}{}_i \, S_{\alpha}{}^i
	- \tfrac{1}{2} \bar\Phi_M{}_{\dalpha}{}^i\, \bar S^{\dalpha}{}_i
	- F_M{}^a K_a
	\Big)~.
\end{align}
There is no explicit gravitino connection as it has been absorbed into the supervielbein.
To recover the action of $\nabla_a$ on a component field $\phi$, one identifies
\begin{align}
\rd z^M \nabla_M \phi \Big\vert_{\q=0, \,\,\rd \q=0} = 
    \rd x^\mu \, \nabla_\mu \phi 
\end{align}
with
\begin{align}
E^A \nabla_A \phi \Big\vert_{\q=0, \,\,\rd \q=0}
    = \rd x^\mu e_\mu{}^a\, \nabla_a \phi 
    + \frac{1}{2} \rd x^\mu \psi_\mu{}^{\alpha i} \nabla_{\alpha i} \phi
    + \frac{1}{2} \rd x^\mu \psi_\mu{}_{\dalpha i} \nabla^{\dalpha i} \phi~.
\end{align}
Here we have chosen $E_\mu{}^{\alpha i} \vert_{\theta=0} \equiv \tfrac{1}{2} \psi_\mu{}^{\alpha i}$
as the definition of the gravitino. The factor of $\tfrac{1}{2}$ is to match conventions.
From this expression, one finds
\begin{align}
e_\mu{}^a \nabla_a \phi \equiv \nabla_\mu \phi
        - \frac{1}{2} \psi_\mu{}^{\alpha i} \nabla_{\alpha i} \phi
        - \frac{1}{2} \psi_\mu{}_{\dalpha i} \nabla^{\dalpha i} \phi\label{eq:covdd}
\end{align}
where $\nabla_\mu$ carries all the connections other than the gravitino.
It is $\nabla_a$ rather than $\nabla_\mu$ that corresponds to the supercovariant
derivative $D_a$ in the component formalism \cite{Bergshoeff:1980is}.

The gauge transformations of the various connections are defined so that
$\nabla_A$ transforms under Lorentz, dilatation, and 
$\SU(4) \times \gU(1)$ R-symmetry transformations as
\begin{gather} 
[M_{ab}, \nabla_c] = -\eta_{bc} \nabla_a + \eta_{ac} \nabla_b ~,\quad 
[M_{ab}, \nabla_{\gamma i}] =  \tfrac{1}{2} {(\gamma_{ab})_\gamma}^{\beta} \nabla_{\beta i}~, \quad
[M_{ab}, \bar \nabla^{\dgamma i}] = -\tfrac{1}{2} {(\gamma_{ab})^\dgamma}_{\dbeta} \bar \nabla^{\dbeta i}~, \eol{}
[\mathbb D,\nabla_a] = \nabla_a, \quad 
[\mathbb D, \nabla_{\alpha i}] = \tfrac{1}{2} \nabla_{\alpha i}, \quad 
[\mathbb D, \bar \nabla^{\dalpha i}] = \tfrac{1}{2} \bar \nabla^{\dalpha i}~, \eol{}
[\mathbb A, \nabla_{\alpha i}] = \tfrac{1}{2} \ri \,\nabla_{\alpha i},\quad 
[\mathbb A, \bar \nabla^{\dalpha i}] = -\tfrac{1}{2} \ri \,\bar \nabla^{\dalpha i}~, \eol{}
[I^j{}_i, \nabla_{\alpha k}] = - \delta_k^j \nabla_{\alpha i} + \tfrac{1}{4} \delta_i^j \nabla_{\alpha k}, \quad 
[I^j{}_i, \bar \nabla^{\dalpha k}] = +\delta_k^i \bar \nabla^{\dalpha j} - \tfrac{1}{4} \delta^i_j \bar \nabla^{\dalpha k}~.
\end{gather}
The (anti)commutators involving the special conformal and S-supersymmetry generators include
\begin{gather}
\{S_{\alpha}{}^i, \bar S_{\dalpha j} \} = -\delta^i_j\,(\gamma^a)_{\alpha \dalpha} \,K_a~, \qquad
[K_a, \nabla_b] = - \eta_{ab} \mathbb D - M_{ab}~, \eol{} 
\{S_{\alpha}{}^i, \nabla_{\beta j}\} = -\delta^i_j \epsilon_{\alpha \beta} \mathbb D 
	+ 2 \delta^i_j M_{\alpha\beta}
	- 2 \eps_{\alpha \beta} I^i{}_j~, \eol{}
\{\bar S^{\dalpha}{}_i, \bar \nabla^{\dbeta j} \} = -\delta_i^j \epsilon^{\dalpha \dbeta} \mathbb D 
	+ 2 \delta_i^j \bar M^{\dalpha\dbeta}
	+ 2 \eps^{\dalpha \dbeta} I^j{}_i~, \eol{}
[K_a, \nabla_{\alpha i}] = (\gamma_a)_{\alpha \dbeta} \,\bar S^{\dbeta}{}_i, \;\;\; 
[K_a, \bar \nabla^{\dalpha i}] = (\gamma_a)^{\dalpha \beta} \,S_{\beta}{}^i~,
\end{gather}
where $M_{\alpha\beta} = \frac{1}{4} (\gamma^{ab})_{\alpha\beta} M_{ab}$ and 
$\bar M^{\dalpha\dbeta} = \frac{1}{4} (\gamma^{ab})^{\dalpha \dbeta} M_{ab}$ 
are the anti-self-dual and self-dual parts of the Lorentz generator.
These coincide with the commutators of the superconformal algebra $\mathfrak{su}(2,2|4)$.
Field-dependent deviations are found in the S-supersymmetry transformations of the
covariant derivatives, or equivalently, in the transformations of the connections, so that
\begin{gather}
[S_{\alpha}{}^i, \nabla_a] = 
	-\tfrac{1}{2} (\gamma_a)_{\alpha \dbeta} \,\bar \nabla^{\dbeta i}
     - \tfrac{1}{12} T_{b c}{}^{i j} (\gamma^{b c} \gamma_{a})_{\alpha \dbeta} \,S^{\dbeta}{}_{j}
     + \tfrac{1}{4} R(Q)_{a b}{}_{\alpha}{}^{i} \,K^{b} \eol{} %
[\bar S^{\dalpha}{}_i, \nabla_a] = 
	-\tfrac{1}{2} (\gamma_a)^{\dalpha \beta} \,\nabla_{\beta i}
     - \tfrac{1}{12} T_{b c}{}_{i j} (\gamma^{b c} \gamma_{a})^{\dalpha \beta} \,S_{\beta}{}^{j}
     + \tfrac{1}{4} R(Q)_{a b}{}^{\dalpha}{}_{i} \,K^{b}~, \eol{}
\{ S_\beta{}^j, \bar\nabla^{\dalpha i} \}
      = - \frac{1}{2} \veps^{j i k l} \Lambda_{\beta k} \,S^{\dalpha}{}_{l}
     + \frac{1}{6} T_{b c}{}^{j i} (\gamma^{b c} \gamma^a)_{\beta \dbeta} \,\eps^{\dbeta \dalpha} \,K_{a} \eol{}
\{ S^\dbeta{}_j, \nabla_{\alpha i} \}
	= - \frac{1}{2} \veps_{j i k l} \Lambda^{\dbeta k} \,S_{\alpha}{}^{l}
     + \frac{1}{6} T_{b c}{}_{j i} (\gamma^{b c} \gamma^a)^{\dbeta \beta} \,\eps_{\beta \alpha} \,K_{a} ~.
\end{gather}
Here we are using covariant superfields $T_{a b}{}^{ij}$ and $R(Q)_{ab\, \alpha}{}^i$ 
whose $\q=0$ pieces correspond to the identically named component fields.
The deformation of the above (anti)commutators from the flat $\mathfrak{su}{(2,2|4)}$
superalgebra is necessary for $\cN=4$, but not for
$\cN \leq 2$.\footnote{Corresponding deviations are also required for the $\cN=(2,0)$ theory in six
dimensions \cite{Bergshoeff:1999db} but not for $\cN=(1,0)$.} 

The anti-commutators of the spinor covariant derivatives are
\begin{align}
\{ \nabla_{\alpha i}, \nabla_{\beta j} \}
	&= \eps_{\a\b} \,\Big(
	\veps_{i j k l} \bar \L_{\dgamma}^k \bar \nabla^{\dgamma l}
	-2 \,T^{a b}{}_{i j} M_{ab}
	+ \veps_{i j k l} E^{p k} I^l{}_p
	- \tfrac{1}{2} \veps_{i j k l} (\gamma^c)^{\gamma}{}_\dgamma \nabla_c \Lambda^{\dgamma l} S_\gamma{}^k
	\eol & \qquad
	+ 2 \,\bar \chi_{\dbeta}{}_{ij}{}^k \bar S^{\dbeta}{}_k
	- \tfrac{2}{3} (E_{k [i} \bar \Lambda_\dgamma{}^k + 2 (\gamma^a)_\dgamma{}^\gamma P_a \Lambda_{\gamma [i}) \bar S^{\dgamma}{}_{j]}
	+ \tfrac{8}{3} \nabla_b T^{ab}{}_{i j} K_a
	\Big)~,\eol{}
\{ \bar \nabla^{\dalpha i}, \bar\nabla^{\dbeta j} \}
	&=
	\eps^{\dalpha \dbeta} \Big(
	\veps^{i j k l} \Lambda^\alpha{}_{k}  \nabla_{\alpha l}
	- 2 \,T^{a b}{}^{i j} M_{a b}
	- \veps^{i j k l} E_{p k} I^{p}{}_{l}
	- \tfrac{1}{2} \veps^{i j k l} (\gamma^{c})_\dgamma{}^{\gamma} 
	\nabla_{c} \Lambda_{\gamma l} \bar S^{\dgamma}{}_{k}
	\eol & \qquad
	+ 2 \chi^\gamma{}^{i j}{}_{k}  S_{\gamma}{}^{k}
	- \tfrac{2}{3} ( E^{k [i} \Lambda^\gamma_{k}  + 2 (\gamma^{c})^\gamma{}_{\dgamma} \bar P_{c} \Lambda^{\dgamma [i}   )  S_{\gamma}{}^{j]}
	+ \tfrac{8}{3} \nabla_{b} T^{a b}{}^{i j} K_a
	\Big)~, \eol{}
\{\nabla_{\alpha i}, \nabla_\dalpha{}^{j} \} &=
(\gamma^{a})_{\alpha \dalpha} \Big\{
- 2 \,\delta_{i}^{j} \,\nabla_{a}
+ \tfrac{i}{2} (
	\bar \Lambda^{j} \gamma_{a}\Lambda_{i}
     - \delta_{i}^{j} \bar\Lambda^{k} \gamma_{a} \Lambda_{k}
    ) \mathbb A
\eol & \qquad
+ \tfrac{1}{2} ( - \bar\Lambda^{k} \gamma_{a} \Lambda_{i} I^{j}{}_{k}
     - \bar\Lambda^{j} \gamma_{a} \Lambda_{k} I^{k}{}_{i}
     + \bar\Lambda^{k} \gamma_{a} \Lambda_{k} I^{j}{}_{i}
     + \delta_{i}^{j} \bar\Lambda^{k} \gamma_{a} \Lambda_{l} I^{l}{}_{k}
   )
\eol & \qquad
+  \eps_{\ddelta \dgamma} \Big(
    \tfrac{1}{2} (\gamma_{a})^{\dgamma \gamma} \chi_{\gamma}{}^{k j}{}_{i}
    - \tfrac{1}{4} \veps^{k j l p} (\gamma^{b c} \gamma_{a} \Lambda_{l})^{\dgamma} \,T_{b c}{}_{i p}
	\eol & \qquad \qquad
    + \tfrac{1}{6} \delta^{[k}_{i} \delta^{j]}_{l} (
         E^{l p} (\gamma_{a} \Lambda_{p})^{\dgamma}
         + 2 \bar P_{c}\, (\gamma_{a} \gamma^{c} \Lambda^l){}^\dgamma ) 
     \Big) \bar S^{\ddelta}{}_{k}
\eol & \qquad
+  \eps^{\delta \gamma} \Big(
    \tfrac{1}{2} (\gamma_{a})_{\gamma \dgamma} \chi^{\dgamma}{}_{k i}{}^{j}
    - \tfrac{1}{4} \veps_{k i l p}
	(\gamma^{b c} \gamma_{a} \Lambda^l)_{\gamma} \,T_{b c}{}^{j p} 
	\eol & \qquad \qquad
    + \tfrac{1}{6} \delta_{k}^{[j} \delta_{i}^{l]} (
         E_{ l p} (\gamma_{a} \Lambda^{p})_{\gamma}
         + 2 P_{c} \,(\gamma_{a} \gamma^{c} \Lambda_l)_{\gamma} 
       \Big) 
) S_{\delta}{}^{k}
\eol & \qquad
+ \Big(\tfrac{1}{3} \veps_{a b c d} R(V)^{c d}{}^{j}{}_{i}
     - \tfrac{i}{6} \delta_{i}^{j} \veps_{a b c d} F^{c d} 
     - 2 ( T_{a d}{}^{j k} T_{b}{}^{d}{}_{k i} + T_{b}{}^{d}{}^{j k} T_{a d}{}_{k i} )
   \Big) K^{b}
   \Big\} ~.
\end{align}
While the first two anti-commutators mirror the constraint structure of super-Yang-Mills \eqref{eq:cSYM},
as in the $\cN\leq 2$ case, the third anti-commutator does not involve only
$\delta_i{}^j \nabla_a$ on the right-hand side. The additional terms can be
decomposed into a singlet and a traceless operator in the $\rep{15}$.
While the singlet operator could be absorbed into a redefinition of $\nabla_a$, the operator
in the $\rep{15}$ cannot be.\footnote{We leave the singlet operator unabsorbed in order to
maintain contact with the conventional constraints chosen in \cite{Bergshoeff:1980is}.
Note that this is in contrast to the choice made in $\cN=2$ where a corresponding
singlet operator was absorbed \cite{Butter:2011sr}.}

The commutators between spinor and vector derivatives are
\begin{align}
[\nabla_{\beta i}, \nabla_a] &=
- \tfrac{1}{4} T_{b c}{}_{i k} \,(\gamma_{a}\gamma^{b c})_{\beta \dgamma} \,\bar\nabla^{\dgamma k}
- \tfrac{1}{2} (\gamma_{a})_{\beta \dbeta} R(Q)^{b c}{}^{\dbeta}{}_{i} \,M_{bc}
\eol & \quad
+ \tfrac{i}{4} E_{i j} \, (\gamma_{a} \Lambda^j)_\beta\, \bbA
+ \tfrac{i}{2} P_{c} \,(\gamma_{a} \gamma^{c} \Lambda_i)_{\beta} \,\bbA
+ \tfrac{i}{8} \veps_{i j k l} T_{b c}{}^{k l}
	(\gamma^{b c} \gamma_{a} \Lambda^j)_{\beta} \, \bbA
\eol & \quad
- (\gamma_{a})_{\beta \dbeta} \chi^{\dbeta}{}_{i j}{}^{k} I^{j}{}_{k}
    + \tfrac{1}{4} \veps_{k i l p} T_{b c}{}^{l j}
		(\gamma^{b c} \gamma_{a} \Lambda^p)_{\beta} I^{k}{}_{j}
    - \tfrac{1}{6} E_{j k} (\gamma_{a} \Lambda^k)_{\beta} \,I^{j}{}_{i}
    - \tfrac{1}{3} P_{c} \,(\gamma_{a} \gamma^{c} \Lambda_{j})_{\beta} I^{j}{}_{i}
\eol & \quad
+ \tfrac{1}{8} T_{bc}{}_{i j} T_{de}{}^{j k} 
	(\gamma^{de} \gamma_{a} \gamma^{bc} )_{\beta \dbeta}
                 S^{\dbeta}{}_{k}
- \tfrac{1}{12} \Big(R(V)_{b c}{}^{j}{}_{i}  - \tfrac{i}{2} F_{b c} \delta^{j}_{i}\Big)
	(\gamma^{b c} \gamma_{a} - 3 \gamma_{a} \gamma^{b c})_{\beta \dgamma} S^{\dgamma}{}_{j}
\eol & \quad
+ \tfrac{1}{24} \,\nabla_{b}{T_{e f}{}_{i j}} \,
	(\gamma_{c} \gamma^{ef})_{\beta \ddelta}
	(\gamma^{b c} \gamma_{a} - 3 \gamma_{a}\gamma^{b c})^{\ddelta \delta} \,
                 \,S_{\delta}{}^{j}
\eol & \quad
+ \tfrac{1}{8} \veps_{i j k l} \, R(Q)_{b c}{}_{\beta}{}^{k} \,\eps^{\alpha \gamma}
	(\tfrac{1}{3} \gamma_{a} \gamma^{bc} \Lambda^l - \gamma^{b c} \gamma_{a} \Lambda^l)_{\gamma} S_{\alpha}{}^{j}
\eol & \quad
+ \Big( \tfrac{1}{8} \veps_{a b c d} R(S)^{c d}{}_{\beta i}
     - (\gamma_{a})_{\beta \dbeta} \nabla^{c} R(Q)_{c b}{}^{\dbeta}{}_{i}
     - 2 \,T_{a}{}^{c}{}_{i j} R(Q)_{c b}{}_{\beta}{}^{j}
    \Big) K^{b}~,\eol[2ex]
[\bar\nabla^{\dbeta i}, \nabla_a] &=
- \tfrac{1}{4} T_{b c}{}^{i k}  \,(\gamma_{a} \gamma^{b c})^{\dbeta \gamma} \nabla_{\gamma k}
- \tfrac{1}{2} (\gamma_{a})^{\dbeta \beta} \,R(Q)^{b c}{}_{\beta}{}^{i} \,M_{bc}
\eol & \quad
- \tfrac{i}{4} E^{i j} (\gamma_{a} \Lambda_j)^{\dbeta} \,\bbA
- \tfrac{i}{2} \bar{P}_{c} \,(\gamma_{a} \gamma^{c} \Lambda^i)^{\dbeta} \, \bbA
- \tfrac{i}{8} \veps^{i j k l} T_{b c}{}_{k l} 
	(\gamma^{b c} \gamma_{a} \Lambda_j)^{\dbeta}\,\bbA
\eol & \quad
+ (\gamma_{a})^{\dbeta \beta} \,\chi_{\beta}{}^{i j}{}_{k} \,I^{k}{}_{j}
- \tfrac{1}{4} \veps^{k i l p} T_{b c}{}_{l j}  \,(\gamma^{b c} \gamma_{a} \Lambda_p)^{\dbeta} \,I^{j}{}_{k}
+ \tfrac{1}{6} E^{j k} (\gamma_{a} \Lambda_k)^{\dbeta} \,I^{i}{}_{j}
+ \tfrac{1}{3} \bar P_{c} (\gamma_{a} \gamma^{c} \Lambda^j)^{\dbeta} \,I^{i}{}_{j}
\eol & \quad
+ \tfrac{1}{8}
	T_{b c}{}^{i j} T_{d e}{}_{j k}
	(\gamma^{d e} \gamma_{a} \gamma^{b c})^{\dbeta \beta}
               S_{\beta}{}^{k}
+ \tfrac{1}{12} \Big(R(V)_{b c}{}^{i}{}_{j}  - \tfrac{i}{2} F_{b c} \delta^{i}_{j}\Big) 
	(\gamma^{b c} \gamma_{a} - 3 \gamma_{a} \gamma^{b c})^{\dbeta \gamma}
	S_{\gamma}{}^{j}
\eol & \quad
+ \tfrac{1}{24} \nabla_{b} T_{e f}{}^{i j}
	(\gamma_{c} \gamma^{e f})^{\dbeta \delta}
	(\gamma^{b c} \gamma_{a} - 3 \gamma_{a} \gamma^{b c})_{\delta \dgamma} S^{\dgamma}{}_{j}
\eol & \quad
+ \tfrac{1}{8} \veps^{i j k l} \,
	R(Q)_{b c}{}^{\dbeta}{}_{k} \eps_{\dalpha \dgamma}
	(\tfrac{1}{3} \gamma_{a} \gamma^{b c} \Lambda_{l}- \gamma^{b c} \gamma_{a} \Lambda_{l})^{\dgamma}
       S^{\dalpha}{}_{j}
\eol & \quad
+ \Big(- \tfrac{1}{8} \veps_{a b c d} R(S)^{c d}{}^{\dbeta i}
     - (\gamma_{a})^{\dbeta \beta} \nabla^{c} R(Q)_{c b}{}_{\beta}{}^{i}
     - 2 \,T_{a}{}^{c}{}^{i j} R(Q)_{c b}{}^{\dbeta}{}_{j}
    \Big) K^{b}~.
\end{align}
Finally, the purely vector commutator defines the various curvatures
\begin{align}
[\nabla_a , \nabla_b] &=
     - \tfrac{1}{2} R(Q)_{a b}{}^{\alpha i} \nabla_{\alpha i}
     - \tfrac{1}{2} R(Q)_{a b}{}_{\dalpha i} \bar\nabla^{\dalpha i}
     - \tfrac{1}{2} R(M)_{a b}{}^{c d} M_{c d}
     - R(V)_{a b}{}^{i}{}_{j} \,I^{j}{}_{i}
     \eol & \quad
     - F_{a b} \,\bbA
     - \tfrac{1}{2} R(S)_{a b}{}^\alpha{}_{i} S_{\alpha}{}^{i}
     - \tfrac{1}{2} R(S)_{a b}{}_{\dalpha}{}^{i} S^{\dalpha}{}_{i}
     - R(K)_{a b}{}^c K_{c}~.
\end{align}
The torsion tensor $T_{a b}{}^c$ has been constrained to vanish.

The superfield content exactly mirrors the component field content of
\cite{Bergshoeff:1980is} summarized in Table \ref{table:WeylMultiplet}. Their superconformal transformations are
exactly the same in superspace. Below we give these transformations in
our conventions.

Recall the lowest dimension covariant superfields of the $\cN=4$ Weyl multiplet
are a doublet of superfields $\phi_\balpha$ describing an element of $\SU(1,1)$.
$\phi_\balpha$ carries $\gU(1)$ charge $-1$. It is natural to introduce
three vielbeins for the group manifold
\begin{align}\label{eq:PDef1}
P \equiv \veps_{\balpha \bbeta} \phi^{\balpha} \rd \phi^{\bbeta}~, \qquad
\bar P \equiv -\veps^{\balpha \bbeta} \phi_{\balpha} \rd \phi_{\bbeta}~, \qquad
\cA \equiv i\, \phi^\balpha \rd \phi_\balpha
	= - i\, \rd \phi^\balpha \phi_\balpha~,
\end{align}
so that the exterior derivative on $\SU(1,1)$ becomes 
$\rd = i \cA \cD^0 + P \cD^{--} + \bar P \cD^{++}$.
The one-forms obey
\begin{align}
\rd P = 2 i P \wedge \cA~, \qquad \rd \bar P = -2i \bar P \wedge \cA~.
\end{align}
On the coset space $\SU(1,1)/\gU(1)$, $P$ and $\bar P$ are the vielbeins and $\cA$ is the $\gU(1)$ connection. The pullback of these one-forms to superspace gives superconnections
obeying identical equations. For $P$ and $\bar P$, the expressions are unchanged if
one replaces the exterior differential $\rd$ with the covariant one $\nabla$ so that
the expansion in tangent space reads
\begin{align}\label{eq:PDef2}
P_M \equiv E_M{}^A P_A = E_M{}^A \veps_{\balpha \bbeta} \phi^{\balpha} \nabla_A \phi^{\bbeta}~, \qquad
\bar P_M \equiv E_M{}^A \bar P_A = -E_M{}^A \veps^{\balpha \bbeta} \phi_{\balpha} \nabla_A \phi_{\bbeta}~.
\end{align}
However, in order to match the conventions for the $\gU(1)$ connection $A$ used in
\cite{Bergshoeff:1980is}, it is necessary to shift $\cA$ by a fermion bilinear,
\begin{align}
\cA_M = A_M - \frac{i}{4} E_M{}^a \bar\Lambda^i \gamma_a \Lambda_i~.
\end{align}
From a superspace perspective this complicates the basic equations above
but simplifies the gravitino torsion tensor. Note that the standard $\cN=4$
superspace reference \cite{Howe:1981gz} does not make this redefinition, along
with other differences.

The superspace one-form $P$ in \eqref{eq:PDef2} is constrained so that
\begin{align}
\bar P^{\dalpha i} = 0~, \qquad \bar P_{\alpha i} \equiv -\Lambda_{\alpha i}~, \eol
P_{\alpha i} = 0~, \qquad P^{\dalpha i} \equiv -\Lambda^{\dalpha i}~.
\end{align}
By choosing the spinor component of
the $\gU(1)$ connection appropriately, one can ensure that $\phi_\balpha$ is chiral, with
\begin{align}
\nabla^{\dalpha i} \phi_{\balpha} = 0~, \qquad
\nabla_{\alpha i} \phi_{\balpha} = - \Lambda_{\alpha i}\, \veps_{\balpha \bbeta} \phi^{\bbeta}~.
\end{align}
The remaining vector component is defined (as in the component formalism) to be
\begin{align}\label{eq:def.Pa}
P_a \equiv \veps_{\balpha \bbeta}\, \phi^\balpha \nabla_a \phi^\bbeta~, \qquad
\bar P_a \equiv -\veps^{\balpha \bbeta}\, \phi_\balpha \nabla_a \phi_\bbeta~.
\end{align}

The spinor superfield $\Lambda_{\alpha i}$ is $S$-invariant. It transforms
under supersymmetry as
\begin{align}
\nabla_{\alpha i} \Lambda_{\beta j} &=
	- \eps_{\alpha  \beta }  E_{i j} 
	- \tfrac{1}{2} \veps_{i j k l} T_{b c}{}^{k l} (\gamma^{b c})_{\alpha \beta}~, \eol
\nabla_{\beta j} \Lambda^{\dalpha i} &=
	2 \,\delta^{i}_{j}\, (\gamma^{a})_\beta{}^\dalpha \, P_{a} ~.
\end{align}
$E_{i j}$ is symmetric in its SU(4) indices while $T_{a b}{}^{ij}$ is antisymmetric
in its SU(4) indices and anti-self-dual in its Lorentz indices. 
These and their complex conjugates transform as
\begin{align}
\nabla_{\alpha k} E_{i j} &= -2 \,\chi_{\alpha}{}^{rs}{}_{(i} \,\veps_{j) k rs}~, \eol
\nabla_{\alpha k} E^{i j} &=
	2 \,\delta_{k}^{(i} (\gamma^{c})_{\alpha \dbeta} \nabla_{c}{\Lambda^{\dbeta j)}}
	- \Lambda_\dgamma{}^{i} \Lambda^{\dgamma j} \,\Lambda_{\alpha k}
	+ 2 \,\Lambda_\dgamma{}^{l} \Lambda^{\gamma (i} \Lambda_{\dalpha l} \delta_{k}^{j)}~, \eol
\nabla_{\alpha k} T_{a b}{}^{i j} &=	
	2 \delta^{[i}_{k} R(Q)_{a b}{}_{\alpha}{}^{j]}
	+ (\gamma_{a b})_{\alpha}{}^{\beta} \Big(
		\tfrac{1}{2} \chi_{\beta}{}^{i j}{}_{k}
		- \tfrac{1}{6} E^{l [i} \delta^{j]}_{k} \Lambda_{\beta l}
		+ \tfrac{1}{3} \delta^{i}_{k} (\gamma^{c})_{\beta \dbeta} \bar P_{c} \Lambda^{\dbeta j}
	\Big)~, \eol
\nabla_{\alpha k} T_{a b}{}_{i j} &=
	\tfrac{1}{4} \veps_{k i j l} (\gamma^{c} \gamma_{a b})_{\alpha \dbeta} \nabla_{c}{\Lambda^{\dbeta l}}~.
\end{align}
Complex conjugation gives the other transformations. $\chi_\alpha{}^{ij}{}_k$ is in the $\rep{20}$
of SU(4). $R(Q)_{a b}{}_\alpha{}^j$ is anti-self-dual and gamma-traceless, meaning it can be written
\begin{align}
R(Q)_{a b}{}_\alpha{}^j = \frac{1}{2} (\gamma_{ab})^{\beta \gamma} R(Q)_{\gamma \beta \alpha}{}^j
\end{align}
in terms of a totally symmetric $R(Q)_{\gamma \beta \alpha}{}^j$. Their supersymmetry transformations are
given by
\begin{align}
\nabla_{\beta l} \chi_\alpha{}^{ij}{}_{k}
	&=
	(\gamma^{a b})_{\alpha \beta} R(V)_{a b}{}^{[i}{}_{k} \delta_{l}^{j]}
	+ \eps_{\alpha \beta} D^{i j}{}_{k l}
	+ \tfrac{1}{6} \veps_{k rst} (\gamma^{a b})_{\alpha \beta}\, E^{r [i} 
	      ( T_{a b}{}^{j] t} \delta_{l}^{s} + \delta_{l}^{j]} T_{a b}{}^{st} )
	\eol & \quad
	+ \tfrac{1}{2} \eps_{\alpha \beta} \,\delta_{l}^{[j} E^{i] m} E_{k m} 
	+ \tfrac{1}{4} \eps_{\alpha \beta} \,\delta_{l}^{[i} (
	     2 \,\Lambda_\dgamma{}^{j]} (\gamma_{c})^{\dgamma \gamma} \nabla_{c}{\Lambda_{\gamma k}}
	     + \Lambda^\gamma{}_{k} (\gamma_{c})_{\gamma \dgamma} \nabla_{c}{\Lambda^{\dgamma j]}}
	    )
	\eol & \quad
	+ \tfrac{1}{4} (\gamma^{a b})_{\alpha \beta} \,\delta_{l}^{[i} (
		2 \,\Lambda_\dgamma{}^{j]} (\gamma_{a})^{\dgamma \gamma} \nabla_{b}{\Lambda_{\gamma k}}
		- \Lambda^\gamma{}_{k} (\gamma_{a})_{\gamma \dgamma} \nabla_{b}{\Lambda^{\dgamma j]}}
		)
	\eol & \quad
	+ \tfrac{1}{2} \eps_{\alpha \beta} \,\delta_{l}^{[i}
		\,\Lambda_\dgamma{}^{j]} \Lambda^{\dgamma m}
		\,\Lambda^\gamma{}_{k} \Lambda_{\gamma m}
	- \text{traces}~, \eol[2ex]
\nabla_{\beta l} \chi^{\dalpha}{}_{ij}{}^k &=
	- \tfrac{1}{2} \delta_{l}^{k} (\gamma^{a b} \gamma^c)^{\dalpha \gamma} \eps_{\gamma \beta} \nabla_{c}{T_{a b}{}_{i j}}
	+ \tfrac{1}{2} \veps_{i j m l} (\gamma^{c})_{\beta}{}^{\dalpha} \nabla_{c}{E^{k m}}
	\eol & \quad
	- \tfrac{1}{2} \veps_{i j m l} (\gamma^{c} \gamma^{ab})^{\dalpha \gamma} \eps_{\gamma \beta} \,P_c\,T_{a b}{}^{k m}
	- \tfrac{1}{4} (\gamma^{a})_{\beta}{}^{\dalpha}
	     ( 2 \veps_{i j m l} \chi^{\gamma}{}^{m k}{}_{n} 
	     - \veps_{i j m n} \chi^{\gamma}{}^{m k}{}_{l})
	     (\gamma_{a})_{\gamma \dgamma} \Lambda^{\dgamma n}
	\eol & \quad
	+ \tfrac{5}{12} \delta^k_p \veps_{i j l m} \Lambda^{\dalpha m} 
	     ( E^{p n} \Lambda_{\beta n} + 2 (\gamma^{c})_{\beta \dgamma} \bar P_{c} \Lambda^{\dgamma p} )
	\eol & \quad
	- \tfrac{1}{12} \delta^{k}_{l} \veps_{i j p m} \Lambda^{\dalpha m} 
	     ( E^{p n} \Lambda_{\beta n} + 2 (\gamma^{c})_{\beta \dgamma} \bar P_{c} \Lambda^{\dgamma p} )
	\eol & \quad
	- \tfrac{1}{4} (\gamma^{a b} \gamma^c)^{\dalpha \alpha} \eps_{\alpha \beta}
	T_{a b}{}_{i j} 
     ( \delta^{k}_{l}  \Lambda^\gamma{}_{m} (\gamma_{c})_{\gamma \dgamma} \Lambda^{\dgamma m}
      - \delta^{m}_{l}  \Lambda^\dgamma{}_{m} (\gamma_{c})_{\gamma \dgamma} \Lambda^{\dgamma k})
	\eol & \quad
	- \tfrac{1}{4} (\gamma^{a b} \gamma^c)^{\dalpha \alpha} \eps_{\alpha \beta}
	T_{a b}{}_{m i} 
		( \delta^{k}_{l}  \Lambda^\dgamma{}_{j} (\gamma_{c})_{\gamma \dgamma} \Lambda^{\dgamma m}
		- \delta^{m}_{l}  \Lambda^\dgamma{}_{j} (\gamma_{c})_{\gamma \dgamma} \Lambda^{\dgamma k})
	\eol & \quad
	- \text{traces}~, \eol[2ex]
\nabla_{\beta j} R(Q)_{a b}{}_{\alpha}{}^{i} &=
	\tfrac{1}{2} \delta^{i}_{j} (\gamma_{c d})_{\alpha \beta}  \, R(M)_{a b}{}^{cd}
	+ \tfrac{1}{4}
		(\gamma^{cd}\gamma_{ab} + \tfrac{1}{3} \gamma_{ab}\gamma^{cd})_{\alpha}{}^\gamma \eps_{\gamma \beta}
		( R(V)_{c d}{}^{i}{}_{j} - \tfrac{i}{2} F_{c d} \delta^{i}_{j} )~, \eol[2ex]
\nabla_{\beta j} R(Q)_{a b}{}^{\dalpha}{}_{i} &=
	\tfrac{1}{4} 
	(\gamma^{cd}\gamma_{ab} \gamma^e + \tfrac{1}{3} \gamma_{ab}\gamma^{cd} \gamma^e)^{\dalpha \gamma}
	\eps_{\gamma \beta} \nabla_{e} T_{c d}{}_{i j}~.
\end{align}

The superfield $D^{ij}{}_{kl}$ is pseudo-real and in the $\rep{20'}$ of SU(4). 
It transforms as
\begin{align}
\nabla_{\alpha m} D^{ij}{}_{kl} &= \delta_m{}^{[i} \,\Big\{
	- 4 \,(\gamma^{c})_{\alpha \dbeta} \nabla_{c}{\chi^{\dbeta}{}_{k l}{}^{j]}}
	+ 2 (\gamma^{a})_{\alpha \dbeta} \chi^{\dbeta}{}_{k l}{}^{r}  
		\Lambda_\dgamma{}^{j]} (\gamma_{a})^{\dgamma \gamma} \Lambda_{\gamma r}
	\eol & \qquad
	+ \veps_{k l r s} (
	   -2 \,E^{j] t} \chi_{\alpha}{}^{rs}{}_{t}
	   + \tfrac{1}{2} (\gamma^{a b} \gamma^c)_{\alpha \dbeta} \,\nabla_{c}{\Lambda^{\dbeta j]}} \,T_{a b}{}^{rs} 
	   - \tfrac{1}{2} (\gamma^{a b} \gamma^c)_{\alpha \dbeta} \,\Lambda^{\dbeta j]}\, \nabla_{c}{T_{a b}{}^{rs}} 
	   \eol & \qquad \qquad
	   + \tfrac{1}{3} E^{j] r} E^{s t} \Lambda_{\alpha t}
	   - \tfrac{2}{3} (\gamma^{c})_{\alpha \dbeta} \bar P_{c} \Lambda^{\dbeta r} E^{j] s}
	   + \tfrac{1}{2} \Lambda_\dbeta{}^{j]} \Lambda^{\dbeta t}\, (\gamma^{a b})_{\alpha}{}^{\beta} T_{a b}{}^{rs} \Lambda_{\beta t}
	   )
	\eol & \qquad
	  + 2 \bar P_{c} (\gamma^{c} \gamma^{ab})_{\alpha \dgamma} \,T_{a b}{}_{k l} \,\Lambda^{\dgamma j]}
	  + \tfrac{1}{3} \Lambda_{\alpha k} E_{l r} \Lambda_\dgamma{}^{j]} \Lambda^{\dgamma r}
	  - \tfrac{1}{3} \Lambda_{\alpha l} E_{k r} \Lambda_\dgamma{}^{j]} \Lambda^{\dgamma r}
	\eol & \qquad
	  - \tfrac{1}{6} (\gamma^{a b} \gamma^c)_{\alpha \dbeta} P_{c} \Lambda^{\dbeta j]}
	          (\gamma_{a b})^{\delta \epsilon} \Lambda_{\delta k} \Lambda_{\epsilon l}
	- 2 \veps^{ j] m r s} T^{a b}{}_{k l} T_{a b}{}_{r s} \Lambda_{\alpha m}
	\Big\}
	- \text{traces}~.
\end{align}
The SU(4) curvature $R(V)_{ab}{}^i{}_j$ and the Lorentz curvature $R(M)_{a b}{}^{cd}$
transform as
\begin{align}
\nabla_{\alpha k} R(V)_{a b}{}^{i}{}_{j} &=
	- 2\, T_{a b k l} \,\chi_{\alpha}{}^{i l}{}_{j} 
	+ \delta^{i}_{k} \, R(S)_{a b}{}_{\alpha j} 
	+ 2 \,(\gamma_{[a})_{\alpha \dalpha} \nabla_{b]} \chi^{\dalpha}{}_{j k}{}^{i}
	- \tfrac{1}{3} E^{i l} T_{a b}{}_{j k} \Lambda_{\alpha l} 
	\eol & \quad
	+ \tfrac{1}{2} \veps_{j k l m} \,E^{i l} R(Q)_{a b}{}_{\alpha}{}^{m} 
	- \tfrac{1}{3} \delta^{i}_{k}  E_{j l} (\gamma_{[a})_{\alpha \dalpha} \nabla_{b]} \Lambda^{\dalpha l}
	- \tfrac{2}{3} T_{a b}{}_{j k} \bar P_{c} (\gamma^{c})_{\alpha \dalpha} \Lambda^{\dalpha i} 
	\eol & \quad
	+ \tfrac{1}{2} \Lambda_{\alpha j} \Lambda_\dbeta{}^{i} R(Q)_{a b}{}^{\dbeta}{}_{k} 
	- \tfrac{1}{2} \Lambda_{\alpha k} \Lambda_{\dbeta}{}^{i} R(Q)_{a b}{}^{\dbeta}{}_{j}
	- \tfrac{1}{3} \delta^{i}_{k} (\gamma_{[a})_{\alpha \dalpha} \nabla_{b]} E_{j l} \Lambda^{\dalpha l} 
	\eol & \quad
	- \tfrac{1}{2} \delta^{i}_{k} \Lambda_{\alpha j} \Lambda_\dbeta{}^{l} R(Q)_{a b}{}^{\dbeta}{}_{l} 
	+ \tfrac{1}{2} \delta^{i}_{k} \Lambda_{\alpha l} \Lambda_\dbeta{}^{l} R(Q)_{a b}{}^{\dbeta}{}_{j} 
	- \tfrac{2}{3} \delta^{i}_{k} (\gamma_{[a} \gamma^c)_{\alpha}{}^\beta \,\Lambda_{\beta j} \,\nabla_{b]}{P_{c}} 
	\eol & \quad
	+ \tfrac{1}{2} \veps_{j k l m} \,(\gamma^{c d}\gamma_{[a})_{\alpha \dalpha} \nabla_{b]} T_{c d}{}^{i m}  \Lambda^{\dalpha l} 
	- \tfrac{1}{8} \veps^{i l m n}\, T_{c d \,j l} T_{e f \,k m} 
		(\gamma_{[a} \gamma^{e f} \gamma^{c d} \gamma_{b]})_\alpha{}^\beta
		\Lambda_{\beta n} 
	\eol & \quad
	- \tfrac{2}{3} \delta^{i}_{k} P_{c} (\gamma_{[a} \gamma^c)_{\alpha}{}^\beta \nabla_{b]} \Lambda_{\beta j}
	- \tfrac{1}{2} \veps_{j k l m} T_{c d}{}^{i l} (\gamma^{c d} \gamma_{[a})_{\alpha \dalpha}  \nabla_{b]}{\Lambda^{\dalpha m}} 
	- \text{trace} ~,\eol[2ex]
\nabla_{\alpha i} R(M)_{abcd} &=
	- \tfrac{1}{4} (\gamma_{a b})_{\alpha}{}^{\beta} R(S)^-_{c d}{}_{\beta i}
	- \tfrac{1}{4} (\gamma_{c d})_{\alpha}{}^{\beta} R(S)^-_{a b}{}_{\beta i}
	\eol & \quad
	+ \tfrac{1}{4} (\gamma^{e} \gamma_{a b})_{\alpha \dbeta} \nabla_{e}{R(Q)_{c d}{}^{\dbeta}{}_{i}}
	+ \tfrac{1}{4} (\gamma^{e} \gamma_{c d})_{\alpha \dbeta} \nabla_{e}{R(Q)_{a b}{}^{\dbeta}{}_{i}}~.
\end{align}
The Lorentz curvature is purely self-dual or anti-self-dual and traceless in its indices, meaning
\begin{align}
R(M)_{a b c d} = \frac{1}{4} (\gamma_{a b})^{\alpha \beta} (\gamma_{c d})^{\gamma\delta} R(M)_{\alpha\beta\gamma\delta} + \HC
\end{align}
where $R(M)_{\alpha\beta\gamma\delta}$ is totally symmetric.

The S-supersymmetry curvature decomposes as
\begin{align}
R(S)_{a b}{}_{\alpha i} &= R(S)^-_{a b}{}_{\alpha i} 
    + (\gamma^{c})_{\alpha \dalpha} \nabla_{c}{R(Q)_{a b}{}^{\dalpha}{}_{i}},
\end{align}
where its anti-self-dual part is gamma-traceless,
\begin{align}
R(S)^-_{a b}{}_{\alpha i}  = \frac{1}{2} (\gamma_{a b})^{\beta \gamma} R(S)_{\gamma\beta\alpha\, i}~.
\end{align}
It transforms as
\begin{align}
\nabla_{\delta j} R(S)^-_{a b}{}_{\alpha i} &=
	\tfrac{1}{4} \eps_{\gamma \delta} 
	(\gamma^{cd} \gamma_{ab} + \tfrac{1}{3} \gamma_{ab} \gamma^{cd})_{\alpha}{}^\gamma
      ( 4 \,\nabla_{c} \nabla^{e} T_{e d}{}_{i j}
        + 2 \,T_{c d}{}^{l k} T^{ef}{}_{i l} T_{ef}{}_{k j} )
	\eol & \quad
	- \tfrac{1}{2} \veps_{i j k l} (\gamma^{c})_{\alpha \ddelta} \nabla_{c} \Lambda^{\ddelta k}  R(Q)_{a b}{}_{\delta}{}^{l}
	- \tfrac{1}{2} \veps_{i j k l}\, \veps_{\delta \alpha}\,
		\Lambda_{\dgamma}{}^{k} (\gamma^{c})^{\dgamma \gamma} 
		\nabla_{c}{R(Q)_{a b}{}_{\gamma}{}^{l}}
	\eol & \quad
	+ \tfrac{1}{8} \veps_{i j k l} 
		(\gamma^{de} \gamma_{a b} \gamma^c)_{\alpha \ddelta}
		\nabla_{c} \Big(\Lambda^{\ddelta k} \, R(Q)_{de}{}_{\delta}{}^{l}\Big)~, \eol[2ex]
\nabla_{\alpha j} R(S)^+_{a b}{}^{\dalpha i} &=
	2\delta_{j}^{i} (\gamma_{d}) _\alpha{}^{\dalpha} \,
	\nabla_{c}{R(M)^+{}_{a b}{}^{c d}} 
	\eol & \quad
	+ \tfrac{1}{4} (\gamma^{cd} \gamma_{ab} \gamma^e + \tfrac{1}{3} \gamma_{ab} \gamma^{cd} \gamma^e)^{\dalpha \beta}
	\eps_{\beta \alpha} (
         \nabla_{e}{R(V)_{c d}{}^{i}{}_{j}} 
         - \tfrac{i}{2} \nabla_{e}{F_{c d}} \delta^{i}_{j}
         \eol & \qquad \qquad
         - 4 \nabla^{f} T_{f e}{}^{i k} T_{c d}{}_{k j}
         - 2 \nabla^{f} T_{c d}{}_{k j} T_{f e}{}^{i k}
     )
	\eol & \quad
	- R(Q)_{a b}{}^{\dalpha}{}_{k} \chi_{\alpha}{}^{i k}{}_{j}
	- \tfrac{1}{6} E^{i k} \Lambda_{\alpha k} R(Q)_{a b}{}^{\dalpha}{}_{j} 
	+ \tfrac{1}{6} \delta^{i}_{j} E^{k l} \Lambda_{\alpha k} R(Q)_{a b}{}^{\dalpha}{}_{l} 
	\eol & \quad
	- \tfrac{1}{3} (\gamma^{c})_{\alpha \dbeta} \bar P_{c} \Lambda^{\dbeta i} R(Q)_{a b}{}^{\dalpha}{}_{j} 
	+ \tfrac{1}{3} \delta^{i}_{j} (\gamma^{c})_{\alpha \dbeta} \bar P_{c} \Lambda^{\dbeta k} R(Q)_{a b}{}^{\dalpha}{}_{k} 
	\eol & \quad
	+ \tfrac{1}{3} \veps^{i k l m} \, (\gamma_{a b})^{\dalpha}{}_{\dbeta} \, T^{c d}{}_{j k} \Lambda_{\alpha l} R(Q)_{c d}{}^{\dbeta}{}_{m} 
	\eol & \quad
	- \tfrac{1}{8} \veps^{i k l m}\, (\gamma_{a b})^{\ddelta}{}_{\dbeta} 
		(\gamma^{c d})_{\dgamma \ddelta} (\gamma^{e f})^{\dbeta \dgamma} \,\, T_{c d}{}_{j k} \Lambda_{\alpha l} R(Q)_{e f}{}^{\dalpha}{}_{m} ~.
\end{align}

For reference, we give the supersymmetry transformations of $P_a$,
\begin{align}
\nabla_{\alpha i} P_a &= 
	- \tfrac{1}{4} (\gamma_a \gamma^{bc})_{\alpha \dbeta} \Lambda^{\dbeta j} \,T_{b c}{}_{j i} ~, \qquad
\nabla_{\alpha i} \bar P_{a} =
	- \nabla_{a}{\Lambda_{\alpha i}} 
	- \tfrac{1}{2} \Lambda_{\alpha i} \, \Lambda_\dbeta{}^{j} (\gamma_{a})^{\dbeta \beta} \Lambda_{\beta j}~.
\end{align}
The $S$-supersymmetry transformations of all the independent fields are 
\begin{align}
S_{\alpha}{}^k E_{i j} &= 2 \delta^{k}_{(i} \Lambda_{\alpha j)}~, \eol
S_{\beta}{}^k T_{a b}{}^{i j} &= -\tfrac{1}{4} \veps^{i j k l} (\gamma_{a b})_{\beta}{}^{\gamma} \Lambda_{\gamma l}~, \eol
S_{\beta}{}^l \chi_{\alpha}{}^{i j}{}_{k} &=
    - \tfrac{1}{2} (\gamma^{a b})_{\alpha \beta} \delta_{k}^{l} T_{a b}{}^{i j} 
    - \tfrac{1}{3} (\gamma^{a b})_{\alpha \beta} \delta^{[i}_{k} T_{a b}{}^{j] l} 
    - \tfrac{1}{2} \eps_{\alpha \beta} \veps^{i j m l} E_{k m} ~, \eol
S_{\beta}{}^l  \chi^{\dalpha}{}_{i j}{}^{k} &=
    \tfrac{1}{2} \delta_{[i}^{l} \Lambda_\beta{}_{j]}  \Lambda^\dalpha{}^{k} 
    - \text{traces}~, \eol
S_{\beta}{}^k T_{a b}{}_{i j} &= 0~, \qquad\qquad
S_{\alpha}{}^k E^{i j} = 0~, \qquad\qquad
S_{\alpha}{}^m D^{ij}{}_{kl} = 0~,
\end{align}
while the transformations of the curvatures are
\begin{align}
S_{\alpha}{}^i P_a &= \tfrac{1}{2} (\gamma_{a})_{\alpha \dalpha} \Lambda^{\dalpha i}~, \qquad\qquad
S_{\alpha}{}^i \bar P_a = 0~, \eol[2ex]
S_{\beta}{}^j R(Q)_{a b}{}_{\alpha}{}^{i} &=
	\tfrac{1}{2} (\gamma_{a b} \gamma^{c d}
      + \tfrac{1}{3} \gamma^{c d} \gamma_{a b})_{\beta}{}^\delta \eps_{\alpha \delta} T_{c d}{}^{i j},\qquad
S_{\beta}{}^j R(Q)_{a b}{}^{\dalpha}{}_{i} = 0 ~, \eol[2ex]
S_{\alpha}{}^k R(V)_{a b}{}^i{}_j &=
	- \delta^{k}_{j} R(Q)_{a b}{}_{\alpha}{}^{i}
	+ \veps^{i k l m} T_{a b}{}_{j l} \Lambda_{\alpha m}
	+ (\gamma_{a b})_{\alpha}{}^{\beta} \chi_{\beta}{}^{k i}{}_{j}
	\eol & \quad
	+ \tfrac{1}{6} \delta^{k}_{j} (\gamma_{a b})_{\alpha}{}^{\beta} \Big(
		2 \bar P_{c} (\gamma^{c})_{\beta \dbeta} \Lambda^{\dbeta i}
		+ E^{i l} \Lambda_{\beta l}
    \Big)
    - \text{trace}~, \eol[2ex]
S_{\alpha}{}^i R(M)_{a b c d} &=
     - \tfrac{3}{4} (\gamma_{a b})_{\alpha}{}^{\beta} R(Q)_{c d}{}_{\beta}{}^{i}
     - \tfrac{3}{4} (\gamma_{c d})_{\alpha}{}^{\beta} R(Q)_{a b}{}_{\beta}{}^{i}~, \eol[2ex]
S_{\alpha}{}^j R(S)^+_{a b}{}^{\dalpha}{}^{i} &=
     \tfrac{3}{2} \veps^{i j k l} \Lambda_{\alpha k} R(Q)_{a b}{}^{\dalpha}{}_{l}~, \eol
S_{\beta}{}^j R(S)^-_{a b}{}_{\alpha}{}_{i} &=
     - \tfrac{3}{4} (\gamma^{c d} \gamma_{a b} + \tfrac{1}{3} \gamma_{a b} \gamma^{cd})_{\alpha}{}^\gamma \eps_{\gamma \beta} 
           (R(V)_{c d}{}^{j}{}_{i} - \tfrac{i}{2} F_{c d} \delta^{j}_{i})
     \eol & \quad
     + \tfrac{1}{2} \delta^{j}_{i} (\gamma^{c d})_{\alpha \beta} R(M)_{a b c d}
\end{align}
The $K$-curvature $R(K)$ is given by
\begin{align}
R(K)_{a b}{}^c  &= - \nabla_{d} R(M)_{a b}{}^{c d}~.
\end{align}

In analyzing the superspace Bianchi identities, we have corrected some minor
typos that have appeared earlier in the literature. 
In \cite{Bergshoeff:1980is}, the definition of $a_\mu$ in eq. (4.8)
should have $-1/4$ for the coefficient of the fermion bilinear rather than $-1/2$.
Also in eq. (4.13), $\delta_Q R(Q)_{ab}{}^i$, the ``+ h.c.'' appearing in the $\Lambda$
bilinear should be ``-h.c.''
A few minor typos in \cite{Ciceri:2015qpa} have also been corrected. In eq. (2.8),
the sign of $\delta_S \bar P_a$ was incorrect. In eq. (B.2),
$\delta_Q R(S)$ and $\delta_S R(S)$ did not include terms quadratic in fermions.

\section{Analysis of the Bianchi identities}\label{app:BianchiAnalysis}

\subsection{Higher Bianchi identities from SUSY closure: an explicit example}
Below we use the closure of the superconformal algebra to derive the full supersymmetry transformations of the fields $\rho$ and
$\kappa$ of the abstract multiplet defining the action principle. The resulting transformations can be
shown to obey the various supersymmetry constraints we encountered in section \ref{sec:2.2}.

Let us start with $\nabla_{\beta m} \rho_{\alpha i j}{}^k$. This decomposes
into a singlet and triplet of the left-handed part of the Lorentz group
and into the $\rep{6}$, $\rep{10}$ and $\rep{64}$ of $\SU(4)$.
Schematically,
\begin{align}
\nabla_{\beta m} \rho_{\alpha i j}{}^k
	&= (\rep{6}, \rep{1}) + (\rep{10}, \rep{1}) + (\rep{64}, \rep{1})
	+ (\rep{6}, \rep{3}) + (\rep{10}, \rep{3}) + (\rep{64}, \rep{3})~.
\end{align}
Now observe that the condition for closure of $\nabla_{\alpha i}$ on $C^{ij}{}_{kl}$ decomposes as
\begin{align}
\{\nabla_{\beta r}, \nabla_{\alpha s}\} C^{ij}{}_{kl}
	&= \Big((\rep{6}, \rep{1}) + (\brep{10}, \rep{3}) \Big) \otimes (\rep{20'}, \rep{1}) \eol
	&= (\rep{6}, \rep{1}) + (\rep{50}, \rep{1}) + (\rep{64}, \rep{1})
	+ (\rep{10}, \rep{3}) + (\rep{64}, \rep{3}) + (\brep{126}, \rep{3})~.
\end{align}
When one actually computes the explicit terms in
$\{\nabla_{\beta r}, \nabla_{\alpha s}\} C^{ij}{}_{kl}$,
the leading terms will be of the form $\nabla_{\beta m} \rho_{\alpha i j}{}^k$.
The terms therein corresponding to the $(\rep{10}, \rep{1})$ and the $(\rep{6}, \rep{3})$
are undetermined by supersymmetry -- they involve the new fields
$\cE^-_{ab}{}^{ij}$ and $\cE^{ij}$.
All the others are determined.
Solving for these and recombining all the various representations gives:
\begin{align}
\nabla_{\beta l} \rho_{\alpha i j}{}^k
	&= 
	- \frac{1}{16} \cE_{a b}{}^{m n} \delta^{k}_{l} (\gamma^{ab})_{\alpha \beta} \veps_{i j m n}
	+ \frac{1}{2} \cE^{k m} \eps_{\alpha \beta} \veps_{l i j m}
	- \frac{8}{3} A^{k m}{}_{i j} \cE_{l m} \eps_{\alpha \beta}
	+ \frac{8}{3} A^{k m}{}_{l [i} \cE_{j] m} \eps_{\alpha \beta}
	\eol & \quad
	- \frac{4}{3} A^{k m}{}_{i j} T_{a b}{}^{n p} (\gamma^{ab})_{\alpha \beta} \veps_{l m n p}
	+ \frac{4}{3} A^{k m}{}_{l [i} T_{a b}{}^{n p} (\gamma^{ab})_{\alpha \beta} \veps_{j] m n p}
	\eol & \quad
	+ \frac{4}{15} A^{m n}{}_{i j} T_{a b}{}^{p q} \delta^{k}_{l} (\gamma^{ab})_{\alpha \beta} \veps_{m n p q}
	+ \frac{2}{5} A^{m n}{}_{p q} T_{a b}{}^{p q} \delta^{k}_{l} (\gamma^{ab})_{\alpha \beta} \veps_{i j m n}
	\eol & \quad
	+ \frac{4}{3} C^{m n}{}_{i j} \cE^{k p} \eps_{\alpha \beta} \veps_{l m n p}
	- \frac{4}{3} C^{m n}{}_{l [i} \cE^{k p} \eps_{\alpha \beta} \veps_{j] m n p}
	+ \frac{4}{15} C^{m n}{}_{p[i} \veps_{j] m n q} \cE^{p q} \delta^{k}_{l} \eps_{\alpha \beta} 
	\eol & \quad
	- \frac{14}{3} \bar C^{k m}{}_{i j} \Lambda_{\alpha l} \Lambda_{\beta m}
	- \frac{2}{3} \bar C^{k m}{}_{i j} \Lambda_{\alpha m} \Lambda_{\beta l}
	- \frac{4}{3} \bar C^{k m}{}_{l [i} \Lambda_{\alpha j]} \Lambda_{\beta m}
	+ \frac{20}{3} \bar C^{k m}{}_{l [i|} \Lambda_{\alpha m} \Lambda_{\beta |j]}
	\eol & \quad
	+ 2 \bar C^{m n}{}_{i j} \Lambda_{\alpha m} \Lambda_{\beta n} \delta^{k}_{l}
	- \frac{2}{3} \Lambda_{\alpha [i|} \kappa{}_{\beta l |j]}{}^{k}
	- \frac{2}{3} \Lambda_{\alpha l} \kappa{}_{\beta i j}{}^{k}
	+ \frac{8}{3} \Lambda_{\beta [i|} \kappa{}_{\alpha l |j]}{}^{k}
	+ \frac{2}{3} \Lambda_{\beta l} \kappa{}_{\alpha i j}{}^{k}
	\eol & \quad
	+ \frac{1}{3} \Lambda_{\alpha m} \delta^{k}_{l} \kappa{}_{\beta i j}{}^{m}
	- \frac{1}{3} \Lambda_{\beta m} \delta^{k}_{l} \kappa{}_{\alpha i j}{}^{m}
	+ \frac{1}{3} \Lambda^{\dalpha k} \eps_{\alpha \beta} \eps_{\dalpha \dbeta} \Upsilon{}^{\dbeta m n}{}_{[i} \veps_{j] l m n}
	\eol & \quad
	- \frac{1}{3} \Lambda^{\dalpha k} \eps_{\alpha \beta} \eps_{\dalpha \dbeta} \Upsilon{}^{\dbeta m n}{}_{l} \veps_{i j m n}
	- \frac{1}{15} \Lambda^{\dalpha m} \delta^{k}_{l} \eps_{\alpha \beta} \eps_{\dalpha \dbeta} \Upsilon{}^{\dbeta n p}{}_{[i} \veps_{j] m n p}
	\eol & \quad
	+ \frac{13}{60} \Lambda^{\dalpha m} \delta^{k}_{l} \eps_{\alpha \beta} \eps_{\dalpha \dbeta} \Upsilon{}^{\dbeta n p}{}_{m} \veps_{i j n p}
	-\text{traces}\;({}^k{}_{ij})\,.
\end{align}
An interesting feature is that $\Upsilon{}^{\dalpha i j}{}_k$
explicitly appears, even though this fermion does not appear in the action principle. This is acceptable because when invariance of the action principle is checked, these
terms cancel against similar terms in the spinor derivative of $\kappa^{\dalpha i j}{}_k$.

Next, let us analyze  $\bar\nabla^{\dbeta l} \rho_{\alpha i j}{}^k$. This is
in a vector representation of the Lorentz group, so let us focus just
on the $\SU(4)$ group structure:
$\rep{4} \times \brep{20} = \rep{15} + \rep{20'} + \rep{45}$.
To evaluate this, we use
$\{\nabla_{\alpha i}, \bar\nabla^{\dbeta j}\} C^{rs}{}_{kl}$, which is
generically in the
$\rep{15} + \rep{20'} + \rep{20'} + \rep{45} + \brep{45} + \rep{175}$.
We have already checked that the $\rep{45}$ gets set to zero.
The $\rep{20'}$ can be solved straightforwardly. The $\rep{15}$ is complicated:
it gives a linear combination between
the $\rep{15}$ of $\bar\nabla^{\dbeta l} \rho_{\alpha i j}{}^k$
and the $\rep{15}$ of 
$\nabla_{\alpha i} \Upsilon{}^{\dbeta k l}{}_j$. In other words,
we must introduce a new field in the $\rep{15}$ into which
$\rho_{\alpha i j}{}^k$ transforms and also declare that is the same
$\rep{15}$ (up to additional terms) into which $\Upsilon{}^{\dbeta i j}{}_k$ transforms.
Schematically, this leads to
\begin{align}
\bar \nabla^{\dbeta l} \rho_{\alpha i j}{}^k
	&= 2 \Lambda_{\alpha [i} \kappa{}^{\dbeta l k}{}_{j]}
	+ \delta^{l}_{[i|} \Lambda_{\alpha m} \kappa{}^{\dbeta k m}{}_{|j]}
	- 4 (\gamma^{a})_\alpha{}^{\dbeta} \nabla_{a}{C^{l k}{}_{i j}}
	+ \rep{15}
	\eol & \quad
	+ \frac{8}{3} C^{k m}{}_{i j} \Lambda_{\alpha m} \Lambda^{\dbeta l}
	- \frac{8}{3} C^{l m}{}_{i j} \Lambda_{\alpha m} \Lambda^{\dbeta k}
	+ \frac{8}{3} C^{l k}{}_{i j} \Lambda_{\alpha m} \Lambda^{\dbeta m}
	\eol & \quad
	- \frac{16}{3} C^{l k}{}_{[i |m} \Lambda_{\alpha |j]} \Lambda^{\dbeta m}
	+ \frac{16}{3} \delta^{l}_{[i} C^{k n}{}_{j] m} \Lambda_{\alpha n} \Lambda^{\dbeta m}	
	-\text{traces}\;({}^k{}_{ij})\,.
\end{align}
We have left the $\rep{15}$ unspecified. Note that the two terms
involving $\delta^{l}_{i}$ are actually in the $\rep{15}$ and could be absorbed into this
term. They correspond to traces that are subtracted out from other terms, when the
$\rep{20'}$ was built.
Since this $\rep{15}$ does not appear in the action, we will not worry about
how to precisely define it.

Now let us analyze the supersymmetry transformation of $\kappa{}^{\dalpha i j}{}_k$.
The $\bar\nabla^{\dalpha i}$ transformation is very similar to the calculation of $\nabla_{\alpha i}$ on
$\rho$. Decomposing under $\SU(4)$ and the right-handed part of the Lorentz group gives
\begin{align}
\bar\nabla^{\dbeta m} \bar\kappa{}^{\dalpha i j}{}_k
	&= (\rep{6}, \rep{1}) + (\brep{10}, \rep{1}) + (\rep{64}, \rep{1})
	+ (\rep{6}, \rep{3}) + (\brep{10}, \rep{3}) + (\rep{64}, \rep{3})
\end{align}
and
\begin{align}
\{\bar\nabla^{\dbeta i}, \bar\nabla^{\dalpha j}\} A^{rs}{}_{kl}
	&= \Big((\rep{6}, \rep{1}) + (\rep{10}, \rep{3}) \Big) \otimes (\rep{20'}, \rep{1}) \eol
	&= (\rep{6}, \rep{1}) + (\rep{50}, \rep{1}) + (\rep{64}, \rep{1})
	+ (\brep{10}, \rep{3}) + (\rep{64}, \rep{3}) + (\rep{126}, \rep{3})~.
\end{align}
The calculation is essentially identical to the prior one. The difference is that while we determine
the $(\rep{6}, \rep{3})$ to be essentially defined as the self-dual part of $\cE_{a b}{}^{i j}$,
the $(\brep{10}, \rep{1})$ does not appear in the action. We denote it $X_{i j}$.
We find
\begin{align}
\bar\nabla^{\dbeta m} \bar\kappa{}^{\dalpha i j}{}_k
	&=
	- \frac{3}{16} \delta^{m}_{k} \cE_{a b}{}^{i j} (\gamma^{a b})^ {\dalpha \dbeta}
	- X_{k l} \eps^{\dalpha \dbeta} \veps^{m l i j}
	+ \frac{8}{3} C^{i j}{}_{l k} E^{m l} \eps^{\dalpha \dbeta}
	- \frac{8}{3} C^{m [i}{}_{l k} E^{j]l} \eps^{\dalpha \dbeta}
	\eol & \quad
	+ 2 A^{i j}{}_{rs} E_{k p} \eps^{\dalpha \dbeta} \veps^{rsmp}
	+ 2 C^{i j}{}_{rs} T^{ab}{}_{k p} (\gamma_{a b})^{\dalpha \dbeta} \veps^{rs mp}
	+ \frac{1}{2} \delta^{m}_{k} C^{i j}{}_{rs} T^{ab}{}_{n p} (\gamma_{a b})^{\dalpha \dbeta} \veps^{rs n p}
	\eol & \quad
	- 2 \bar C^{i j}{}_{rs} \Lambda_{\alpha k} \Lambda_{\beta p} \eps^{\alpha \beta} \eps^{\dalpha \dbeta} \veps^{rs m p}
	- \frac{7}{6} \Lambda^{\dalpha m} \Upsilon{}^{\dbeta i j}{}_{k}
	- \frac{1}{3} \Lambda^{\dalpha [i} \Upsilon{}^{\dbeta j] m}{}_{k}
	+ \frac{5}{6} \delta^{m}_{k} \Lambda^{\dalpha p} \Upsilon{}^{\dbeta i j}{}_{p}
	\eol & \quad
	+ \frac{1}{6} \Lambda^{\dbeta m} \Upsilon{}^{\dalpha i j}{}_{k}
	- \frac{5}{3} \Lambda^{\dbeta [i} \Upsilon{}^{\dalpha j]m}{}_{k}
	- \frac{1}{3} \delta^{m}_{k} \Lambda^{\dbeta p} \Upsilon{}^{\dalpha i j}{}_{p}
	- \frac{1}{2} \Lambda_{\alpha k} \eps^{\alpha \beta} \eps^{\dalpha \dbeta} \kappa_{\beta rs}{}^{m} \veps^{i j rs}
	\eol & \quad
	+ \frac{1}{4} \delta^{m}_{k} \Lambda_{\alpha p} \eps^{\alpha \beta} \eps^{\dalpha \dbeta} \kappa_{\beta rs}{}^{p} \veps^{i j rs}
	-\text{traces}\;({}^{ij}{}_k)\,.
\end{align}

Now we analyze $\nabla_{\alpha i}$ on $\bar\kappa$. The steps are quite similar again.
We expect something in the
$\brep{4} \times \rep{20} = \rep{15} + \rep{20'} + \brep{45}$.
To evaluate this, we use
$\{\nabla_{\alpha i}, \bar\nabla^{\dbeta j}\} A^{rs}{}_{kl}$,
generically in the
$\rep{15} + \rep{20'} + \rep{20'} + \rep{45} + \brep{45} + \rep{175}$,
which is self-conjugate.
We will be able to fix the $\brep{45}$ in this way
and presumably the $\rep{20'}$ since we have two $\rep{20'}$ identities
to use between $\nabla_{\alpha i} \bar\kappa$ and $\nabla^{\dalpha i} \kappa$.
Only one real combination of the $\rep{15}$ should be fixed
between $\nabla_{\alpha i} \bar\kappa$ and $\nabla^{\dalpha i} \kappa$.
The remaining $\rep{15}$ should be the new field $\cE_a{}^i{}_j$.
The calculation leads to
\begin{align}
\nabla_{\alpha l} \kappa{}^{\dalpha \,i j}{}_k
	&=
	- \frac{3i}{8} \delta^{[i}{}_{l} \cE_{a}{}^{j]}{}_{k}  (\gamma^{a})_{\alpha}{}^{\dalpha}
	+ \frac{i}{8} \delta^{[i}{}_{k} \cE_{a}{}^{j]}{}_{l} (\gamma^{a})_{\alpha}{}^{\dalpha}
	- 4 (\gamma^{a})_{\alpha}{}^{\dalpha} \nabla_{a}{A^{i j}{}_{l k}}
	\eol & \quad
	+ 4 A^{i j}{}_{l k} \Lambda_{\alpha m} \Lambda^{\dalpha m}
	+ 4 C^{i j}{}_{l k} P_{a} (\gamma^{a})_{\alpha}{}^{\dalpha}
	- 4 \bar C^{i j}{}_{l k} \bar P_{a} (\gamma^{a})_{\alpha}{}^{\dalpha}
	\eol & \quad
	+ \Lambda_{\alpha l} \rho^{\dalpha i j}{}_{k}
	- \delta^{[i}_{k} \Lambda_{\alpha m} \rho^{\dalpha j] m}{}_{l}
	+ 3 \delta^{[i}_{l} \Lambda_{\alpha m} \rho^{\dalpha j] m}{}_{k}
	\eol & \quad
	- 2 \Lambda^{\dalpha [i} \rho_{\alpha k l}{}^{j]}
	- \Lambda^{\dalpha m} \delta^{[i}_{k} \rho_{\alpha l m}{}^{j]}
	+  \Lambda^{\dalpha m} \delta^{[i}_{l} \rho_{\alpha k m}{}^{j]}~, \\
\nabla^{\dalpha l} \kappa{}_{\alpha \,i j}{}^k
	&=
	\frac{3i}{8} \delta_{[i}{}^{l} \cE_{a}{}^{k}{}_{j]}  (\gamma^{a})_{\alpha}{}^{\dalpha}
	- \frac{i}{8} \delta_{[i}{}^{k} \cE_{a}{}^{l}{}_{j]} (\gamma^{a})_{\alpha}{}^{\dalpha}
	- 4 (\gamma^{a})_{\alpha}{}^{\dalpha} \nabla_{a}{A^{l k}{}_{i j}}
	\eol & \quad
	+ 4 A^{l k}{}_{i j} \Lambda^{\dalpha m} \Lambda_{\alpha m}
	+ 4 \bar C^{l k}{}_{i j} \bar P^{a} (\gamma_{a})_{\alpha}{}^{\dalpha}
	- 4 C^{l k}{}_{i j} P^{a} (\gamma_{a})_{\alpha}{}^{\dalpha}
	\eol & \quad
	+ \Lambda^{\dalpha l} \rho_{\alpha i j}{}^{k}
	- \delta_{[i}^{k} \Lambda^{\dalpha m} \rho_{\alpha j] m}{}^{l}
	+ 3 \delta_{[i}^{l} \Lambda^{\dalpha m} \rho_{\alpha j] m}{}^{k}
	\eol & \quad
	- 2 \Lambda_{\alpha [i} \rho^{\dalpha k l}{}_{j]}
	- \Lambda_{\alpha m} \delta_{[i}^{k} \rho^{\dalpha l m}{}_{j]}
	+  \Lambda_{\alpha m} \delta_{[i}^{l} \rho^{\dalpha k m}{}_{j]}~.
\end{align}
One can proceed in a similar way to compute the full supersymmetry transformations
of the fields appearing in the action principle using closure of the algebra.

However, the explicit form of the supersymmetry transformations is not particularly useful.
Aside from the relations \eqref{eq:tr1} that define the higher dimension components from the lower ones, the conditions for supersymmetric invariance must be direct consequences of the basic constraints on $C^{ij}{}_{kl}$ and $A^{ij}{}_{kl}$. To see this without explicitly building the transformations, we employ
a standard technique in superspace: the Bianchi identity of the Bianchi identity. 
We discuss this below.

\subsection{The Bianchi identity of the Bianchi identity}
It is commonly the case in superspace that checking Bianchi identities
involves imposing only a few independent constraints. Here we will demonstrate this by showing that the higher dimension constraints
are automatically satisfied once the lowest ones are fulfilled.

Let $J$ be a gauge-invariant super-four-form, that is a scalar under Lorentz
and $R$-symmetry transformations as well as a conformal primary.
It follows that $I \equiv \nabla J = \rd J$ is a gauge-invariant super-five-form,
and its various components correspond to the Bianchi identities that we wish to check.
Our goal is to show that if the lowest dimension components of $I$ vanish,
the higher ones necessarily do as well.
The key to this computation is to exploit that $I$ must itself be closed,
$\nabla I = \rd I = \rd^2 J = 0$. The latter equation is the Bianchi identity
of the Bianchi identity.

Our starting assumption is that $I_{\psi^5} = 0$. This is the basic supersymmetry constraint \eqref{eq:Bigconstraint}. One easily sees that
\begin{align}
0 &= (\nabla I)_{\psi_L \psi_R^5} = t_0 I_{e \psi_R^4}~, \eol
0 &= (\nabla I)_{\psi_L^2 \psi_R^4} = t_0 I_{e \psi_L \psi_R^3}~, \eol
0 &= (\nabla I)_{\psi_L^3 \psi_R^3} = t_0 I_{e \psi_L^2 \psi_R^2}~,
\end{align}
with the others following from complex conjugation. The key ingredient
is that for $\cN=4$ supersymmetry, the $t_0$ cohomology is
(almost) empty. That is, there are no forms that are $t_0$ closed that are not also
$t_0$ exact, except for terms that only contain gravitini
(so that $t_0$ immediately annihilates it).
This is a technical proof found in \cite{Brandt:2010tz}.
Now $I_{e \psi_R^4}$ clearly is not $t_0$ exact as it possesses gravitini
of only one chirality; therefore, it must vanish. $I_{e \psi_L \psi_R^3}$
and $I_{e \psi_L^2 \psi_R^2}$ may both be $t_0$ exact. However, if they are
$t_0$ exact, one can always choose a different $J$ to make them vanish.
That is, $I_{e \psi_L \psi_R^3}$ involves a term $t_0 J_{e \psi_R^2}$
and $J_{e \psi_R^2}$ can be chosen to eliminate the $t_0$-exact piece
$I_{e \psi_L \psi_R^3}$. In fact, this is what we did when we 
solved Bianchi identities to this order: we used the $t_0$-exact piece to
determine the next part of $J$ and required all the other pieces to vanish. The same argument applies to $I_{e \psi_L^2 \psi_R^2}$, and
one concludes that $I_{e \psi^4}$ must vanish. Iterating this argument ultimately leads to the conclusion that $I=0$, and therefore $J$ is closed.

\bibliography{library.bib}
\bibliographystyle{utphys_mod}

\end{document}